\documentclass[11pt]{article}
\def\hybrid{\topmargin 0pt      \oddsidemargin 0pt
        \headheight 0pt \headsep 0pt
       \voffset-1cm
       \hoffset=-0.15in
         \textwidth 6.5in
       \textheight 9.5in       % A4 paper
        \marginparwidth 0.0in
        \parskip 5pt plus 1pt   \jot = 1.5ex}
\catcode`\@=11
\def\marginnote#1{}

\newcount\hour
\newcount\minute
\newtoks\amorpm
\hour=\time\divide\hour by60
\minute=\time{\multiply\hour by60 \global\advance\minute by-\hour}
\edef\standardtime{{\ifnum\hour<12 \global\amorpm={am}%
        \else\global\amorpm={pm}\advance\hour by-12 \fi
        \ifnum\hour=0 \hour=12 \fi
        \number\hour:\ifnum\minute<10 0\fi\number\minute\the\amorpm}}
\edef\militarytime{\number\hour:\ifnum\minute<10 0\fi\number\minute}

\def\draftlabel#1{{\@bsphack\if@filesw {\let\thepage\relax
   \xdef\@gtempa{\write\@auxout{\string
      \newlabel{#1}{{\@currentlabel}{\thepage}}}}}\@gtempa
   \if@nobreak \ifvmode\nobreak\fi\fi\fi\@esphack}
        \gdef\@eqnlabel{#1}}
\def\@eqnlabel{}
\def\@vacuum{}
\def\draftmarginnote#1{\marginpar{\raggedright\scriptsize\tt#1}}

\def\draftlabel#1{{\@bsphack\if@filesw {\let\thepage\relax
   \xdef\@gtempa{\write\@auxout{\string
      \newlabel{#1}{{\@currentlabel}{\thepage}}}}}\@gtempa
   \if@nobreak \ifvmode\nobreak\fi\fi\fi\@esphack}
        \gdef\@eqnlabel{#1}}
\def\@eqnlabel{}
\def\@vacuum{}
\def\draftmarginnote#1{\marginpar{\raggedright\scriptsize\tt#1}}

\def\draft{\oddsidemargin -.5truein
        \def\@oddfoot{\sl preliminary draft \hfil
        \rm\thepage\hfil\sl\today\quad\militarytime}
        \let\@evenfoot\@oddfoot \overfullrule 3pt
        \let\label=\draftlabel
        \let\marginnote=\draftmarginnote
   \def\@eqnnum{(\theequation)\rlap{\kern\marginparsep\tt\@eqnlabel}%
\global\let\@eqnlabel\@vacuum}  }

\def\numberbysection{\@addtoreset{equation}{section}
        \def\theequation{\thesection.\arabic{equation}}}

\def\underline#1{\relax\ifmmode\@@underline#1\else
        $\@@underline{\hbox{#1}}$\relax\fi}

\def\titlepage{\@restonecolfalse\if@twocolumn\@restonecoltrue\onecolumn
     \else \newpage \fi \thispagestyle{empty}\c@page\z@
        \def\thefootnote{\fnsymbol{footnote}} }

\def\endtitlepage{\if@restonecol\twocolumn \else  \fi
        \def\thefootnote{\arabic{footnote}}
        \setcounter{footnote}{0}}  %\c@footnote\z@ }
\relax

\numberbysection
\hybrid

\newfont{\Bbb}{msbm10 scaled 1\@ptsize00}
\newfont{\Bbbb}{msbm7 scaled 1\@ptsize00}
\newcommand{\CC}{\mbox{\Bbb C}}

\newcommand{\DDD}{\raise-1pt\hbox{$\mbox{\Bbbb D}$}}

\newcommand{\UUU}{\raise-1pt\hbox{$\mbox{\Bbbb U}$}}

\newcommand{\ZZ}{\mbox{\Bbb Z}}
\newcommand{\z}{\raise-1pt\hbox{$\mbox{\Bbbb Z}$}}

\def\beq{\begin{equation}}
\def\eeq{\end{equation}}
\def\p{\partial}

\newtheorem{lemma-definition}{Lemma-Definition}[section]

\newcommand{\tr}{{\rm tr}}
\newcommand{\ti}[1]{\tilde{#1}}

\newcommand{\vf}{\varphi}
\newcommand{\al}{\alpha}
\newcommand{\be}{\beta}

\newcommand{\om}{\omega}
\newcommand{\vth}{\vartheta}

\newcommand{\bfe}{{\bf{e}}}

\newcommand{\mL}{{\mathcal L}}

\newcommand{\Mat}{ {\rm Mat}(N,\CC) }

\def\res{\mathop{\hbox{Res}}\limits}

\begin{document}

\begin{center}

\

\vspace{8mm}

{\LARGE{ Field analogue of the Ruijsenaars-Schneider model  }}

 \vspace{18mm}

{\Large {A. Zabrodin}}\,\footnote{Skolkovo Institute of Science and Technology,
143026, Moscow, Russia;
National Research University Higher School of Economics,
20 Myasnitskaya Ulitsa,
Moscow 101000, Russia and
ITEP NRC KI, 25
B.Cheremushkinskaya, Moscow 117218, Russia;
e-mail: zabrodin@itep.ru
}
 \qquad\qquad\quad
 {\Large {A. Zotov}}\,\footnote{Steklov Mathematical Institute of Russian
Academy of Sciences, Gubkina str. 8, 119991, Moscow, Russia;
Institute for Theoretical and Mathematical Physics, Lomonosov Moscow State University, Moscow, 119991, Russia;
 %Institute for Theoretical and Experimental Physics of NRC ''Kurchatov Institute'',
 %B.Cheremushkinskaya 25, Moscow 117218, Russia;
 % National Research
 % University Higher School of Economics, Russian Federation;
%  Moscow
% Institute of Physics and Technology, Inststitutskii per.  9,
% Dolgoprudny, Moscow region, 141700, Russia;
 e-mail: zotov@mi-ras.ru.}

\end{center}

\vspace{8mm}

%\begin{titlepage}

%\title{Elliptic families of solutions to the 2D Toda equation and the field
%elliptic Ruijsenaars-Schneider system}

%\title{Field analogue of the Ruijsenaars-Schneider model}

%\author{V.~Akhmedova\thanks{National Research University Higher School of
%Economics,
%20 Myasnitskaya Ulitsa, Moscow 101000, Russia,
%e-mail: valeria-58@yandex.ru}
%\and T.~Takebe
%\thanks{
%National Research University Higher School of Economics,
%20 Myasnitskaya Ulitsa,
%Moscow 101000, Russia, e-mail: ttakebe@hse.ru}
%\and A.~Zabrodin
%\thanks{
%National Research University Higher School of Economics,
%20 Myasnitskaya Ulitsa,
%Moscow 101000, Russia and
%ITEP, 25
%B.Cheremushkinskaya, Moscow 117218, Russia
%e-mail: zabrodin@itep.ru}}

%\date{April 2021}
%\maketitle

\vspace{-65mm} \centerline{ \hfill ITEP-TH-17/21}\vspace{65mm}

\begin{abstract}

We suggest a field extension of the classical elliptic Ruijsenaars-Schneider model.
The model is defined in two different ways which lead to the same result.
The first one is via the trace of a chain product of $L$-matrices
which allows one to introduce the Hamiltonian of the model and to show that the
model is gauge equivalent to a classical elliptic spin chain. In this way, one obtains
a lattice field analogue of the Ruijsenaars-Schneider model with continuous time.
The second method is based on investigation of general elliptic families of solutions
to the 2D Toda equation. We derive equations of motion for their poles, which
turn out to be difference equations in space with a lattice spacing
$\eta$, together with a zero curvature representation for them. We also
show that the equations of motion are Hamiltonian.
The obtained system of equations can be naturally regarded as
a field generalization of the Ruijsenaars-Schneider system.
Its lattice version coincides with the model
introduced via the first method. The limit $\eta \to 0$ is shown
to give the field extension of the Calogero-Moser model known in the literature.
The fully discrete version of this construction is also discussed.

\end{abstract}

%\end{titlepage}

%\vspace{5mm}

%

%\newpage
\tableofcontents

%\vspace{5mm}

%\newpage
\section{Introduction}
\setcounter{equation}{0}

Our main purpose in this paper is to introduce a (1+1)-dimensional field theory
generalization of the elliptic $N$-body Ruijsenaars-Schneider model
\cite{RS86,Ruijs12} which is usually regarded as a relativistic extension of the
Calogero-Moser system. This is done
in two different ways, so the paper consists of two main parts.
In the first part (sections 2--4) we define a discrete space
classical Ruijsenaars-Schneider chain starting from the classical
homogeneous elliptic ${\rm GL}_N$ spin chain on $n$ sites. Assuming
periodic boundary conditions in discrete space, it is a finite-dimensional
integrable system of classical mechanics.
By construction, it is gauge equivalent to the elliptic spin chain
(or lattice version of the generalized Landau-Lifshitz model)
with some special choice of the classical spin matrix at each site.
In the second part (sections 5 and 6) we introduce a
(1+1)-dimensional field analogue of the
Ruijsenaars-Schneider model with continuous space variable whose natural
finite-dimensional reduction turns out to be
equivalent to the Ruijsenaars-Schneider chain introduced in the first part.
Our method is based on investigation of
general elliptic solutions (called elliptic families)
to the difference version of the 2D Toda equation.
We derive equations of motion for their
poles together with a zero-curvature representation for them and show that they are
Hamiltonian.
The limit to differential equations in space rather than difference is shown
to give the field extension of the Calogero-Moser model introduced in
\cite{AKV02} via analyzing elliptic families
of solutions to the Kadomtsev-Petviashvili
equation.

Regarding the motivation of our work, we should say that the
field analog of the Ruijsenaars-Schneider system is a new integrable model
which is valuable by itself. Its role in modern mathematical physics can
be compared with the role of the field extension of the Calogero-Moser system, which
is known to be gauge equivalent to continuous 1+1 integrable Landau-Lifshitz model \cite{LOZ,AtZ}.

Below we describe the contents of both parts of the paper in more details.

The classical homogeneous elliptic ${\rm GL}_N$ spin
chain on $n$ sites is a widely known integrable system. It is also called an integrable ${\rm GL}_N$-generalization of the lattice
Landau-Lifshitz equation \cite{Skl}--\cite{FTbook}.
It is defined via the (classical) monodromy matrix
depending on a spectral parameter $z$
as a product of the Lax matrices at each site:
  \beq\label{qa01}
  \begin{array}{c}
    \displaystyle{
 T(z)={\mL}^{1}(z)\mL^{2}(z)\ldots \mL^{n}(z)\,, \quad \mL^i(z)\in\Mat .
 }
 \end{array}
 \eeq
Each Lax matrix depends on a set of dynamical variables (coordinates in the phase space),
which are combined into a matrix $S^i\in\Mat$, so that $\mL^i(z)=\mL^i(z,S^i)$.
The trace of the monodromy matrix
$t(z)=\tr \, T(z)$ is a generating function of Hamiltonians. They
are in involution with respect to
the classical quadratic $r$-matrix structure (with the Belavin-Drinfeld elliptic
classical $r$-matrix \cite{BD} and $c$ be an arbitrary constant)
 \beq\label{qb22}
 \begin{array}{c}
  \displaystyle{
 \{\mL_1^{i}(z),\mL_2^{j}(w)\}=\frac{1}{c}\,
 \delta^{ij}[\mL_1^{i}(z)\mL_2^{i}(w),r_{12}(z-w)]\,,
 }
 \end{array}
 \eeq
which is equivalent to $n$ copies of the classical
generalized
Sklyanin algebras at each site. In this paper we use a modified description
of the classical Sklyanin's elliptic Lax matrix. Namely, following \cite{LOZ8}
we define $\mL^i(z,S^i)$ as
 \beq\label{qb23}
 \begin{array}{c}
  \displaystyle{
 \mL^i(z,S^i)=\tr_2(R_{12}^\eta(z)S_2^i)\,, \quad S_2^i=1_{N}
 \otimes S^i\in{\rm Mat}(N,\CC)^{\otimes 2}\,,
 }
 \end{array}
 \eeq
where $R_{12}^\eta(z)\in{\rm Mat}(N,\CC)^{\otimes 2}$ is the {\it quantum}
Baxter-Belavin elliptic $R$-matrix \cite{Baxter,Belavin} and we use the standard convention
on numbering the spaces where the matrices act.
Let us stress that although $R_{12}^\eta(z)$ is quantum, the Lax matrix
(\ref{qb23}) is classical. The parameter $\eta$ usually plays the role of the Planck constant since
in the classical limit $\eta\rightarrow 0$ we have $R_{12}^\eta(z)=
(N\eta)^{-1}1_{N}\otimes 1_N+r_{12}(z)+\ldots $, where $r_{12}(z)$ is the classical
$r$-matrix entering (\ref{qb22}). At the same time in (\ref{qb23}) $\eta$ is regarded
as the relativistic deformation parameter, similarly to what happens
in the Ruijsenaars-Schneider
model\footnote{An explanation of the presence of quantum $R$-matrix in a classical model comes from
associative Yang-Baxter equation
$R^{\hbar}_{12} R^{\eta}_{23} = R^{\eta}_{13} R^{\hbar-\eta}_{12} + R^{\eta-\hbar}_{23} R^{\hbar}_{13}$ ($R^u_{ab} = R^u_{ab}(x_a-x_b)$),
which is fulfilled by the quantum Baxter-Belavin elliptic $R$-matrix in the fundamental representation of ${\rm GL}_N$. This equation (and its degenerations)
provides the classical Lax equation for the Lax matrix (\ref{qb23}). In this respect the associative Yang-Baxter equation
unifies classical and quantum integrable structures.
%{\bf A reference is required!}
See \cite{KZ19,GSZ,Z18} and references therein.
}. In fact, the explicit dependence on $\eta$ can be removed by some
simple
re-definitions. However, we keep it since the form (\ref{qb23}) has the following important property \cite{Hasegawa} (see also \cite{Hasegawa1}--\cite{VZ},\cite{KZ19}).
In the case when $S^i$ is a rank 1 matrix ($S^i=\xi^i\otimes\psi^i$, $\xi^i,\psi^i\in\CC^N$), the Lax matrix can be represented
in the factorized form
 \beq\label{qb24}
 \begin{array}{c}
  \displaystyle{
 \mL^i(z,S^i)=g(z+N\eta,q^i)e^{P^i/c}g^{-1}(z,q^i)\in\Mat\,,\quad
 P^i={\rm diag}(p_1^i,\ldots ,p_N^i)
 }
 \end{array}
 \eeq
(up to a scalar factor),
where $q^i$ denotes a set of $N$ coordinate
variables $q^i=\{q_1^i,\ldots ,q_N^i\}$ and the explicit form of the matrix
$g(z, q^i)$ is given below
in the main text. It
is known as the intertwining matrix entering the IRF-Vertex
correspondence \cite{Baxter2}--\cite{Pasquier}.
The factorization (\ref{qb24}) provides an explicit parametrization
of the matrix $S^i$ through the canonical variables $p^i_k$, $q^i_k$, $i=1,\ldots ,n$,
$k=1,\ldots ,N$, thus providing the classical analogue for representation of the
generalized Sklyanin algebra by difference operators. Moreover, the gauge transformed
Lax matrix
 \beq\label{qb25}
 \begin{array}{c}
  \displaystyle{
 g^{-1}(z,q^i)\mL^i(z,S^i)g(z,q^i)=g^{-1}(z,q^i)g(z+N\eta,q^i)e^{P^i/c}:=L^{RS}(z,p^i,q^i)
 }
 \end{array}
 \eeq
is equal to the Lax matrix of the classical $N$-body elliptic Ruijsenaars-Schneider model with momenta $p^i$, coordinates of particles $q^i$ and the relativistic deformation
parameter $\eta$. The ``velocity of light'' $c$ enters  (\ref{qb25})
as a normalization factor.
% behind momenta.
On the spin chain side it is the constant in the r.h.s. of  (\ref{qb22}).

We restrict ourselves to the case when all matrices of dynamical variables $S^i$ are of rank one. Then,
taking into account
(\ref{qb24})--(\ref{qb25}), we represent the transfer matrix $t(z)$ in the form
  \beq\label{qa43}
  \begin{array}{l}
  \displaystyle{
 t(z)=
 \tr\Big( {\ti L}^1(z) {\ti L}^2(z)\ldots {\ti L}^n(z) \Big)\,,
 }
 \end{array}
 \eeq
 where %(with the identification $q^0=q^n$)
  \beq\label{qa440}
  \begin{array}{l}
  \displaystyle{
 {\ti L}^i(z)=g^{-1}(z,q^{i-1})g(z+N\eta,q^i)\,e^{P^i/c}\,,\quad i=1,\ldots ,n\ \
 \hbox{and}\ \ q^0=q^n\,.
 }
 \end{array}
 \eeq
The Lax matrices ${\ti L}^i(z)$ can be found explicitly.
Then we derive equations of motion generated by a special Hamiltonian flow. It is the one which has continuous limit to the (1+1)-dimensional field theory (the generalized
Landau-Lifshitz equation)
for the elliptic spin chain.

More precisely, we prove that
%
%\begin{theorem}
the transfer-matrix (\ref{qa43}) provides the Hamiltonian
\beq\label{qb4433}
 \begin{array}{c}
  \displaystyle{
 H=c\sum\limits_{k=1}^n \log h_{k-1,k}\,,\qquad h_{k-1,k}=\sum\limits_{j=1}^N b^k_j\,,
 \qquad b_j^k=
 \frac{\prod\limits_{l=1}^N\vth({\bar q}^k_j-{\bar q}^{k-1}_l-\eta) }
 {\vth(-\eta)\prod\limits_{l: l\neq j}^N\vth({ q}^{k}_j-{ q}^{k}_l) }\,e^{p^k_j/c}\,,
 }
 \end{array}
  \eeq
  where $\vth (z)$ is the odd Jacobi theta-function (\ref{a25}), and ${\bar q}^k_j=q^k_j-\sum_{i=1}^N q^k_i/N$ are coordinates
  ``in the center of masses frame'' at each site\footnote{Notice that ${ q}^{k}_j-{ q}^{k}_l={\bar q}^{k}_j-{\bar q}^{k}_l$.}.
This Hamiltonian generates equations of motion (in the Newtonian form)
   \beq\label{qb706}
 \begin{array}{c}
  \displaystyle{
  \frac{{\ddot q}^k_i }{ {\dot q}^k_i }=
   -\sum\limits_{l=1}^N {\dot { q}}_l^{k+1}E_1({\bar q}_i^k-{\bar q}_l^{k+1}+\eta)
  -\sum\limits_{l=1}^N {\dot { q}}_l^{k-1}E_1({\bar q}_i^k-{\bar q}_l^{k-1}-\eta)
  +2\sum\limits_{l:l\neq i}^N {\dot {q}}_l^{k}E_1({q}_i^k-{q}_l^{k})+
 }
 \\
   \displaystyle{
   +\sum\limits_{m,l=1}^N {\dot { q}}_m^{k}{\dot { q}}_l^{k+1}E_1({\bar q}_m^k-{\bar q}_l^{k+1}+\eta)
  -\sum\limits_{m,l=1}^N {\dot { q}}_l^{k}{\dot { q}}_m^{k-1}E_1({\bar q}_m^{k-1}-{\bar q}_l^k+\eta)\,,
   }
 \end{array}
  \eeq
where $E_1(z)$ is the logarithmic derivative of the odd Jacobi theta-function (\ref{a093}).
With some simple normalization factor the Lax matrix (\ref{qa440}) turns into
 \beq\label{qa5333}
 \displaystyle{
 {L'}^k_{ij}(z)=
 \phi(z,{\bar q}^{k-1}_i-{\bar q}^{k}_j+\eta){\dot q}_j^k\,,
 }
 \eeq
where $\phi (z, q)$ is the Kronecker elliptic function (\ref{a09}).
Equations of motion (\ref{qb706}) are equivalently written in the form of the
semi-discrete zero curvature
(Zakharov-Shabat) equation
\beq\label{qb4733}
 \begin{array}{c}
  \displaystyle{
  \frac{d}{dt}\,{{ L'}^k}(z)={L'}^k(z){M'}^k(z)-{ M'}^{k-1}(z){ L'}^k(z)\,.
 }
 \end{array}
  \eeq
with $M$-matrices
 \beq\label{qb6733}
 \begin{array}{c}
  \displaystyle{
  {M'}^k_{ij}(z)=-(1-\delta_{ij})\phi(z,q_i^k-q_j^k)\,
  {\dot q}_j^k+\delta_{ij}\sum\limits_{m,l=1}^N {\dot { q}}_l^{k+1}{\dot { q}}_m^{k}
  E_1({\bar q}_m^{k}-{\bar q}_l^{k+1}+\eta)+
  }
 \\
   \displaystyle{
  +\delta_{ij}\Big(-E_1(z){\dot q}^k_i+
  \sum\limits_{m: m\neq i}^N{\dot q}_m^kE_1(q_i^k-q_m^k)
  -\sum\limits_{m=1}^N{\dot q}_m^{k+1}E_1({\bar q}^k_i-{\bar q}^{k+1}_m+\eta)
    \Big)\,.
 }
 \end{array}
  \eeq
%
%\end{theorem}
%
%
%Here $E_1(z)$ is the logarithmic derivative of the odd Jacobi theta-function and
%$\phi (z, q)$ is the Kronecker function given by (\ref{a09}).
At this stage the model is discrete and finite-dimensional. To proceed to field generalization
we use another approach. We will see that the 1+1 version corresponds to straightforward
field extension of the described Ruijsenaars-Schneider chain.

The idea of the other approach is to exploit the close
connection between elliptic solutions to nonlinear integrable equations and many-body systems.
The investigation of dynamics of poles of singular solutions to nonlinear integrable
equations was initiated in the seminal paper \cite{AMM77}, where elliptic and rational
solutions to the Korteweg-de Vries and Boussinesq equations were studied.
As it was proved later in \cite{Krichever78,CC77}, poles of rational solutions to the
Kadomtsev-Petviashvili (KP) equation as functions of
the second hierarchical time $t_2$ move as particles of the integrable
Calogero-Moser system \cite{Calogero71,Calogero75,Moser75,OP81}.
The method suggested by Krichever \cite{Krichever80} for elliptic solutions
of the KP equation consists in substituting the solution not in the KP equation itself but
in the auxiliary linear problem for it (this implies a suitable pole ansatz for the wave
function). This method allows one to obtain the equations of motion together with their
Lax representation.

Dynamics of poles of elliptic solutions to the 2D Toda lattice and
modified KP (mKP) equations was
studied in \cite{KZ95}, see also \cite{Z19}.
It was proved that the poles move as particles of the
integrable Ruijsenaars-Schneider many-body system
which is a relativistic generalization
of the Calogero-Moser system.

In the paper \cite{AKV02} elliptic families of solutions to the
KP equation were studied. In this more general case the solution is assumed to be
an elliptic function not of $x=t_1$, as it was assumed before, but of a general linear
combination of higher times of the KP hierarchy.
It was shown that poles of such solutions as functions of $x=t_1$ and $t_2$ move
according to equations of motion of the field generalization of the Calogero-Moser system.
In this paper we extend this result to elliptic families of solutions to the 2D Toda
hierarchy. We derive equations of motion for such solutions.
These equations of motion can be naturally
thought of as a field generalization of the Ruijsenaars-Schneider system.
In the limit when the parameter
$\eta$ having the meaning of the inverse velocity of light tends to 0, the obtained
equations of motion become those dealt with in the paper \cite{AKV02}. We also consider
elliptic families of solutions to the fully difference
integrable version of the 2D Toda lattice equation and derive equations of motion for the poles.

Let us say a few words about the nature of the elliptic families. These solutions
belong to a particular class of algebraic-geometrical solutions associated with an algebraic curve
$\Gamma$ of genus $g$ with some additional data. An algebraic-geometrical solution is elliptic
with respect to some variable $\lambda$ if there exists a $g$-dimensional
vector ${\bf W}$ such that it spans an elliptic curve ${\cal E}$
embedded in the Jacobian
of the curve $\Gamma$. The tau-function of such solution is
\cite{Krichever77,Dubrovin81}
\beq\label{toda8aa}
\tau (x, {\bf t}, \lambda )=e^{Q(x, {\bf t})}
\Theta \Bigl (
{\bf V}_0 x/\eta +\sum_{k\geq 1}{\bf V}_k t_k +{\bf W}\lambda
+{\bf Z}\Bigr ),
\eeq
where $\Theta$ is the Riemann theta-function and $Q$ is a quadratic form in the
hierarchical times ${\bf t}=\{t_1, t_2, t_3, \ldots \}$ of the 2D Toda hierarchy
(we put all ``negative'' times equal to zero for simplicity). The vectors ${\bf V}_k$ are
$b$-periods of certain normalized meromorphic differentials on $\Gamma$.
The existence of a $g$-dimensional
vector ${\bf W}$ such that it spans an elliptic curve ${\cal E}$
embedded in the Jacobian
is a nontrivial transcendental constraint.
If such a vector ${\bf W}$ exists, then the theta-divisor intersects
the shifted elliptic curve $\displaystyle{{\cal E}+{\bf V}_0 x/\eta +
\sum_k {\bf V}_k t_k}$ at
a finite number of points $\lambda_i =\lambda_i(x,{\bf t})$. Therefore,
for elliptic families we have:
\beq\label{toda8bb}
\Theta \Bigl (
{\bf V}_0 x/\eta +\sum_{k\geq 1}{\bf V}_k t_k +{\bf W}\lambda
+{\bf Z}\Bigr )=f(x, {\bf t})e^{\gamma_1\lambda +\gamma_2\lambda^2}\prod _{i=1}^N
\sigma (\lambda -\lambda_i(x, {\bf t})).
\eeq
with a function $f(x, {\bf t})$ and
some constants $\gamma_1, \gamma_2$. Here $\sigma (\lambda )$ is the
Weierstrass $\sigma$-function defined in the Appendix A. The zeros $\lambda_i$ of the
tau-function are poles of the elliptic solutions.

We show that the equations of motion of the poles $\lambda_i =\lambda_i(x,t)$, where
$t=t_1$, are given by
\beq\label{el131}
\begin{array}{c}
\displaystyle{
\ddot \lambda_i(x)+\sum\limits_{k=1}^N \Bigl (
\dot \lambda_i(x)\dot \lambda_k(x-\eta )\zeta (\lambda_i(x)-\lambda_k(x-\eta ))+
\dot \lambda_i(x)\dot \lambda_k(x+\eta )\zeta (\lambda_i(x)-\lambda_k(x+\eta ))\Bigr )}
\\ \\
\displaystyle{
-\, 2 \sum_{k:k\neq i}^N\dot \lambda_i(x)\dot \lambda_k(x )
\zeta (\lambda_i(x)-\lambda_k(x))+(c(x-\eta , t)-c(x, t))\dot \lambda_i(x)=0\,.}
\end{array}
\eeq
Here
\beq\label{el14a1}
c(x, t)=\frac{1}{\beta}\sum\limits_{i,k=1}^N\dot \lambda_i(x)\dot \lambda_k(x+\eta )
\zeta (\lambda_i(x)-\lambda_k(x+\eta )), \qquad \beta =\sum\limits_{i=1}^N \dot \lambda_i(x),
\eeq
and $\zeta (\lambda )$ is the Weierstrass $\zeta$-function (a close relative of the
function $E_1$ in (\ref{qb706})).

Equations (\ref{el131}) are represented in the zero-curvature form
\beq\label{el90}
\dot L(x)+L(x)M(x)-M(x+\eta )L(x)=0\,,
\eeq
and the Lax pair is obtained explicitly in section 5.3.
Then, the equations (\ref{el131}) can be naturally
restricted to the lattice by setting $\lambda_i^k=\lambda_i(x_0+k\eta )$ and rewritten as
\beq\label{el13a1}
\begin{array}{c}
\displaystyle{
\ddot \lambda_i^k+\sum\limits_{j=1}^N \Bigl (
\dot \lambda_i^k\dot \lambda_j^{k-1}\zeta (\lambda_i^k-\lambda_j^{k-1})+
\dot \lambda_i^k\dot \lambda_j^{k+1}\zeta (\lambda_i^k-\lambda_j^{k+1})\Bigr )-
}
\\
\displaystyle{
-\, 2 \sum\limits_{j:j\neq i}^N\dot \lambda_i^k\dot \lambda_j^k
\zeta (\lambda_i^k-\lambda_j^k)+(c^{k-1}(t)-c^k(t))\dot \lambda_i^k=0}
\end{array}
\eeq
with
\beq\label{el14b1}
c^k(t)=\frac{1}{\beta}\sum\limits_{i,j=1}^N\dot \lambda_i^k\dot \lambda_j^{k+1}
\zeta (\lambda_i^k-\lambda_j^{k+1}),
\eeq
in which form they can be shown to be equivalent to equations (\ref{qb706}). Details
of the equivalence between (\ref{el131})-(\ref{el14b1}) and (\ref{qb706})-(\ref{qb6733})
are given in section 5.4.

In section 5.6 we also describe the $\eta \to 0$ limit
to the (1+1)-dimensional
Calogero-Moser field theory discussed
in \cite{AKV02}. The fully discrete version of the equations (\ref{el13a1}) is obtained
in section 6.  In Appendix A the necessary
definitions and properties of elliptic functions are given. In Appendix B we
describe properties of the
elliptic $R$-matrix which are used in the
derivation of the Ruijsenaars-Schneider spin chain. In Appendix C,
using the factorization formulae for the Lax matrix, we obtain the explicit
change of variables between the Ruijsenaars-Schneider model and the relativistic top.

%%%%%%%%%%%%%%%%%%%%%%%%%%%%%%%%%%%%%%%%%%%%%%%%%%%%%%%%%%%%%%%%%%%%%%%%%%%%%%%%%%%%%%%%%%%%%%%%%%%%%%%%
%%%%%%%%%%%%%%%%%%%%%%%%%%%%%%%%%%%%%%%%%%%%%%%%%%%%%%%%%%%%%%%%%%%%%%%%%%%%%%%%%%%%%%%%%%%%%%%%%%%%%%%%

\section{Ruijsenaars-Schneider model in the form of relativistic top}
\setcounter{equation}{0}

In this section we recall the necessary preliminaries related to the
Ruijsenaars-Schneider model and relativistic top. From the point of view of the next sections
this case corresponds to the ``Ruijsenaars-Schneider chain'' on one site.

  \subsection{Classical Ruijsenaars-Schneider model}
  \paragraph{The standard Hamiltonian and equations of motion.}
  The elliptic Ruijsenaars-Schneider model is defined by the Lax matrix \cite{Ruijs12}
  \beq\label{a12}
  \begin{array}{l}
  \displaystyle{
 L^{\rm RS}_{ij}(z)=\phi(z,q_{ij}+\eta)\,b_j\,,\ i,j=1,\ldots ,N\,,
 }
 \end{array}
 \eeq
 where $\phi (z, q)$ is the
 Kronecker function defined in (\ref{a09}) and
  \beq\label{a13}
  \begin{array}{l}
  \displaystyle{
 b_j=\prod_{k:k\neq j}^N\frac{\vth(q_{j}-q_k-\eta)}{\vth(q_{j}-q_k)}\,
 e^{p_j/c}\,,\quad c={\rm const}\in\CC\,.
 }
 \end{array}
 \eeq
 Here $\vth (z)$ is the odd Jacobi theta-function (\ref{a25}).
 Note that original definition of $b_j$ in \cite{Ruijs12} is different from (\ref{a13}).
 This is due to a freedom in the definition of (\ref{a12})--(\ref{a13})
 coming from the canonical map
  \beq\label{a17}
  \begin{array}{c}
  \displaystyle{
p_j\ \rightarrow\ p_j+c_1\log \prod\limits_{k:k\neq j}^N
\frac{\vth(q_{j}-q_k+\eta)}{\vth(q_{j}-q_k-\eta)}
 }
 \end{array}
 \eeq
with arbitrary constant $c_{1}$.

 The Hamiltonian
  \beq\label{a14}
  \begin{array}{l}
  \displaystyle{
 H^{\rm RS}=c\frac{\tr L^{\rm RS}(z)}{\phi(z,\eta)}=c\sum\limits_{j=1}^N b_j(p,q)
 }
 \end{array}
 \eeq
 with the canonical Poisson brackets
  \beq\label{a15}
  \begin{array}{c}
    \displaystyle{
\{p_i,q_j\}=\delta_{ij}\,,\quad \{p_i,p_j\}=\{q_i,q_j\}=0
 }
 \end{array}
 \eeq
 provides the following equations of motion:
  \beq\label{a131}
  \begin{array}{l}
  \displaystyle{
 {\dot q}_j=\{H^{\rm RS},q_j\}=\p_{p_j}H^{\rm RS}=b_j=\prod^N_{k:k\neq j}\frac{\vth(q_{j}-q_k-\eta)}{\vth(q_{j}-q_k)}\,e^{p_j/c}\,.
 }
 \end{array}
 \eeq
We see that the Hamiltonian is proportional to the sum of velocities:
  \beq\label{a132}
  \begin{array}{l}
  \displaystyle{
 \frac{1}{c}\,H^{\rm RS} =\sum\limits_{j=1}^N{\dot q}_j
 }
 \end{array}
 \eeq
 and the Lax matrix (\ref{a12}) takes the form
  \beq\label{a135}
  \begin{array}{l}
  \displaystyle{
 L^{\rm RS}_{ij}(z)=\phi(z,q_{ij}+\eta)\,{\dot q}_j\,.
 }
 \end{array}
 \eeq
 The Hamiltonian equations for momenta are as follows:
  \beq\label{a133}
  \begin{array}{c}
  \displaystyle{
 \frac{1}{c}\,{\dot p}_i=\frac{1}{c}\,\{H^{\rm RS},p_i\}=-\frac{1}{c}\,\p_{q_i}H^{\rm RS}=
 }
 \\ \ \\
  \displaystyle{
  =\sum\limits_{l:l\neq i}^N ({\dot q}_i+{\dot q}_l)E_1(q_{il})-{\dot q}_i E_1(q_{il}-\eta)
  -{\dot q}_l E_1(q_{il}+\eta)\,,
  }
 \end{array}
 \eeq
  where $q_{ij}=q_i-q_j$, and $E_1(w)=\vth'(w)/\vth(w)$ (\ref{a091}). By differentiating both parts of
  (\ref{a131}) with respect to time we get
  \beq\label{a134}
  \begin{array}{c}
  \displaystyle{
 \frac{{\ddot q}_i}{{\dot q}_i}=\frac{1}{c}\,{\dot p}_i+\sum\limits_{l:l\neq i}^N
 ({\dot q}_i-{\dot q}_l)(E_1(q_{il}-\eta)-E_1(q_{il}))\,.
  }
 \end{array}
 \eeq
 Plugging (\ref{a133}) into (\ref{a134}) we get the well known
 equations of motion of the elliptic Ruijsenaars-Schneider model in the Newtonian form:
 \beq\label{a16}
  \begin{array}{c}
  \displaystyle{
 {\ddot q}_i=\sum\limits_{k:k\neq i}^N{\dot q}_i{\dot q}_k
 (2E_1(q_{ik})-E_1(q_{ik}+\eta)-E_1(q_{ik}-\eta))\,,\quad i=1, \ldots ,
 N\,.
 }
 \end{array}
 \eeq
Equations of motion (\ref{a16}) are equivalent to the Lax equation
  \beq\label{a18}
  \begin{array}{c}
  \displaystyle{
 {\dot L}^{\rm RS}(z)\equiv\{H^{\rm RS},L^{\rm RS}(z)\}=[L^{\rm RS}(z),M^{\rm
 RS}(z)]
 }
 \end{array}
 \eeq
  with the $M$-matrix
  \beq\label{a181}
  \begin{array}{c}
  \displaystyle{
 M^{\rm RS}_{ij}(z)=
 -(1-\delta_{ij})\phi(z,q_i-q_j)\,{\dot q}_j
 }
 \\ \ \\
   \displaystyle{
 -\delta_{ij}\Big({\dot q}_i\,(E_1(z)+E_1(\eta)) +
 \sum\limits_{k:k\neq i}^N {\dot q}_k\,(E_1(q_{ik}+\eta)-E_1(q_{ik}))
 \Big).
 }
 \end{array}
 \eeq
This follows from a direct calculation with the help of (\ref{b151}) and (\ref{a092}).

\paragraph{Logarithm of the Hamiltonian.} Alternatively, one can use the following Hamiltonian:
  \beq\label{a171}
  \begin{array}{c}
  \displaystyle{
  H'=c\log (H^{\rm RS}/c)=c\log\sum\limits_{j=1}^N b_j-c\log c\,, \qquad H^{\rm RS}=c\sum\limits_{j=1}^N b_j\,.
 }
 \end{array}
 \eeq
Then
  \beq\label{a172}
  \begin{array}{c}
  \displaystyle{
  {\dot q}_j=\frac{\p H'}{\p p_j}=\frac{c b_j}{H^{\rm RS}}=\frac{b_j}{\sum_k b_k}\,,
\qquad
  \frac{1}{c}\,{\dot p}_i=-\frac{1}{c}\frac{\p H'}{\p q_i}=-\frac{1}{H^{\rm RS}}\frac{\p H^{\rm RS}}{\p q_i}\,,
 }
 \end{array}
 \eeq
 so that (cf. (\ref{a132}))
  \beq\label{a173}
  \begin{array}{c}
  \displaystyle{
  \sum\limits_{j=1}^N{\dot q}_j=1\,.
 }
 \end{array}
 \eeq
 The Lax matrix (\ref{a12}) becomes now
  \beq\label{a174}
  \begin{array}{c}
 L^{\rm RS}_{ij}(z)=\phi(z,q_{ij}+\eta)\,{\dot q}_jH^{\rm RS}/c
 \end{array}
 \eeq
instead of (\ref{a135}) but this makes no difference since $H^{\rm RS}$ is a conserved quantity.
 It is easy to see that the equations of motion in the Newtonian form (\ref{a16})
 remain the same with
 the Hamiltonian (\ref{a171}). We will use the Hamiltonian description similar to the presented above (with logarithm of the Hamiltonian) in the Ruijsenaars chain.

 \subsection{Classical relativistic top}

 Let us consider the elliptic ${\rm GL}_N$ spin chain on a single site.
 It is an integrable system called relativistic top \cite{LOZ8}. The Lax matrix is as follows:
 \beq\label{b040}
 \begin{array}{c}
  \displaystyle{
  \mL^\eta(z)=\sum\limits_{a\in\,\z_{ N}\times\z_{ N}} T_a { S}_a\vf_a(z,\om_a+\eta)\,,
 }
 \end{array}
 \eeq
 where $S_a$ are dynamical variables (classical spins)
 numbered by the index $a=(a_1,a_2)\in\ZZ_N\times\ZZ_N$ in the special
 matrix basis $T_a\in\Mat$ (\ref{a07}), which is often used for elliptic quantum
 $R$-matrices. The set of functions
 $\vf_a(z,\om_a+\eta)$ and the quantities $\omega_a$
 are given in (\ref{a08}). The dynamical variables are combined into a matrix $S\in\Mat$:
 \beq\label{b0401}
 \begin{array}{c}
  \displaystyle{
  S=\sum\limits_{i,j=1}^N S_{ij}E_{ij}=\sum\limits_{a_1,a_2=0}^{N-1} S_{a}T_a\,,
 }
 \end{array}
 \eeq
 where $E_{ij}$ are the usual matrix units.
 Using the property (\ref{a072}) let us rewrite the Lax matrix (\ref{b040})
 in terms of the Baxter-Belavin
 elliptic $R$-matrix $R_{12}^\eta(z)$
 \cite{Baxter,RT} in the form (\ref{d07}):
 \beq\label{a10}
 \begin{array}{c}
  \displaystyle{
 R_{12}^\eta(z)=\frac{1}{N}\sum\limits_{a\in\,\z_{ N}\times\z_{ N}}
 T_a\otimes T_{-a} \vf_a(z,\om_a+\eta)
 \in{\rm Mat}(N,\CC)^{\otimes 2}\,.
 }
 \end{array}
 \eeq
 Alternative equivalent forms are given in the Appendix B. In terms of the $R$-matrix,
 the Lax matrix (\ref{b040}) acquires the following compact form:
 \beq\label{b0402}
 \begin{array}{c}
  \displaystyle{
  \mL^\eta(z)=\tr_2(R_{12}^\eta(z)S_2)\,,\quad S_2=1_N\otimes S,
 }
 \end{array}
 \eeq
 where the trace is over the second tensor component.

We emphasis that the $R$-matrix is quantum, while the Lax matrix is classical. The parameter $\eta$
 plays the role of the Planck constant in the $R$-matrix since
 the classical $r$-matrix comes from (\ref{a10}) in the classical limit\footnote{The classical $r$-matrix is used for construction of the non-relativistic top similarly to (\ref{b0402}). See \cite{GSZ,LOZ}.}
 \beq\label{b05}
 \begin{array}{c}
  \displaystyle{
R_{12}^\eta(z)=\frac{1}{N\eta}+r_{12}(z)+O(\eta)\,,
 }
 \end{array}
  \eeq
   \beq\label{b06}
 \begin{array}{c}
  \displaystyle{
 r_{12}(z)=\frac{1}{N}\,1_N\otimes 1_N\, E_1(z)
 +\frac{1}{N}\sum\limits_{a\in\ZZ_N^{\times 2},a\neq 0} T_a\otimes T_{-a} \vf_a(z,\om_a)
 \in{\rm Mat}(N,\CC)^{\otimes 2}\,.
 }
 \end{array}
  \eeq
%
% Having a single site the integrable system of classical mechanics corresponding to the Lax matrix (\ref{a06}) %or (\ref{a11}) is the relativistic elliptic top \cite{LOZ8}. The parameter $\eta$ is the Planck constant in the %quantum $R$-matrix (\ref{a10}).
At the same time $\eta$ plays the role of the relativistic deformation parameter in the relativistic top model similarly to the Ruijsenaars-Schneider model. In the standard approach
\cite{Skl,Skl2} the parameter
$\eta$ is absent in the Lax matrix and so it does not enter
the classical (Poisson) Sklyanin algebra. Below we explain how this parameter
can be eliminated by some re-definitions. However, it is
important for us to keep it
in the Lax matrix because in this form the latter has a nice property of factorization.

  \paragraph{Classical Sklyanin algebra.} The classical quadratic
$r$-matrix Poisson structure is\footnote{The coefficient $1/c$ in (\ref{b04})
is introduced here in order to match the relation with the
Ruijsenaars-Schneider model.}:
 \beq\label{b04}
 \begin{array}{c}
  \displaystyle{
 \{\mL_1^{\eta}(z),\mL_2^{\eta}(w)\}=\frac{1}{c}\,[\mL_1^{\eta}(z)\mL_2^{\eta}(w),r_{12}(z-w)]\,,
 }
 \end{array}
 \eeq
 where the classical $r$-matrix is given by (\ref{b06}).
 Plugging the Lax matrix (\ref{b040}) into (\ref{b04}) and using identity (\ref{b153}) one gets
 %
% \beq\label{b062}
% \begin{array}{c}
%   \displaystyle{
% \{S_\al,S_\be\}=\!\frac{1}{c}\sum\limits_{\xi\neq 0} \kappa_{\al-\be,\xi}S_{\al-\xi}S_{\be+\xi}
% \Big( E_1(\om_\xi)\!-\!E_1(\om_{\al-\be-\xi})\!+\!E_1(\om_{\al-\xi}+\eta)\!-\!E_1(\om_{\be+\xi}+\eta) \Big)\,,
% }
% \end{array}
%  \eeq
 \beq\label{b062}
 \begin{array}{c}
   \displaystyle{
 \{S_\al,S_\be\}=\frac{1}{Nc}\times
 }
 \\ \ \\
  \displaystyle{
 \sum\limits_{\xi\in\ZZ_N^{\times 2},\,\xi\neq 0,\al-\be} \kappa_{\al-\be,\xi}S_{\al-\xi}S_{\be+\xi}
 \Big( E_1(\om_\xi)-E_1(\om_{\al-\be-\xi})+E_1(\om_{\al-\xi}+\eta)-E_1(\om_{\be+\xi}+\eta) \Big)\,,
 }
 \end{array}
  \eeq
which is the classical Sklyanin algebra. The constants $\kappa_{\al,\be}$ are defined in (\ref{a071}).

\paragraph{Eliminating the parameter $\eta$.}
Let us comment on the form of the classical Lax matrix $\mL^{\eta}(z)$
(\ref{b040}), (\ref{b0402}). Usually \cite{Skl,Skl2} the classical
 Lax matrix of the top is written as
 \beq\label{b33}
 \begin{array}{c}
  \displaystyle{
 \mL(z,{\ti S})=1_N {\ti S}_0+\sum\limits_{a\in\ZZ_N^{\times 2}, a\neq 0} T_a {\ti S}_a\vf_a(z,\om_a)\,.
 }
 \end{array}
 \eeq
 It is known to satisfy (\ref{b04}), which provides the classical Sklyanin Poisson algebra for $N^2$ generators ${\ti S}_a$.
 Writing (\ref{b04}), we assume that it is also fulfilled for $\mL^{\eta}(z)$. It happens for
 the following reason
 \cite{CLOZ,LOZ8}. First of all, this can be verified by a direct calculation, so that the classical Sklyanin algebra for $S_a$ contains additional parameter $\eta$. However, this dependence is artificial. Using the relation
 \beq\label{b34}
 \begin{array}{c}
  \displaystyle{
  \frac{\vf_a(z-\eta,\om_a+\eta)}{\phi(z-\eta,\eta)}=\frac{\vf_a(z,\om_a)}{\vf_a(\eta,\om_a)},
 }
 \end{array}
 \eeq
 one easily obtains
 \beq\label{b35}
 \begin{array}{c}
  \displaystyle{
   \frac{1}{\phi(z-\eta,\eta)}\,\mL^\eta(z-\eta,S)=\mL(z,{\ti S})
 }
 \end{array}
 \eeq
 if
 \beq\label{b36}
 \begin{array}{c}
  \displaystyle{
   S=\mL(\eta,{\ti S})\,.
 }
 \end{array}
 \eeq
 Using (\ref{b33}) in the basis $T_\al$ (\ref{a07}) we may write (\ref{b36}) explicitly:
 \beq\label{b361}
 \begin{array}{c}
  \displaystyle{
   S_0={\ti S}_0\,,\qquad S_\al={\ti S}_\al\vf_\al(\eta,\om_\al)\ {\rm for}\ \al\neq 0\,.
 }
 \end{array}
 \eeq
 Let us remark that a similar phenomenon with the same change of variables
 take place in quantum Sklyanin
 algebra generated by exchange relation
 $R_{12}^\hbar(z-w){\hat \mL}^\eta_1(z){\hat \mL}^\eta_2(w)=
 {\hat \mL}^\eta_2(w){\hat \mL}^\eta_1(z)R_{12}^\hbar(z-w)$.
 Then it contains two parameters $\hbar$ and $\eta$, but the
 latter can be removed by (\ref{b36}) or fixed somehow.
 For example, in the case $\hbar=\eta$ the Sklyanin algebra has
 representation ${\hat S}_a=T_{-a}$ since the
 exchange relation turns into the Yang-Baxter equation in this case. So that the second parameter is artificial.

We see that the
two Lax matrices $\mL^\eta(z)$ and (\ref{b33}) are related by the
explicit change of variables (\ref{b36}) or (\ref{b361}), and the shift of the spectral parameter
 $z\rightarrow z-\eta$ does not effect (\ref{b04}) because $r_{12}(z-w)$ depends on the difference of spectral parameters. In what follows we need the explicit dependence on $\eta$ in $\mL^\eta(z)$ for establishing its relation to the Ruijsenaars-Schneider model. For this purpose we will consider $S$ to be a rank one matrix (this is not true for $\ti S$). The possibility to fix the rank of matrix $S$ follows from equations of motion (see (\ref{b14}) below). The eigenvalues of $S$ are conservation laws on these equations.

\paragraph{Lax pair.} The Lax equation follows from (\ref{b04}) in the following way. Since $S=\res\limits_{w=0}\mL^\eta(w)$, then the residue at $w=0$ of both parts of (\ref{b04}) yields
 \beq\label{b10}
 \begin{array}{c}
  \displaystyle{
 \{\mL_1^{\eta}(z),S_2\}=[\mL_1^{\eta}(z)S_2,\frac{1}{c}\,r_{12}(z)]\,.
 }
 \end{array}
 \eeq
 Taking trace of both parts of (\ref{b10}) in the second tensor component, we get the Lax equation
 \beq\label{b11}
 \begin{array}{c}
  \displaystyle{
 {\dot \mL}^{\eta}(z)=\{H^{top},\mL_1^{\eta}(z)\}=[\mL_1^{\eta}(z),M(z)]\,,\qquad M(z)=-\tr_2\Big(r_{12}(z)S_2\Big)\,,
 }
 \end{array}
 \eeq
 where the Hamiltonian is
 \beq\label{b12}
 \begin{array}{c}
  \displaystyle{
 H^{top}=c\,\tr\, S=c\,\frac{\tr \mL^\eta(z)}{\phi(z,\eta)}\,.
 }
 \end{array}
 \eeq
 More precisely,
  \beq\label{b13}
 \begin{array}{c}
  \displaystyle{
M(z)=-S_0 1_N E_1(z)-\sum\limits_{\al\in\ZZ_N^{\times 2}, \al\neq
0}T_\al S_\al\vf_\al(z,\om_\al)\,.
 }
 \end{array}
 \eeq
 Equations of motion take the form:
 \beq\label{b14}
 \begin{array}{c}
  \displaystyle{
\dot S=[S,J^\eta(S)]\,,
 }
 \end{array}
 \eeq
 \beq\label{b15}
 \begin{array}{c}
  \displaystyle{
 J^\eta(S)=1_N S_0 E_1(\eta)+\sum\limits_{\al\in\ZZ_N^{\times 2}, \al\neq
0}T_\al S_\al J_\al^\eta\,,\quad
 J_\al^\eta=E_1(\eta+\om_\al)-E_1(\om_\al)\,.
 }
 \end{array}
 \eeq
 They follow from the Lax equation (\ref{b11}) under the substitution (\ref{b040}), (\ref{b13})
 and usage of
 (\ref{a071}) and the identity (\ref{b151}).

%  \subsection{The Lax equations and factorization of Lax matrices}

%%%%%%%%%%%%%%%%%%%%%%%%%%%%%%%%%%%%%%%%%%%%%%%%%%%%%%%%%
%
 \subsection{Factorization of Lax matrices and relation between the models}
 %
 %\paragraph{.}
   Following \cite{Baxter2,Jimbo,Pasquier}, we introduce the intertwining matrix
  \beq\label{a21}
  \begin{array}{l}
  \displaystyle{
 g(z,q)=\Xi(z,q)\left(d^{0}\right)^{-1}
 }
 \end{array}
 \eeq
 with
 \beq\label{a22}
 \begin{array}{c}
  \displaystyle{
\Xi_{ij}(z,q)=
 \vth\left[  \begin{array}{c}
 \frac12-\frac{i}{N} \\ \frac N2
 \end{array} \right] \left(z-Nq_j+\sum\limits_{m=1}^N
 q_m\left.\right|N\tau\right)\,,
 }
 \end{array}
 \eeq
and the diagonal matrix
 \beq\label{a23}
 \begin{array}{c}
  \displaystyle{
d^0_{ij}(z,q)=\delta_{ij}d^0_{j}=\delta_{ij}
 {\prod\limits_{k:k\neq j}^N\vth(q_j-q_k)}\,,
 }
 \end{array}
 \eeq
 where the theta function with characteristics is defined in (\ref{a24}).
 The matrix (\ref{a21}) is the intertwining matrix entering the relations of the
 IRF-Vertex correspondence. Its properties are described in the Appendix B.

 \paragraph{Factorization formula.}It was observed in \cite{Hasegawa} (at quantum level) that the Lax matrix
(\ref{a12})--(\ref{a13}) can be represented in the factorized form
  \beq\label{a31}
  \begin{array}{l}
  \displaystyle{
 L^{\rm RS}_{ij}(z)=\frac{\vth'(0)}{\vth(\eta)}\sum\limits_{k=1}^N
 g^{-1}_{ik}(z,q)g_{kj}(z+N\eta,q)\,e^{p_j/c}\,,
 }
 \end{array}
 \eeq
or
  \beq\label{a32}
  \begin{array}{l}
  \displaystyle{
 L^{\rm RS}(z)=\frac{\vth'(0)}{\vth(\eta)}\,
 g^{-1}(z,q)g(z+N\eta,q)\,e^{P/c}\,,\quad P={\rm diag}(p_1,\ldots ,p_N)\,.
 }
 \end{array}
 \eeq
 Moreover, the gauge transformed Lax matrix
  \beq\label{a33}
  \begin{array}{l}
  \displaystyle{
 \mL^\eta(z)=g(z,q)L^{\rm RS}(z)g^{-1}(z,q)=\frac{\vth'(0)}{\vth(\eta)}\,
 g(z+N\eta,q)\,e^{P/c}g^{-1}(z,q)
 }
 \end{array}
 \eeq
  is the Lax matrix of type (\ref{b040}) since it
 has the same quasi-periodic properties (see (\ref{a03}) below) and
 a simple pole at $z=0$. In contrast to (\ref{b040})
 %is the Lax matrix of type (\ref{a06}) ({\bf The references to the next section are not good!})
 a special choice of the residue $S$ is assumed in (\ref{a33}). It is a rank one matrix.
 We may fix the rank since the eigenvalues of $S$ are conserved on the dynamics given by (\ref{b14}).
 %, i.e. it
 %corresponds to some special choice of values of the Casimir functions
 %(in the classical Sklyanin algebra),
 likewise the spinless Calogero-Moser model is related to the coadjoint orbit of minimal dimension. Relation (\ref{a33}) can be viewed as the classical version of the IRF-Vertex relation (\ref{a26}). It provides the change $S=S(p,q,\eta,c)$ from canonical variables to spin variables,
which will be discussed in detail in the next subsection. Let us
compute the residue of both parts of (\ref{a33}) at $z=0$. For this purpose we need
the properties of the matrix $g(z,q)$ (\ref{a50})--(\ref{b32}).
In particular, it is degenerated at $z=0$,
and the residue  $\breve{g}(0,q)=\res\limits_{z=0}g^{-1}(z,q)$ is a rank one matrix.
In this way we get parametrization of $S$ matrix in the form $S=\xi\otimes\psi$:
  \beq\label{b37}
  \begin{array}{l}
  \displaystyle{
 S=\frac{\vth'(0)}{\vth(\eta)}\,
 g(N\eta,q)\,e^{P/c} \breve{g}(0,q)
 }
 \end{array}
 \eeq
 or
  \beq\label{b38}
  \begin{array}{c}
  \displaystyle{
 S=\xi\otimes\psi\,,\qquad \xi=\frac{\vth'(0)}{\vth(\eta)}\,
 g(N\eta)\,e^{P/c}\,\rho\,,\qquad \psi=\frac1N\,\rho^T{\breve g}(0)\,.
 }
 \end{array}
 \eeq
 with $\rho$ from (\ref{b30}) and $\breve{g}(0,q)$
 % ({\bf I do not like this notation})
 from (\ref{b31}).

 \paragraph{Factorization from IRF-Vertex relations.} Notice that on one
 hand we deal with the Lax matrix $\mL^\eta(z)$ in the form (\ref{b0402}), and on the other hand we
 use its factorized form (\ref{a33}) for a special  choice of $S$. A connection between these two representations come
 from the IRF-Vertex relation, which includes both $R$-matrix and the matrix
 $g(z,q)$. We review it in Appendix B.
 The easiest way is to use  the identity (\ref{b01}) in $\Mat^{\otimes 2}$, which includes a special matrix
 $\mathcal O_{12}\in\Mat^{\otimes 2}$ (\ref{b03}) with the property (\ref{b07}).
%
 %Let us also remark that the factorized structure of the Lax matrix (\ref{a33}) or (\ref{a31}) directly
 %follows from the relation (\ref{b01}).
 Following \cite{VZ},
 multiply both parts of (\ref{b01}) by $S_2=1_N\otimes S$ with $S$-matrix presented in the form (\ref{b38})
  \beq\label{b381}
  \begin{array}{c}
  \displaystyle{
 \frac{\vth'(0)}{\vth(\eta)}\,g_2(N\eta)e^{P_2/c}{\breve g}_2(0,q)\,R^\eta_{12}(z)=
  }
  \\
   \displaystyle{
 =\frac{\vth'(0)}{\vth(\eta)}\,g_2(N\eta)e^{P_2/c}g_1(z+N\eta,q)\,
 \mathcal O_{12}\, g_2^{-1}(N\eta,q)\,
 g_1^{-1}(z,q)\,.
 }
 \end{array}
 \eeq
 %
% ({\bf Too many notations which are not defined! Undefined $\mathcal O_{12}$ here is too much!})
 Next, compute the trace over the second tensor component of both parts (\ref{b381}). The property  (\ref{b07}) simplifies the r.h.s. of
 (\ref{b381}) since
  $\tr_2(\mathcal O_{12}e^{P_2/c})=e^{P/c}$, and therefore
  \beq\label{b39}
  \begin{array}{c}
  \displaystyle{
 \mL^\eta(z,S)=\tr_2(R^\eta_{12}(z)S_2)=\frac{\vth'(0)}{\vth(\eta)}\,
 g(z+N\eta,q)\,e^{P/c}g^{-1}(z,q)\,.
 }
 \end{array}
 \eeq
 In what follows we also need degeneration of the factorized form. By comparing (\ref{a31}) and (\ref{a12})
 in the $\eta\rightarrow 0$ limit we get
  \beq\label{b391}
  \begin{array}{l}
  \displaystyle{
 \left(g^{-1}(z)g'(z)\right)_{ij}=\frac{1}{N}\,\delta_{ij}\left(E_1(z)-\sum\limits_{k:k\neq i}^NE_1(q_{ik})\right)
 +\frac{1}{N}\,(1-\delta_{ij})\phi(z,q_{ij})\,.
 }
 \end{array}
 \eeq
 In Section 4 we also use degeneration of (\ref{b381}) coming from (\ref{b011}) to derive the accompany $M$-matrix.

%%%%%%%%%%%%%%%%%%%%%%%%%%%%%%%%%%%%%%%%%%
  \paragraph{Explicit change of variables.}
The explicit change of variables
 $S_a=S_a(p,q,\eta,c)$ can be found
 in \cite{Hasegawa} (see also \cite{Chen1,Chen})
 in the elliptic case. (The trigonometric and rational cases
  were addressed in \cite{LOZ8}.) In Appendix C we derive this formula in the elliptic case.
  In our notation it takes the form
  \beq\label{b40}
  \begin{array}{c}
  \displaystyle{
 S_a=\frac{(-1)^{a_1+a_2}}{N}\,e^{\pi\imath a_2\om_a}
 \sum\limits_{m=1}^N e^{p_m/c} e^{2\pi\imath a_2(\eta-{\bar q}_m)}
 \frac{\vth(\eta+\om_{a})}{\vth(\eta)}
 \prod\limits_{l:\,l\neq m}^N\frac{\vth(q_m-q_l-\eta-\om_a)}{\vth(q_m-q_l)}\,,
 }
 \end{array}
 \eeq
 where $\bar q_m$ is the coordinate in the center of masses frame.
The classical Sklyanin generators $S_a$ are dynamical variables in the
relativistic top model described above, and (\ref{b40}) provides its relation to the
Ruijsenaars-Schneider model in the special case ${\rm rk}(S)=1$ generated by the
gauge equivalence (\ref{a33}). Put it differently, (\ref{b40}) is a classical analogue
of the representation of the generalized Sklyanin algebra by difference operators
(in the classical limit the shift operators are substituted by exponents of momenta).

Therefore, we have the following statement.
%\begin{proposition}
 The set of functions $S_a=S_a(p,q,\eta,c)$ satisfy the classical Sklyanin algebra Poisson brackets (\ref{b062})
 computed by means of the canonical Poisson brackets (\ref{a15}).
 The Lax matrix (\ref{b39}) satisfy the classical exchange relations (\ref{b04}).
%\end{proposition}
The proof of a similar statement was proposed in \cite{Chen} by a
direct gauge transformation relating
the $r$-matrix structure (\ref{b04}) with the dynamical one known for the Ruijsenaars-Schneider model \cite{NKSR}.

We also claim that
%\begin{proposition}
the matrix of dynamical variables $S$ with components (\ref{b40})
is represented in the form (\ref{b38}).
%\end{proposition}
%
The proof and explicit expressions for $\xi$ and $\psi$ are given in Appendix C.

%%%%%%%%%%%%%%%%%%%%%%%%%%%%%%%%%%%%%%%%%%%%%%%%%%%%%%%%%%%%%%%%%%%%%%%%%%%%%%%%%%%%%%%%%%%%%%%%%%%
%%%%%%%%%%%%%%%%%%%%%%%%%%%%%%%%%%%%%%%%%%%%%%%%%%%%%%%%%%%%%%%%%%%%%%%%%%%%%%%%%%%%%%%%%%%%%%%%%%%

%\section{Introduction: brief review and summary}
\section{Classical ${\rm GL}_N$ elliptic spin chain}
\setcounter{equation}{0}

%\subsection{Lax operators and factorization}

Here we review properties of the Lax matrices for elliptic classical spin chains and describe the
Hamiltonian flow which is then used for constructing the Ruijsenaars-Schneider
chain in the next section.

We deal with the classical version \cite{Skl,Skl2} of the generalized
elliptic (anisotropic) homogeneous spin chain on $n$ sites associated with ${\rm GL}_N$.
It is described by the
elliptic Baxter-Belavin $R$-matrix \cite{Baxter,RT}.
The generating function of the Hamiltonians is given by
trace $t(z)$ of the monodromy matrix $T(z)$:
  \beq\label{a01}
  \begin{array}{c}
    \displaystyle{
 t(z)=\tr T(z)\,,\qquad T(z)=\mL^{1}(z)\mL^{2}(z)...\mL^{n}(z)\,,
 }
 \end{array}
 \eeq
where $\mL^i(z)$ is the classical Sklyanin's Lax matrix (\ref{b040}) on the $i$-th site of the chain.
It is fixed by the
 quasi-periodic properties on the lattice $\ZZ\oplus\tau\ZZ$
 in the complex plane
 (defining the elliptic curve $\Sigma_\tau={\CC}/(\ZZ\oplus\tau\ZZ)$)
 and the residue at a simple pole $z=0$:
 \beq\label{a02}
 \begin{array}{c}
  \displaystyle{
 \res\limits_{z=0}\mL^i(z)=S^i=\sum\limits_{k,j=1}^N E_{kj}S_{kj}^i\in{\rm Mat}(N,\CC)
 }
 \end{array}
 \eeq
(it is the only pole in the fundamental domain).
Here $E_{kj}$ is the standard matrix basis in ${\rm Mat}(N,\CC)$
(matrix units) and $S^i_{kj}$ are the classical Sklyanin's
generators at $i$-th site\footnote{The latter means that the variables $S^i_{\al}$ satisfy the quadratic Poisson brackets (\ref{b062}) for any fixed $i$.}. The monodromy properties are as follows:
 \beq\label{a03}
 \begin{array}{c}
  \displaystyle{
 \mL^i(z+1)= Q_1^{-1}\mL^{i}(z)Q_1\,,\quad\quad \mL^{i}(z+\tau)=\exp(-2\pi\imath\eta)
  Q_2^{-1} \mL^{i}(z)Q_2\,,
 }
 \end{array}
 \eeq
 where $Q_{1,2}\in{\rm Mat}(N,\CC)$ are finite-dimensional representations for generators of the Heisenberg group given by (\ref{a04}).
 More explicitly,
 \beq\label{a06}
 \begin{array}{c}
  \displaystyle{
 \mL^{i}(z)=\sum\limits_{a\in\,\z_{ N}\times\z_{ N}} S^i_a T_a \vf_a(z,\om_a+\eta)\,,
 }
 \end{array}
 \eeq
 where $T_a$ is the special basis (\ref{a07})
 in ${\rm Mat}(N,\CC)$ constructed by means of the
 matrices $Q_1$, $Q_2$. Similarly to (\ref{b0402}), we can
 write the Lax matrices (\ref{a06}) in the compact form:
  \beq\label{a11}
 \begin{array}{c}
  \displaystyle{
 \mL^i(z)=\tr_2(R_{12}^\eta(z)S_2^i)\,,
 \qquad S_2^i=1_{N\times N}\otimes S^i\in{\rm Mat}(N,\CC)^{\otimes 2}\,.
 }
 \end{array}
 \eeq
 %

%\subsection{Classical spin chain} Let us now proceed to the spin chain.
Consider the lattice
version of the generalized
Landau-Lifshitz model, i.e. the classical elliptic
spin chain \cite{Skl}. It is defined by
the monodromy matrix $T(z)$ (\ref{a01}) with the Lax matrices $\mL^i(z)$ (\ref{a06}) or (\ref{a11}).
The Poisson structure is given by
 $n$ copies of (\ref{b04}):
 \beq\label{b22}
 \begin{array}{c}
  \displaystyle{
 \{\mL_1^{i}(z),\mL_2^{j}(w)\}=\frac{1}{c}\,\delta^{ij}[\mL_1^{i}(z)\mL_2^{i}(w),r_{12}(z-w)]\,.
 }
 \end{array}
 \eeq
In order to have a local Hamiltonian (when only neighbouring sites interact), the residues $S^i$
   \beq\label{b08}
 \begin{array}{c}
  \displaystyle{
 S^i=\res\limits_{z=0}\mL^i(z)\,,\quad i=1\,,\ldots \,,n
 }
 \end{array}
  \eeq
should be rank one matrices:
   \beq\label{b09}
 \begin{array}{c}
  \displaystyle{
 S^i=\xi^i\otimes \psi^i\,,
 }
 \end{array}
  \eeq
 where $\xi^i\in\CC^N$ are column-vectors and $\psi^i\in\CC^N$ are row-vectors.
 Then the local Hamiltonian  is defined as follows.
 Let us compute the coefficient of $t(z)$ (\ref{a01})
 in front of $1/z^n$. It equals
    \beq\label{b16}
 \begin{array}{c}
  \displaystyle{
 \exp(H/c)=\res\limits_{z=0}z^{n-1}t(z)=\tr(S^1S^2...S^n).
 }
 \end{array}
  \eeq
  Plugging (\ref{b09}) into (\ref{b16}) and taking its logarithm, we get
      \beq\label{b17}
 \begin{array}{c}
  \displaystyle{
 H=c\log\tr(S^1S^2...S^n)=c\sum\limits_{k=1}^n \log h_{k,k+1}\,,
 }
 \end{array}
  \eeq
        \beq\label{b18}
 \begin{array}{c}
  \displaystyle{
 h_{k,k+1}=(\psi^k,\xi^{k+1})=\sum\limits_{l=1}^N \psi^k_l\xi^{k+1}_l\,,
 }
 \end{array}
  \eeq
 where $\xi^{n+1}=\xi^{1}$ and the
 notation $(\psi^k,\xi^{k+1})$ means the standard scalar product. To get equations of motion, consider
    \beq\label{b23}
 \begin{array}{c}
  \displaystyle{
  \tr_2\{\mL_1^k(z),T_2(w)\}\stackrel{(\ref{b22})}{=}-\mL^k(z) M^k(z,w)+M^{k-1}(z,w)\mL^k(z)\,,
 }
 \end{array}
  \eeq
 where
     \beq\label{b24}
 \begin{array}{c}
  \displaystyle{
  M^k(z,w)=-\frac{1}{c}\,\tr_2\Big( \mL^1_2(w)\,...\,
  \mL_2^k(w)r_{12}(z-w)\mL_2^{k+1}(w)\,...\,\mL_2^n(w) \Big)
 }
 \end{array}
  \eeq
By taking the coefficient of
the $n$th order pole at $w=0$ in (\ref{b23})--(\ref{b24}) and dividing both parts of
(\ref{b23}) by $\exp(H/c)$ (\ref{b16}), we see that the Lax matrices $\mL^k(z)$ satisfy
 a set
 of the semi-discrete Zakharov-Shabat equations\footnote{The Lax
 equation holds for the monodromy matrix $T(z)$. From
 (\ref{b19}) it follows that ${\dot T}(z)=[T(z),M^n(z)]$.}
  \beq\label{b19}
 \begin{array}{c}
  \displaystyle{
  {\dot \mL}^k(z)=\{H,\mL^k(z)\}=\mL^k(z)M^k(z)-M^{k-1}(z)\mL^k(z)\,,
 }
 \end{array}
  \eeq
 where
         \beq\label{b20}
 \begin{array}{c}
  \displaystyle{
  M^k(z)=-\tr_2\Big( r_{12}(z) {\mathcal S}^{k+1,k}_2\Big)\,,
  \quad \res\limits_{z=0}M^k(z)=-{\mathcal S}^{k+1,k}\,,
  }
 \end{array}
  \eeq
and
 \beq\label{b201}
 \begin{array}{c}
  \displaystyle{
  \quad {\mathcal S}^{k+1,k}=%\frac{1}{c}\,
  \frac{\xi^{k+1}\otimes \psi^{k}}{h_{k,k+1}}\,.
 }
 \end{array}
  \eeq
The second order pole at $z=0$ in the r.h.s. of (\ref{b19})
is cancelled out since $S^k{\mathcal S}^{k+1,k}=S^k$, i.e.
          \beq\label{b21}
 \begin{array}{c}
  \displaystyle{
   \frac{1}{h_{k,k+1}} (\xi^k\otimes \psi^{k})(\xi^{k+1}\otimes \psi^{k})
   -\frac{1}{h_{k-1,k}} (\xi^{k}\otimes \psi^{k-1})  (\xi^k\otimes \psi^{k})=0\,.
 }
 \end{array}
  \eeq
 Similarly to (\ref{b13}) we have the following explicit expression for $M^k(z)$:
  \beq\label{b25}
 \begin{array}{c}
  \displaystyle{
M^k(z)=-{\mathcal S}^{k+1,k}_0 1_N E_1(z)-\sum\limits_{\al\in\ZZ_N^{\times 2}, \al\neq
0}T_\al {\mathcal S}^{k+1,k}_\al\vf_\al(z,\om_\al)\,.
 }
 \end{array}
 \eeq
 Introduce notations $I^\eta(S)=\sum\limits_{\al}T_\al S_\al (E_1(\eta+\om_\al)+2\pi\imath \al_2/N)$
 and $I^0(S)=\sum\limits_{\al\neq 0}T_\al S_\al (E_1(\om_\al)+2\pi\imath \al_2/N)$. Then
  the equations of motion are of the form:
 \beq\label{b26}
 \begin{array}{c}
  \displaystyle{
\dot S^k=
%S^kJ^\eta({\mathcal S}^{k+1,k})-{\mathcal S}^{k+1,k}J^\eta(S^k)\,,
{\mathcal S}^{k,k-1}I^\eta(S^k)-I^\eta(S^k){\mathcal S}^{k+1,k}
+I^0({\mathcal S}^{k,k-1}){S}^k-{S}^k I^0({\mathcal S}^{k+1,k})\,.
 }
 \end{array}
 \eeq
 %
% where ${\bar S}=\sum\limits_{\al\neq 0}T_\al S_\al$ is the traceless part of $S$.
Notice that
 for the linear map (\ref{b15}) we have $J^\eta(S)=I^\eta(S)-I^0(S)$.

%%%%%%%%%%%%%%%%%%%%%%%%%%%%%%%%%%%%%%%%%%%%%%%%%%%%%%%%%%%%%%%%%%%%%%%%%%%%%%%%%%%%%%%%%%%%%%%%%%%
%%%%%%%%%%%%%%%%%%%%%%%%%%%%%%%%%%%%%%%%%%%%%%%%%%%%%%%%%%%%%%%%%%%%%%%%%%%%%%%%%%%%%%%%%%%%%%%%%%%

\section{The Ruijsenaars-Schneider chain}

This section is organized as follows. First, we define the lattice finite-dimensional analogue
of the Ruijsenaars-Schneider model
(the Ruijsenaars spin chain) and find its Lax matrix.
In subsection 4.2 the Hamiltonian and equations of motion
 are derived similarly to those for the elliptic spin chain described in the previous section. In subsection 4.3, using a set of IRF-Vertex type relations, we compute the $M$-matrices
 entering the semi-discrete zero curvature (Zakharov-Shabat) equations. Finally,
 we explain how the obtained Lax pair can be modified in order to have a form similar to the ordinary Ruijsenaars-Schneider model.
 %in
 %subsection 4.4 it is explained how the obtained equations of motion and the
%$L$-$M$ pair can be modified. One way is to remove dependence on
% the center of mass coordinates ${\bar q}^i$. Another one is to change normalization of the Lax matrix. These %modifications allow one to identify the description of the Ruijsenaars-Schneider chain with the results of the next %section obtained in a very different way.

 \subsection{Classical $L$-matrix}

 Let us parameterize all the $L$-matrices of the elliptic spin chain in (\ref{a01})
 by $n$ sets of canonical variables $p^{k}_i,q^{k}_j$, $i,j=1,\ldots ,N$, $k=1,\ldots ,n$
  \beq\label{a410}
  \begin{array}{l}
  \displaystyle{
 \{p^k_i,q_j^l\}=\delta^{kl}\delta_{ij}
 }
 \end{array}
 \eeq
 as in (\ref{a33}), so that
  \beq\label{a41}
  \begin{array}{l}
  \displaystyle{
 \mL^k(z)=\frac{\vth'(0)}{\vth(\eta)}\,
 g(z+N\eta,q^k)\,e^{P^k/c}g^{-1}(z,q^k)\,,\quad P^k={\rm diag}(p^k_1,\ldots ,p^k_N)
 }
 \end{array}
 \eeq
 and
  \beq\label{a411}
  \begin{array}{l}
  \displaystyle{
 S^k=S^k(p^k,q^k)=\xi^k\otimes\psi^k\,,\ \ k=1,\ldots ,n
 }
 \end{array}
 \eeq
 with
  \beq\label{a412}
  \begin{array}{l}
  \displaystyle{
 \xi^k=\xi^k(p^k,q^k)=\frac{\vth'(0)}{\vth(\eta)}\,
 g(N\eta,q^k)\,e^{P^k/c}\,\rho\,,\qquad \psi^k=\psi^k(q^k)=\frac1N\,\rho^T{\breve g}(0,q^k)\,,
 }
 \end{array}
 \eeq
 where $\rho$ is the column vector (\ref{b30}).
Plugging the Lax matrices being written in the factorized form (\ref{a41}) into the monodromy matrix (\ref{a01}), we get
  \beq\label{a42}
  \begin{array}{l}
  \displaystyle{
 T(z)=\Big(\frac{\vth'(0)}{\vth(\eta)}\Big)^n
 g(z+N\eta,q^1)\,e^{P^1/c}g^{-1}(z,q^1)g(z+N\eta,q^2)\,e^{P^2/c}g^{-1}(z,q^2)
 \ldots \,,
 }
 \end{array}
 \eeq
 and, therefore, the transfer-matrix $t(z)$ (\ref{a01}) can be equivalently rewritten in the form
  \beq\label{a43}
  \begin{array}{l}
  \displaystyle{
 t(z)=
 \tr\Big( {\ti L}^1(z) {\ti L}^2(z)\ldots {\ti L}^n(z) \Big)
 }
 \end{array}
 \eeq
(by identifying $q^0=q^n$),
 where
  \beq\label{a440}
  \begin{array}{l}
  \displaystyle{
 {\ti L}^k(z)=g^{-1}(z,q^{k-1}){\mL}^k(z)g(z,q^{k})
 }
 \end{array}
 \eeq
 or (from (\ref{a41}))
  \beq\label{a44}
  \begin{array}{l}
  \displaystyle{
 {\ti L}^k(z)=\frac{\vth'(0)}{\vth(\eta)}\,g^{-1}(z,q^{k-1})g(z+N\eta,q^k)\,e^{P^k/c}\,.
 }
 \end{array}
 \eeq
 To obtain explicit an
 expression for ${\ti L}^k(z)$, we need to compute the matrix $g^{-1}(z,q^{k-1})g(z+N\eta,q^k)$:
  \beq\label{a45}
  \begin{array}{l}
  \displaystyle{
 g^{-1}(z,q^{k-1})g(z+N\eta,q^k)\stackrel{(\ref{a21})}{=}d^{0}(q^{k-1})\Xi^{-1}(z,{\bar q}^{k-1})
 \Xi(z+N\eta,{\bar q}^k)\Big(d^0(q^k)\Big)^{-1}\,,
 }
 \end{array}
 \eeq
 where we have introduced the notation
  \beq\label{a46}
  \begin{array}{l}
  \displaystyle{
 {\bar q}^k_i=q^k_i-\frac{1}{N}\sum\limits_{j=1}^N q^k_j\,,
 }
 \end{array}
 \eeq
 i.e. each $g$-matrix depends on the coordinates in the center of masses frame (see the definitions
 (\ref{a21})-(\ref{a23}) and notice that $d^0(q^k)=d^0({\bar q}^k)$). It is necessary
 for the following reason. The Lax matrix should have a pole at some fixed point ($z=0$),
 and the latter comes from the inverse of $g(z)$. The pole at $z=0$ then appears from $\det\Xi(z,q)$ (\ref{a50}).
 Finally, the theta function $\vth(z)$ in (\ref{a50}) comes from (\ref{a22}) in the following way:
 $\vth(\frac{1}{N}\sum\limits_{k=1}^N(z-N{\bar q}_k))=\vth(z)$, where we used $\sum\limits_{k=1}^N{\bar q}_k=0$.

Coming back to the calculation (\ref{a45}),
 we use the following formula proved in \cite{Hasegawa}:
 \beq\label{a51}
 \begin{array}{c}
  \displaystyle{
 \Big( -\vth'(0)\, \Xi^{-1}(z,{\bar q}^{k-1})\Xi(z+N\eta,{\bar q}^k)\Big)_{ij}=
 \phi(z,{\bar q}^{k-1}_i-{\bar q}^{k}_j+\eta)
 \frac{\prod\limits_{l=1}^N\vth({\bar q}^k_j-{\bar q}^{k-1}_l-\eta) }
 {\prod\limits_{l: l\neq i}^N\vth({\bar q}^{k-1}_i-{\bar q}^{k-1}_l) }\,.
 }
 \end{array}
 \eeq
 Plugging also the matrices $d^0$ (\ref{a23}) into (\ref{a45}), we get
 \beq\label{a52}
 \begin{array}{c}
  \displaystyle{
 \Big( -\vth'(0)\, g^{-1}(z,{\bar q}^{k-1})g(z+N\eta,{\bar q}^k)\Big)_{ij}=
 \phi(z,{\bar q}^{k-1}_i-{\bar q}^{k}_j+\eta)
 \frac{\prod\limits_{l=1}^N\vth({\bar q}^k_j-{\bar q}^{k-1}_l-\eta) }
 {\prod\limits_{l: l\neq j}^N\vth({\bar q}^{k}_j-{\bar q}^{k}_l) }\,.
 }
 \end{array}
 \eeq
 Note that under the
identification ${\bar q}^{k-1}:={\bar q}^{k}$ the upper product in the r.h.s.
acquires the factor $\vth(-\eta)$. Dividing by it the both sides, we reproduce the Lax matrix of the Ruijsenaars-Schneider model (\ref{a12})--(\ref{a13}) or (\ref{a32}).

 Finally, for the $L$-matrices (\ref{a44}) entering the transfer matrix (\ref{a43}) we have:
 \beq\label{a53}
 \begin{array}{c}
  \displaystyle{
 {\ti L}^k_{ij}(z)=
 \phi(z,{\bar q}^{k-1}_i-{\bar q}^{k}_j+\eta)
 \frac{\prod\limits_{l=1}^N\vth({\bar q}^k_j-{\bar q}^{k-1}_l-\eta) }
 {\vth(-\eta)\prod\limits_{l: l\neq j}^N\vth({\bar q}^{k}_j-{\bar q}^{k}_l) }\,e^{p^k_j/c}\,.
 }
 \end{array}
 \eeq

\subsection{Hamiltonian and equations of motion}

\label{section:EM}

\paragraph{The Hamiltonian.} The Hamiltonian can be obtained from $t(z)$ (\ref{a43})
in the same way as in the spin chain case (see (\ref{b16})--(\ref{b18})).
For this purpose compute the residue of ${\ti L}^k_{ij}(z)$:
 \beq\label{b41}
 \begin{array}{c}
   \displaystyle{
 \res\limits_{z=0}{\ti L}^k(z)=\rho\otimes b^k\,,
 }
 \end{array}
 \eeq
 where $\rho$ is taken from (\ref{b30}) and $b^k$ is a row-vector, so that
 \beq\label{b42}
 \begin{array}{c}
  \displaystyle{
 \res\limits_{z=0}{\ti L}^k_{ij}(z)=b_j^k\,,\quad
 b_j^k=
 \frac{\prod\limits_{l=1}^N\vth({\bar q}^k_j-{\bar q}^{k-1}_l-\eta) }
 {\vth(-\eta)\prod\limits_{l: l\neq j}^N\vth({\bar q}^{k}_j-{\bar q}^{k}_l) }\,e^{p^k_j/c}\,.
 }
 \end{array}
 \eeq
Then
\beq\label{b43}
 \begin{array}{c}
  \displaystyle{
 \exp(H/c)=\res\limits_{z=0}z^{n-1}t(z)=\tr\Big( (\rho\otimes b^1)(\rho\otimes b^2)\ldots
 (\rho\otimes b^n)  \Big)\,.
 }
 \end{array}
  \eeq
Finally, the Hamiltonian is of the form
      \beq\label{b44}
 \begin{array}{c}
  \displaystyle{
 H=c\sum\limits_{k=1}^n \log h_{k,k+1}\,,\quad h_{k,k+1}=(\rho^T,b^{k+1})
 }
 \end{array}
  \eeq
  and
   \beq\label{b45}
 \begin{array}{c}
  \displaystyle{
 h_{k-1,k}=(\rho^T,b^{k})=\sum\limits_{j=1}^N b^k_j=\sum\limits_{j=1}^N
 \frac{\prod\limits_{l=1}^N\vth({\bar q}^k_j-{\bar q}^{k-1}_l-\eta) }
 {\vth(-\eta)\prod\limits_{l: l\neq j}^N\vth({\bar q}^{k}_j-{\bar q}^{k}_l) }\,e^{p^k_j/c}\,.
 }
 \end{array}
  \eeq
By construction, the trace $t(z)$ (\ref{a43}) coincides with the one for the elliptic
spin chain (\ref{a01}) under the substitution
(\ref{a41})--(\ref{a412}). To see this, we mention that the terms $h_{k,k+1}$
entering (\ref{b17})--(\ref{b18})
and those from (\ref{b44})--(\ref{b45}) are equal to each other:
   \beq\label{b46}
 \begin{array}{c}
  \displaystyle{
 h_{k-1,k}=(\rho^T,b^{k})=(\psi^{k-1},\xi^k)
 }
 \end{array}
  \eeq
  for $\xi^k$ and $\psi^k$ defined in (\ref{a412}). In order to verify (\ref{b46}),
  one should compare the trace of the
  residue of ${\ti L}^k(z)$ computed from (\ref{a44}) and (\ref{a53}).

\paragraph{The
Hamiltonian equations of motion.} Let us proceed to the
equations of motion. From (\ref{b44})--(\ref{b45}) we have
   \beq\label{b461}
 \begin{array}{c}
  \displaystyle{
 {\dot q}_i^k=\frac{\p H}{\p p_i^k}=
 \frac{b_i^k}{h_{k-1,k}}.
 }
 \end{array}
  \eeq
    The latter yields
   \beq\label{b462}
 \begin{array}{c}
  \displaystyle{
 \sum\limits_{i=1}^N{\dot q}_i^k=1\quad \hbox{for}\ \hbox{all}\ k\,.
 }
 \end{array}
  \eeq
  With (\ref{b461}) the Lax matrix (\ref{a53}) takes the form:
 \beq\label{b463}
 \begin{array}{c}
  \displaystyle{
 {\ti L}^k_{ij}(z)=
 \phi(z,{\bar q}^{k-1}_i-{\bar q}^{k}_j+\eta)b_j^k\,,\quad b_j^k=h_{k-1,k}\,{\dot q}_j^k\,.
 }
 \end{array}
 \eeq
  Next,
   \beq\label{b464}
 \begin{array}{c}
  \displaystyle{
\frac{1}{c}\, {\dot p}_i^k=-\frac{1}{c}\, \frac{\p H}{\p q_i^k}=
-\frac{1}{h_{k-1,k}}\,\frac{\p}{\p q_i^k }\,h_{k-1,k}-\frac{1}{h_{k,k+1}}\,
\frac{\p}{\p q_i^k }\,h_{k,k+1}.
 }
 \end{array}
  \eeq
Its r.h.s is evaluated from the explicit expression (\ref{b45}):
   \beq\label{b465}
 \begin{array}{c}
  \displaystyle{
\frac{1}{h_{k-1,k}}\,\frac{\p}{\p q_i^k }\,h_{k-1,k}=
{\dot q}_i^k\sum\limits_{l=1}^N E_1({\bar q}^{k}_i-{\bar q}^{k-1}_l-\eta)
-{\dot q}_i^k\sum\limits_{l:l\neq i}^N E_1(q_i^k-q_l^k)
+\sum\limits_{l:l\neq i}^N {\dot q}_l^kE_1(q_l^k-q_i^k)-
 }
 \\ \ \\
   \displaystyle{
   -\frac1N\sum\limits_{l=1}^N {\dot q}_l^k \sum\limits_{m=1}^N
   E_1({\bar q}^{k}_l-{\bar q}^{k-1}_m-\eta)\,,
   }
 \end{array}
  \eeq
   \beq\label{b466}
 \begin{array}{c}
  \displaystyle{
\frac{1}{h_{k,k+1}}\,\frac{\p}{\p q_i^k }\,h_{k,k+1}=-\sum\limits_{l=1}^N {\dot q}_l^{k+1}E_1({\bar q}_l^{k+1}-{\bar q}_i^k-\eta)
+\frac1N\sum\limits_{l=1}^N {\dot q}_l^{k+1}\sum\limits_{m=1}^N E_1({\bar q}_l^{k+1}-{\bar q}_m^k-\eta)\,,
 }
 \end{array}
  \eeq
where the last terms (double sums) come from dependence on the center of masses
coordinates (\ref{a46}). Summing up
(\ref{b465}) and (\ref{b466}), we get the following equation for momenta (\ref{b464}):
$$
  \displaystyle{
\frac{1}{c}\, {\dot p}_i^k=-{\dot q}_i^k\sum\limits_{l=1}^N E_1({\bar q}^{k}_i-{\bar q}^{k-1}_l-\eta)
-\sum\limits_{l=1}^N {\dot q}_l^{k+1}E_1({\bar q}_i^k-{\bar q}_l^{k+1}+\eta)
+\sum\limits_{l:l\neq i}^N({\dot q}_i^k+{\dot q}_l^k)E_1(q_i^k-q_l^k)+
 }
 $$
   \beq\label{b467}
 \begin{array}{c}
   \displaystyle{
   +\frac1N\sum\limits_{l=1}^N {\dot q}_l^k \sum\limits_{m=1}^N
   E_1({\bar q}^{k}_l-{\bar q}^{k-1}_m-\eta)
   -\frac1N\sum\limits_{l=1}^N {\dot q}_l^{k+1}\sum\limits_{m=1}^N
   E_1({\bar q}_l^{k+1}-{\bar q}_m^k-\eta)\,.
   }
 \end{array}
  \eeq
  The second line of this equation is independent of the index $i$.
  It has appeared from the dependence of ${\bar q}_l^k$ on the center of masses
  coordinates $\sum_l q_l^k$ at each ($k$-th) site.

\paragraph{The Newtonian form.} Let us represent the Hamiltonian equations of motion in the Newtonian
form.
By differentiating both parts of (\ref{b461}) with respect to
the time $t$, we get
   \beq\label{b701}
 \begin{array}{c}
  \displaystyle{
 {\ddot q}^k_i=\frac{ {\dot b}^k_i }{h_{k-1,k}}-\frac{ {\dot h}_{k-1,k} }{h_{k-1,k}}\,{\dot q}^k_i
 ={\dot q}^k_i\Big( \frac{ {\dot b}^k_i }{b^k_i}-\frac{ {\dot h}_{k-1,k} }{h_{k-1,k}} \Big)\,,
 }
 \end{array}
  \eeq
where
   \beq\label{b702}
 \begin{array}{c}
  \displaystyle{
  \p_t\log h_{k-1,k}=\frac{ {\dot h}_{k-1,k} }{h_{k-1,k}}=
  \frac{ 1 }{h_{k-1,k}}\sum\limits_{l=1}^N {\dot b}^k_l=
  \sum\limits_{l=1}^N {\dot q}^k_l\,\frac{{\dot b}^k_l }{b^k_l}\,.
 }
 \end{array}
  \eeq
  We see that we need to compute ${{\dot b}^k_i }/{b^k_i}$.
  From the definition of $b^k_i$ (\ref{b42}) we have
   \beq\label{b703}
 \begin{array}{c}
  \displaystyle{
  \frac{{\dot b}^k_i }{b^k_i}=
  \sum\limits_{l=1}^N ({\dot {\bar q}}_i^{k}-{\dot {\bar q}}_l^{k-1})E_1({\bar q}_i^k-{\bar q}_l^{k-1}-\eta)
  -\sum\limits_{l:l\neq i}^N ({\dot { q}}_i^{k}-{\dot {q}}_l^{k})
  E_1({q}_i^k-{q}_l^{k})+\frac{1}{c}\, {\dot p}_i^k\,.
 }
 \end{array}
  \eeq
Note that we can remove ``bar'' from velocities ${\dot {\bar q}}$ in the first sum since
   \beq\label{b704}
 \begin{array}{c}
  \displaystyle{
   {\dot {\bar q}}^k_i-{\dot {\bar q}}^m_j={\dot { q}}^k_i-{\dot { q}}^m_j
 }
 \end{array}
  \eeq
for any values of indices due to (\ref{b462}). Using (\ref{b704}) and plugging (\ref{b467}) into (\ref{b703}),
we get
   \beq\label{b705}
 \begin{array}{c}
  \displaystyle{
  \frac{{\dot b}^k_i }{b^k_i}=
   -\sum\limits_{l=1}^N {\dot { q}}_l^{k+1}E_1({\bar q}_i^k-{\bar q}_l^{k+1}+\eta)
  -\sum\limits_{l=1}^N {\dot { q}}_l^{k-1}E_1({\bar q}_i^k-{\bar q}_l^{k-1}-\eta)
  +2\sum\limits_{l:l\neq i}^N {\dot {q}}_l^{k}E_1({q}_i^k-{q}_l^{k})+
 }
 \\
   \displaystyle{
   +\frac1N\sum\limits_{l=1}^N {\dot q}_l^k \sum\limits_{m=1}^N
   E_1({\bar q}^{k}_l-{\bar q}^{k-1}_m-\eta)
   -\frac1N\sum\limits_{l=1}^N {\dot q}_l^{k+1}\sum\limits_{m=1}^N
   E_1({\bar q}_l^{k+1}-{\bar q}_m^k-\eta)
  \,.
   }
 \end{array}
  \eeq
Therefore, from (\ref{b701}) we obtain the following result:
   \beq\label{b706}
 \begin{array}{c}
  \displaystyle{
  \frac{{\ddot q}^k_i }{ {\dot q}^k_i }=
   -\sum\limits_{l=1}^N {\dot { q}}_l^{k+1}E_1({\bar q}_i^k-{\bar q}_l^{k+1}+\eta)
  -\sum\limits_{l=1}^N {\dot { q}}_l^{k-1}E_1({\bar q}_i^k-{\bar q}_l^{k-1}-\eta)
  +2\sum\limits_{l\neq i}^N {\dot {q}}_l^{k}E_1({q}_i^k-{q}_l^{k})+
 }
 \\
   \displaystyle{
   +\frac1N\sum\limits_{l=1}^N {\dot q}_l^k \sum\limits_{m=1}^N
   E_1({\bar q}^{k}_l-{\bar q}^{k-1}_m-\eta)
   -\frac1N\sum\limits_{l=1}^N {\dot q}_l^{k+1}\sum\limits_{m=1}^N
   E_1({\bar q}_l^{k+1}-{\bar q}_m^k-\eta)
    -\p_t\log h_{k-1,k}\,.
   }
 \end{array}
  \eeq
The last term $\p_t\log h_{k-1,k}$ can be found using (\ref{b702}) and (\ref{b705}). We have:
   \beq\label{b707}
 \begin{array}{c}
  \displaystyle{
  \p_t\log h_{k-1,k}=
   -\sum\limits_{m,l=1}^N {\dot { q}}_m^{k}{\dot { q}}_l^{k+1}E_1({\bar q}_m^k-{\bar q}_l^{k+1}+\eta)
  +\sum\limits_{m,l=1}^N {\dot { q}}_l^{k}{\dot { q}}_m^{k-1}E_1({\bar q}_m^{k-1}-{\bar q}_l^k+\eta)
 }
 \\
   \displaystyle{
   +\frac1N\sum\limits_{l=1}^N {\dot q}_l^k \sum\limits_{m=1}^N
   E_1({\bar q}^{k}_l-{\bar q}^{k-1}_m-\eta)
   -\frac1N\sum\limits_{l=1}^N {\dot q}_l^{k+1}\sum\limits_{m=1}^N
   E_1({\bar q}_l^{k+1}-{\bar q}_m^k-\eta)
   \,,
   }
 \end{array}
  \eeq
where for the last line we also used (\ref{b462}). Note also that the
latter expression can be represented in the form
   \beq\label{b708}
 \begin{array}{c}
  \displaystyle{
  \p_t\log h_{k-1,k}={\ti c}^{k-1}-{\ti c}^{k}\,,
   }
 \end{array}
  \eeq
where
   \beq\label{b709}
 \begin{array}{c}
  \displaystyle{
  {\ti c}^{k-1}=\sum\limits_{m,l=1}^N {\dot { q}}_l^{k}{\dot { q}}_m^{k-1}
  E_1({\bar q}_m^{k-1}-{\bar q}_l^k+\eta)
  +\frac1N\sum\limits_{l=1}^N {\dot q}_l^k \sum\limits_{m=1}^N
  E_1({\bar q}^{k}_l-{\bar q}^{k-1}_m-\eta)\,.
   }
 \end{array}
  \eeq
Finally, we obtain the equations of motion by plugging (\ref{b707}) into (\ref{b706}):
   \beq\label{b7091}
 \begin{array}{c}
  \displaystyle{
  \frac{{\ddot q}^k_i }{ {\dot q}^k_i }=
   -\sum\limits_{l=1}^N {\dot { q}}_l^{k+1}E_1({\bar q}_i^k-{\bar q}_l^{k+1}+\eta)
  -\sum\limits_{l=1}^N {\dot { q}}_l^{k-1}E_1({\bar q}_i^k-{\bar q}_l^{k-1}-\eta)
  +2\sum\limits_{l\neq i}^N {\dot {q}}_l^{k}E_1({q}_i^k-{q}_l^{k})+
 }
 \\
   \displaystyle{
   +\sum\limits_{m,l=1}^N {\dot { q}}_m^{k}{\dot { q}}_l^{k+1}E_1({\bar q}_m^k-{\bar q}_l^{k+1}+\eta)
  -\sum\limits_{m,l=1}^N {\dot { q}}_l^{k}{\dot { q}}_m^{k-1}E_1({\bar q}_m^{k-1}-{\bar q}_l^k+\eta)\,.
   }
 \end{array}
  \eeq

\subsection{Semi-discrete Zakharov-Shabat equation}

\label{section:ZS}

As is seen above, the Ruijsenaars-Schneider chain is the gauge transformed
elliptic spin chain together with the change of variables (\ref{a411})--(\ref{a412}).
With this identification, the Hamiltonians (\ref{b44}) and (\ref{b18}) coincide. From the relation (\ref{a440}) between the Lax matrices and the semi-discrete Zakharov-Shabat
equation (\ref{b19}) we conclude that we also have the semi-discrete
Zakharov-Shabat representation for the Ruijsenaars-Schneider chain
\beq\label{b47}
 \begin{array}{c}
  \displaystyle{
  \frac{d}{dt}\,{{\ti L}^k}(z)=\{H,{\ti L}^k(z)\}={\ti L}^k(z){\ti M}^k(z)-{\ti M}^{k-1}(z){\ti L}^k(z)\,,
 }
 \end{array}
  \eeq
with the Lax matrices (\ref{a53}) and the $M$-matrices ${\ti M}^k(z)$:
 \beq\label{b48}
 \begin{array}{c}
  \displaystyle{
  {\ti M}^k(z)=g^{-1}(z,q^k) M^k(z) g(z,q^k) +g^{-1}(z,q^k) {\dot g}(z,q^k)\,.
 }
 \end{array}
  \eeq
Here $M^k(z)$ is given by (\ref{b25}), and the variables
$\xi^k,\psi^k$ in (\ref{b201}) are taken from (\ref{a412}).

The aim of this subsection is to obtain explicit expression for ${\ti M}^k(z)$ using (\ref{b48}). We follow the strategy used in \cite{VZ} to reproduce the Ruijsenaars-Schneider $M$-matrix (\ref{a181}) from the IRF-Vertex relations.

First, let us express $M^k(z)$ (\ref{b20})--(\ref{b201}) through the canonical variables
 \beq\label{b50}
 \begin{array}{c}
  \displaystyle{
  M^k(z)=-\frac{1}{h_{k,k+1}}\tr_2\Big( \Big( \xi^{k+1}(p^{k+1},q^{k+1})\otimes \psi^{k}(q^k) \Big)_2 r_{12}(z)\Big)\,.
  }
 \end{array}
  \eeq
Plugging $\xi^{k+1}$ and $\psi^{k}$ from (\ref{a412}), we have
 \beq\label{b51}
 \begin{array}{c}
  \displaystyle{
  M^k(z)=-\frac{1}{h_{k,k+1}}\frac{\vth'(0)}{\vth(\eta)}\,\frac1N\,
  \tr_2\Big( (\rho\otimes \rho^T)_2\, {\breve g}_2(0,q^k)\,r_{12}(z)
  \,g_2(N\eta,q^{k+1})\,e^{P_2^{k+1}/c} \Big)\,.
  }
 \end{array}
  \eeq
Next, we substitute ${\breve g}(0,q^k)\,r_{12}(z)$ from (\ref{b011}), where all matrices in the r.h.s. depend on $q^k$. Using $(\rho\otimes \rho^T){\mathcal O}_{12}=N{\mathcal O}_{12}$, we obtain:
 \beq\label{b52}
 \begin{array}{c}
  \displaystyle{
  M^k(z)=-\frac{1}{h_{k,k+1}}\frac{\vth'(0)}{\vth(\eta)}\,
  \tr_2\Big(  \Big(g_1'(z,q^k)\, \mathcal O_{12}\, {\breve g}_2(0,q^k)\, g_1^{-1}(z,q^k)
  }
  \\ \ \\
   \displaystyle{
 +g_1(z,q^k)\, \mathcal O_{12}\,A_2(q^k)\, g_1^{-1}(z,q^k)\Big)
 g_2(N\eta,q^{k+1})\,e^{P_2^{k+1}/c} \Big)\,.
  }
 \end{array}
  \eeq
Then the gauged transformed $M$-matrix (\ref{b48}) takes the form
 \beq\label{b53}
 \begin{array}{c}
  \displaystyle{
  {\ti M}^k(z)=-g^{-1}(z)g'(z)G-F+g^{-1}(z,q^k) {\dot g}(z,q^k)\,,
 }
 \end{array}
  \eeq
where
  \beq\label{b54}
  \begin{array}{c}
  \displaystyle{
 G=\frac{1}{h_{k,k+1}}\, \tr_2\left({\mathcal O}_{12}\frac{\vth'(0)}{\vth(\eta)}\,{\breve
 g}_2(0,q^k)g_2(N\eta,q^{k+1})\,
 e^{P^{k+1}_2/c}\right)
 }
 \end{array}
 \eeq
and
  \beq\label{b55}
  \begin{array}{c}
  \displaystyle{
 F=\frac{1}{h_{k,k+1}}\, \tr_2\left({\mathcal O}_{12}
 \frac{\vth'(0)}{\vth(\eta)}\,A_2(q^k)\,g_2(N\eta,q^{k+1})\,
 e^{P_2^{k+1}/c}\right).
 }
 \end{array}
 \eeq
Let us compute the matrices $F$ and $G$. Consider the residue of ${\ti L}^{k+1}(z)$ at $z=0$. On the one hand
it comes from (\ref{a44}), and on the other hand it can be found from (\ref{b41}):
  \beq\label{b56}
  \begin{array}{c}
  \displaystyle{
 {\ti L}^{k+1}[-1]=\res\limits_{z=0}{\ti L}^{k+1}(z)=\frac{\vth'(0)}{\vth(\eta)}\,
 {\breve g}(0,q^{k})g(N\eta,q^{k+1})\,e^{P^{k+1}/c}
 =\rho\otimes b^{k+1}\,.
 }
 \end{array}
 \eeq
Plugging it into (\ref{b54}) and taking also into account (\ref{b46}), we see that $G$ is the identity matrix:
  \beq\label{b57}
  \begin{array}{c}
  \displaystyle{
 G=1_N\,.
 }
 \end{array}
 \eeq
In order to compute the matrix $F$, consider the $z^0$-term in the expansion
of ${\ti L}^{k+1}(z)$ near $z=0$. Using the factorized form (\ref{a44})
and the expansion (\ref{b31}), we obtain:
   \beq\label{b58}
  \begin{array}{c}
  \displaystyle{
 {\ti L}^{k+1}[0]=\frac{\vth'(0)}{\vth(\eta)}\,
 {\breve g}(0,q^{k})g'(N\eta,q^{k+1})\,e^{P^{k+1}/c}+
 \frac{\vth'(0)}{\vth(\eta)}\,A(q^k)\,g(N\eta,q^{k+1})\,
 e^{P^{k+1}/c}.
 }
 \end{array}
 \eeq
The first term in the r.h.s. of (\ref{b58}) is obtained by differentiating both sides of (\ref{b56}) with respect to $\eta$:
  \beq\label{b59}
  \begin{array}{c}
  \displaystyle{
 \p_\eta{\ti L}^{k+1}[-1]=-E_1(\eta){\ti L}^{k+1}[-1]+N\frac{\vth'(0)}{\vth(\eta)}\,
 {\breve g}(0,q^{k})g'(N\eta,q^{k+1})\,e^{P^{k+1}/c}\,,
 }
 \end{array}
 \eeq
so that
   \beq\label{b60}
  \begin{array}{c}
  \displaystyle{
 \frac{\vth'(0)}{\vth(\eta)}\,A(q^k)\,g(N\eta,q^{k+1})\,
 e^{P^{k+1}/c}=
 {\ti L}^{k+1}[0]-\frac1N\Big(\p_\eta{\ti L}^{k+1}[-1]+E_1(\eta){\ti L}^{k+1}[-1]\Big)\,.
 }
 \end{array}
 \eeq
 Expressions ${\ti L}^{k+1}[-1]$ and ${\ti L}^{k+1}[0]$ are known explicitly from (\ref{a53}) and (\ref{a091}):
   \beq\label{b61}
  \begin{array}{c}
  \displaystyle{
  {\ti L}^{k+1}_{ij}[-1]=b_j^{k+1}\,,\quad {\ti L}^{k+1}_{ij}[0]=b_j^{k+1}
  E_1({\bar q}^k_i-{\bar q}^{k+1}_j+\eta)\,,
 }
 \\ \ \\
  \displaystyle{
  \Big(\p_\eta{\ti L}^{k+1}[-1]+E_1(\eta){\ti L}^{k+1}[-1]\Big)_{ij}
  =-b_j^{k+1}\sum\limits_{l=1}^NE_1({\bar q}^{k+1}_j-{\bar q}^{k}_l-\eta)\,.
  }
 \end{array}
 \eeq
Plugging all this into (\ref{b60}) and dividing both parts by $h_{k,k+1}$, we obtain
 (using also the definition (\ref{b461})):
   \beq\label{b62}
  \begin{array}{c}
  \displaystyle{
 \frac{1}{h_{k,k+1}}\frac{\vth'(0)}{\vth(\eta)}
 \Big(A(q^k)\,g(N\eta,q^{k+1})\,
 e^{P^{k+1}/c}\Big)_{ij}
 }
 \\ \ \\
   \displaystyle{
 ={\dot q}_j^{k+1}E_1({\bar q}^k_i-{\bar q}^{k+1}_j+\eta)
 +\frac{1}{N}\,{\dot q}_j^{k+1}\sum\limits_{l=1}^NE_1({\bar q}^{k+1}_j-{\bar q}^{k}_l-\eta)\,.
 }
 \end{array}
 \eeq
Using the property (\ref{b07}), we find the (diagonal) matrix $F$ (\ref{b55}):
   \beq\label{b63}
  \begin{array}{c}
   \displaystyle{
 F_{ij}=\delta_{ij}\sum\limits_{m=1}^N{\dot q}_m^{k+1}E_1({\bar q}^k_i-{\bar q}^{k+1}_m+\eta)
 +\delta_{ij}\frac{1}{N}\sum\limits_{l,m=1}^N{\dot q}_m^{k+1}
 E_1({\bar q}^{k+1}_m-{\bar q}^{k}_l-\eta)\,.
 }
 \end{array}
 \eeq
To get the final answer for ${\ti M}^k(z)$ (\ref{b53}),
let us simplify its last term $g^{-1}(z,q^k) {\dot g}(z,q^k)$.
Using its definition (\ref{a21})--(\ref{a23}), we have:
 \beq\label{b64}
  \begin{array}{c}
  \displaystyle{
 g^{-1}(z){\dot g}(z)=g^{-1}(z)g'(z)\left(-N\,{\rm diag}(\dot q)+1_{N}\sum\limits_{k=1}^N {\dot
 q}_k\right)-{\dot d}^0(d^0)^{-1}\,.
 }
 \end{array}
 \eeq
Substitute (\ref{b64}) and (\ref{b57}) into (\ref{b53}).
Due to (\ref{b462}) the term  $-g^{-1}g'(z)G$
is canceled with the one proportional to $\sum\limits_k{\dot
 q}_k$ in (\ref{b64}):
 \beq\label{b65}
 \begin{array}{c}
  \displaystyle{
  {\ti M}^k(z)=-F-Ng^{-1}(z)g'(z){\rm diag}(\dot q)-{\dot d}^0(d^0)^{-1}\,.
 }
 \end{array}
  \eeq
The quantity $g^{-1}(z)g'(z)$ is known from (\ref{b391}) and
 \beq\label{b66}
 \begin{array}{c}
  \displaystyle{
 \Big({\dot d}^0(d^0)^{-1}\Big)_{ii}=
 \sum\limits_{m:m\neq i}^N({\dot q}_i^k-{\dot q}_m^k)E_1(q_i^k-q_m^k)\,.
 }
 \end{array}
  \eeq
Therefore,
 \beq\label{b67}
 \begin{array}{c}
  \displaystyle{
  {\ti M}_{ij}^k(z)=-(1-\delta_{ij})\phi(z,q_i^k-q_j^k)\,
  {\dot q}_j^k-\delta_{ij}E_1(z){\dot q}^k_i+
 }
 \end{array}
  \eeq
  $$
   \displaystyle{
  +\delta_{ij}\Big(
  \sum\limits_{m:m\neq i}^N{\dot q}_m^kE_1(q_i^k-q_m^k)
  -\sum\limits_{m=1}^N{\dot q}_m^{k+1}E_1({\bar q}^k_i-{\bar q}^{k+1}_m+\eta)
  -\frac{1}{N}\sum\limits_{l,m=1}^N{\dot q}_m^{k+1}E_1({\bar q}^{k+1}_m-{\bar q}^{k}_l-\eta)
    \Big)\,.
 }
 $$

To summarize, we have proved that
%the following statement.
%\begin{proposition}
 the semi-discrete Zakharov-Shabat equation (\ref{b47}) holds for the matrices (\ref{b463}) and (\ref{b67}) on the equations of motion
 of the Ruijsenaars-Schneider chain (\ref{b461}), (\ref{b467}) or (\ref{b705})--(\ref{b707}).
%\end{proposition}
This can be also verified by direct substitution using identities (\ref{b151}) and (\ref{a092}) similarly to
verification of the Lax pair for the
Ruijsenaars-Schneider model (\ref{a12}), (\ref{a181}).

\paragraph{Modified Lax pair.} All Lax matrices (\ref{b463}) can be
simultaneously divided by $h_{k-1,k}$. Then the resultant Lax matrix depend on the velocities (\ref{b461}). From the point of view of the ordinary Ruijsenaars-Schneider model it is similar to transition to the logarithm of Hamiltonian
(\ref{a172}). Consider
 \beq\label{b46333}
 \begin{array}{c}
  \displaystyle{
 {L'}^k_{ij}(z)={\ti L}^k_{ij}(z)\frac{1}{h_{k-1,k}}=
 \phi(z,{\bar q}^{k-1}_i-{\bar q}^{k}_j+\eta){\dot q}_j^k\,.
 }
 \end{array}
 \eeq
This can be done since the transfer matrix is divided by conserved quantity (\ref{b43}):
  \beq\label{a4333}
  \begin{array}{l}
  \displaystyle{
 t'(z)=
 \tr\Big( { L'}^1(z) { L'}^2(z)...{ L'}^n(z) \Big)=
 \frac{t(z)}{h_{1,2}h_{2,3}...h_{n,1}}=t(z)\exp(-H/c)\,,
 }
 \end{array}
 \eeq
The Lax equation for ${ L'}^k(z)$ is of the form:
\beq\label{b4733}
 \begin{array}{c}
  \displaystyle{
  \frac{d}{dt}\,{{ L'}^k}(z)={L'}^k(z){\ti M}^k(z)-{\ti M}^{k-1}(z){ L'}^k(z)-
  { L'}^k(z)\p_t\log h_{k-1,k}\,,
 }
 \end{array}
  \eeq
Using (\ref{b708}) the last term in (\ref{b4733}) can be removed by redefining ${\ti M}^k(z)$:
\beq\label{b4734}
 \begin{array}{c}
  \displaystyle{
  {M'}^k(z)={\ti M}^k(z)+{\ti c}^k 1_N\,.
 }
 \end{array}
  \eeq
Then
\beq\label{b4735}
 \begin{array}{c}
  \displaystyle{
  \frac{d}{dt}\,{{ L'}^k}(z)={L'}^k(z){M'}^k(z)-{ M'}^{k-1}(z){ L'}^k(z)\,.
 }
 \end{array}
  \eeq
Explicit form of the $M$-matrix (\ref{b4734}) is as follows:
\beq\label{b4736}
 \begin{array}{c}
  \displaystyle{
  {M'}^k_{ij}(z)=-(1-\delta_{ij})\phi(z,q_i^k-q_j^k)\,
  {\dot q}_j^k-\delta_{ij}E_1(z){\dot q}^k_i+
 }
 \end{array}
  \eeq
  $$
   \displaystyle{
  +\delta_{ij}\Big(
  \sum\limits_{m:m\neq i}^N{\dot q}_m^kE_1(q_i^k-q_m^k)
  -\sum\limits_{m=1}^N{\dot q}_m^{k+1}E_1({\bar q}^k_i-{\bar q}^{k+1}_m+\eta)
  +\sum\limits_{m,l=1}^N {\dot { q}}_l^{k+1}{\dot { q}}_m^{k}
  E_1({\bar q}_m^{k}-{\bar q}_l^{k+1}+\eta)
    \Big)\,.
 }
 $$
To summarize, the Lax pair (\ref{b46333}) and (\ref{b4736}) satisfy the semi-discrete zero curvature equation
(\ref{b4735}) and provide equations of motion of the Ruijsenaars-Schneider chain (\ref{b7091}).

%%%%%%%%%%%%%%%%%%%%%%%%%%%%%%%%%%%%%%%%%%%%%%%%%%%%%%%%%%%%%%%%%%%%%%%%%%%%%%%%%%%%%%%%%%%%%%%%%%%%%%%%%%
%%%%%%%%%%%%%%%%%%%%%%%%%%%%%%%%%%%%%%%%%%%%%%%%%%%%%%%%%%%%%%%%%%%%%%%%%%%%%%%%%%%%%%%%%%%%%%%%%%%%%%%%%%
%%%%%%%%%%%%%%%%%%%%%%%%%%%%%%%%%%%%%%%%%%%%%%%%%%%%%%%%%%%%%%%%%%%%%%%%%%%%%%%%%%%%%%%%%%%%%%%%%%%%%%%%%%

\section{Field analogue of the elliptic Ruijsenaars-Schneider system from elliptic families
of solutions to the 2D Toda lattice}

In this section we derive equations of motion for poles of general elliptic solutions
(which we call elliptic families) to the
2D Toda lattice hierarchy and show that
they are Hamiltonian and equivalent to (\ref{b7091}) under some simple
substitutions and re-definitions.

\subsection{The 2D Toda lattice hierarchy}

Following \cite{UT84}, we briefly review the 2D Toda lattice
hierarchy. The sets of independent variables are two infinite
sets of continuous time variables ${\bf t}=\{t_1, t_2, t_3, \ldots \}$,
$\bar {\bf t}=\{\bar t_1, \bar t_2, \bar t_3, \ldots \}$ and a discrete
integer-valued variable $n$ which is sometimes denoted as $t_0$.
The main objects are two pseudo-difference Lax operators
\beq\label{toda1}
{\bf L}=e^{\p_n}+\sum_{k\geq 0}U_{k,n} e^{-k\p_n}, \quad
\bar {\bf L}=a_n e^{-\p_n}+\sum_{k\geq 0}\bar U_{k,n} e^{k \p_n},
\eeq
where $e^{\p_n}$ is the shift operator acting as
$e^{\pm \p_n}f(n)=f(n\pm 1 )$ and
the coefficient functions
are functions of ${\bf t}$, $\bar {\bf t}$.
The equations of the hierarchy are differential
equations for the functions $a_n$, $U_{k,n}$, $\bar U_{k,n}$.
They are encoded in the Lax equations
\beq\label{toda2}
\p_{t_m}{\bf L}=[{\cal B}_m, {\bf L}], \quad
\p_{t_m}\bar {\bf L}=[{\cal B}_m, \bar {\bf L}]
\qquad {\cal B}_m=({\bf L}^m)_{\geq 0},
\eeq
\beq\label{toda2a}
\p_{\bar t_m}{\bf L}=[\bar {\cal B}_m, {\bf L}], \quad
\p_{\bar t_m}\bar {\bf L}=[\bar {\cal B}_m, \bar {\bf L}]
\qquad \bar {\cal B}_m=(\bar {\bf L}^m)_{< 0},
\eeq
where we use the notation
$$\Bigl (\sum_{k\in \z} U_{k,n} e^{k \p_n}\Bigr )_{\geq 0}=
\sum_{k\geq 0} U_{k,n} e^{k\p_n}, \quad
\Bigl (\sum_{k\in \z} U_{k,n} e^{k\p_n}\Bigr )_{< 0}=
\sum_{k<0} U_{k,n} e^{k\p_n}.$$
For example, ${\cal B}_1=e^{\p_n}+b_n$, $\bar {\cal B}_1=a_ne^{-\p_n}$,
where we have denoted $U_{0,n}=b_n$. It can be shown that the zero
curvature (Zakharov-Shabat) equations
\beq\label{toda3}
\p_{t_n}{\cal B}_m -\p_{t_m}{\cal B}_n +[{\cal B}_m, {\cal B}_n]=0,
\eeq
\beq\label{toda3a}
\p_{\bar t_n}{\cal B}_m -\p_{t_m}\bar {\cal B}_n +[{\cal B}_m, \bar {\cal B}_n]=0,
\eeq
\beq\label{toda3b}
\p_{\bar t_n}\bar {\cal B}_m -\p_{\bar t_m}\bar {\cal B}_n +[\bar {\cal B}_m,
\bar {\cal B}_n]=0
\eeq
provide an equivalent formulation of the hierarchy.

The 2D Toda equation is the first member of the hierarchy. It is obtained
from (\ref{toda3a}) at $m=n=1$ which is equivalent to the system of equations
$$
\left \{ \begin{array}{l}
\p_{t_1}\log a_n=b_n-b_{n-1}
\\ \\
\p_{\bar t_1} b_n =a_n -a_{n+1}.
\end{array}
\right.
$$
Excluding $b_n$ from this system, we get the differential equation for $a_n$:
\beq\label{toda4}
\p_{t_1}\p_{\bar t_1}\log a_n=2a_n-a_{n+1}-a_{n-1}.
\eeq
It is one of the forms of the 2D Toda equation. In terms of the function
$\varphi_n$ introduced through the relation $a_n=e^{\varphi_n-\varphi_{n-1}}$
it acquires the familiar form
\beq\label{toda5}
\p_{t_1}\p_{\bar t_1}\varphi_n=e^{\varphi_n-\varphi_{n-1}}-
e^{\varphi_{n+1}-\varphi_n}.
\eeq

The universal dependent variable of the hierarchy is the tau-function
$\tau_n =\tau_n ({\bf t}, \bar {\bf t})$.
The change of the dependent variables from $a_n, b_n$ to the tau-function,
\beq\label{toda6}
a_n=\frac{\tau_{n+1}\tau_{n-1}}{\tau^2_n}, \qquad
b_n=\p_{t_1}\log \frac{\tau_{n+1}}{\tau_n},
\eeq
brings the 2D Toda equation to the form \cite{JM83}
\beq\label{toda7}
\p_{t_1}\p_{\bar t_1}\log \tau_n=-\frac{\tau_{n+1}\tau_{n-1}}{\tau^2_n}.
\eeq

At fixed $n$ and $\bar {\bf t}$ the equations of the
2D Toda lattice hierarchy are reduced
to the Kadom\-tsev-Pet\-vi\-ash\-vi\-li (KP) hierarchy with the
independent variables
${\bf t}=\{t_1, t_2, t_3,\ldots \}$, with the KP equation (the first member
of the hierarchy) being satisfied by
\beq\label{toda9}
u_n =\p^2_{t_1}\log \tau_n.
\eeq

An important class of solutions to the 2D Toda lattice hierarchy is the
algebraic-geometrical solutions constructed from a smooth algebraic curve
$\Gamma$ of genus $g$ with some extra data. The tau-function for such solutions
is given by \cite{Krichever77,Dubrovin81}
\beq
\tau_n ({\bf t}, \bar {\bf t})=e^{Q(n, {\bf t}, \bar {\bf t})}
\Theta \Bigl (
{\bf V}_0 n +\sum_{k\geq 1}{\bf V}_k t_k +\sum_{k\geq 1}\bar {\bf V}_k \bar t_k
+{\bf Z}\Bigr ),
\eeq
where $Q$ is a quadratic form of its variables and $\Theta$ is the Riemann
theta-function with the Riemann's matrix being the period matrix of holomorphic
differentials on the curve $\Gamma$. Components of the
$g$-dimensional vectors ${\bf V}_k$,
$\bar {\bf V}_k$ are $b$-periods of certain normalized meromorphic differentials
on $\Gamma$ with poles at two marked points $P_0, P_{\infty}\in \Gamma$
(see \cite{Krichever77,Dubrovin81} for details).

When one considers algebraic-geometrical solutions, it is natural to treat
$t_0=n$ as a continuous rather than discrete variable.
Namely, let us introduce the continuous
variable $x=x_0+\eta n$, where $\eta$ is a constant (a lattice spacing),
then the Toda equation becomes a difference equation in $x$:
\beq\label{t3}
\p_{t_1}\p_{\bar t_1}\log a(x)=2a(x)-a(x+\eta )-a(x-\eta ).
\eeq
It is equivalent to the zero curvature equation
$\p_{\bar t_1}{\cal B}_1 -\p_{t_1}\bar {\cal B}_1 +[{\cal B}_1, \bar {\cal B}_1]=0$ for the
difference operators
\beq\label{t1}
{\cal B}_1=e^{\eta \p_x}+b(x), \qquad \bar {\cal B}_1 =a(x)e^{-\eta \p_x},
\eeq
which is the compatibility condition of the linear problems
\beq\label{t2}
\begin{array}{l}
\p_{t_1}\psi (x)=\psi (x+\eta )+b(x)\psi (x),
\\ \\
\p_{\bar t_1}\psi (x)=a(x)\psi (x-\eta )
\end{array}
\eeq
for a wave function $\psi$.

\subsection{Elliptic families}

Let us fix the variables $\bar {\bf t}$ and consider the dependent variables
as functions of $x, {\bf t}$.
A general solution to the 2D Toda and KP equations is known to be of the form
\beq\label{t8}
\begin{array}{l}
\displaystyle{a(x,  {\bf t})=
\frac{\tau (x+\eta , {\bf t})
\tau (x-\eta , {\bf t})}{\tau^2 (x, {\bf t})},}
\\ \\
\displaystyle{
b(x, {\bf t})=\p_{t_1}\log \frac{\tau (x+\eta , {\bf t})}{\tau (x, {\bf t})},}
\\ \\
\displaystyle{
u(x, {\bf t})=\p^2_{t_1}\log \tau (x, {\bf t}).}
\end{array}
\eeq
We are going to consider solutions
that are elliptic functions with respect to
some variable $t_k$ or a linear combination
$\displaystyle{\lambda =\beta_0x+\sum_k \beta_k t_k}$.
We call them elliptic families. The elliptic families form a subclass
of algebraic-geometrical solutions. As it was already mentioned
in Introduction, an algebraic-geometrical solution is elliptic
with respect to some direction if there exists a $g$-dimensional
vector ${\bf W}$ such that it spans an elliptic curve ${\cal E}$
embedded in the Jacobian
of the curve $\Gamma$:
\beq\label{toda8a}
\tau (x, {\bf t}, \lambda )=e^{Q(x, {\bf t})}
\Theta \Bigl (
{\bf V}_0 x/\eta +\sum_{k\geq 1}{\bf V}_k t_k +{\bf W}\lambda
+{\bf Z}\Bigr ),
\eeq
This is a nontrivial transcendental constraint. The space
of corresponding algebraic curves has codimension $g-1$ in the moduli space of
all the curves. If such a vector ${\bf W}$ exists, then the theta-divisor intersects
the shifted elliptic curve $\displaystyle{{\cal E}+{\bf V}_0 x/\eta +
\sum_k {\bf V}_k t_k}$ at
a finite number of points $\lambda_i =\lambda_i(x,{\bf t})$. Therefore,
for elliptic families we have:
\beq\label{toda8b}
\Theta \Bigl (
{\bf V}_0 x/\eta +\sum_{k\geq 1}{\bf V}_k t_k +{\bf W}\lambda
+{\bf Z}\Bigr )=f(x, {\bf t})e^{\gamma_1\lambda +\gamma_2\lambda^2}\prod _{i=1}^N
\sigma (\lambda -\lambda_i(x, {\bf t})).
\eeq
Here $\gamma_1, \gamma_2$ are constants and $\sigma (\lambda )$ is the
Weierstrass $\sigma$-function defined in the Appendix. The form of the exponential
factor in the right hand side
of (\ref{toda8b}) follows from monodromy properties of the theta-function.
The function $f(x, {\bf t})$ does not depend on $\lambda$. In the tau-function,
it is modified by the factor $e^{Q(x, {\bf t})}$ which also does not depend on
$\lambda$.
Having in mind the discrete version, we will also denote $\lambda_i^k=\lambda_i (x)$
for $x=x_0+k\eta$.

In what follows we denote $t_1=t$. From (\ref{toda8b}) we conclude that
if $b(x, t, \lambda )$ is an elliptic family of solutions to the
2D Toda equation, then it has the form
\beq\label{el1}
b(x, t, \lambda )=\sum_{i=1}^N \Bigl (\dot \lambda_i (x)\zeta (\lambda -\lambda_i(x))
-\dot \lambda_i (x+\eta )\zeta (\lambda -\lambda_i(x+\eta ))\Bigr ) +c(x,t),
\eeq
where dot means the $t$-derivative and $c(x,t)$ is some function.
Since $a(x,t, \lambda )$,
$b(x, t, \lambda )$ and $u(x,t,\lambda )$ given by (\ref{t8})
are elliptic functions of $\lambda$, one should have
\beq\label{el2a}
\sum\limits_{i=1}^N\Bigl (\lambda_i(x+\eta )+\lambda_i(x-\eta )-2\lambda_i(x)\Bigr )=0,
\eeq
\beq\label{el2}
\sum\limits_{i=1}^N \dot \lambda_i(x+\eta )=\sum\limits_{i=1}^N \dot \lambda_i(x),
\eeq
\beq\label{el2c}
\sum\limits_{i=1}^N \ddot \lambda_i(x)=0.
\eeq
From (\ref{el2a}), (\ref{el2}), (\ref{el2c}) it follows that
\beq\label{el2b}
\sum\limits_{i=1}^N \lambda_i(x)= \alpha x +\beta t+\alpha_0, \qquad \dot \alpha =\dot \beta =0.
\eeq
Here $\alpha_0, \alpha , \beta$ are unspecified
$\eta$-periodic functions of $x$. We can say
that the requirement of ellipticity implies that the ``center of masses'' of the points
$\lambda_i$ moves linearly in time.

A meromorphic function $f(\lambda )$ is called a {\it double-Bloch} function if it satisfies the
following monodromy properties:
\beq\label{bl1}
f(\lambda +2\omega_{\alpha})=B_{\alpha}f(\lambda ), \quad \alpha =1,2.
\eeq
The complex constants $B_{\alpha}$ are called Bloch multipliers. Our goal is to find
$b(x, t, \lambda )$ such that the equation
\beq\label{bl2}
\p_{t}\psi (x)-\psi (x+\eta )-b(x)\psi (x)=0
\eeq
has sufficiently many double-Bloch solutions. The existence of such solutions turn out to be
a very restrictive condition.
The double-Bloch functions with simple poles $\lambda_i$ can be represented in the form
\beq\label{el3}
\psi (x)=\sum\limits_{i=1}^N c_i(x)\Phi (\lambda -\lambda_i(x), z ),
\eeq
where $c_i$ are residues at the poles $\lambda_i$ and the function $\Phi (\lambda , z)$ is defined
in (\ref{w02}). The variable $z$ has the meaning of the spectral parameter.
In what follows we often suppress the second argument of $\Phi$ writing simply
$\Phi (\lambda , z):=\Phi (\lambda )$.

\subsection{Equations of motion of the field analogue of the elliptic
Ruij\-se\-na\-ars-Schnei\-der system}\label{section:53}

Our strategy is similar to that of the work \cite{AKV02}. We are going
to substitute (\ref{el1}), (\ref{el3}) into (\ref{bl2}) and cancel all the
poles which are
at $\lambda =\lambda_i(x)$ and $\lambda =\lambda_i(x+\eta)$.
The substitution gives:
$$
\sum\limits_{i=1}^N \dot c_i(x)\Phi (\lambda -\lambda_i(x))-\sum\limits_{i=1}^N c_i(x)\dot \lambda_i(x)
\Phi ' (\lambda -\lambda_i(x))-\sum\limits_{i=1}^N
c_i(x+\eta )\Phi (\lambda -\lambda_i(x+\eta ))
$$
$$
-\sum\limits_{i=1}^N \Bigl ((\dot \lambda_i (x)\zeta (\lambda -\lambda_i(x))
-\dot \lambda_i (x+\eta )\zeta (\lambda -\lambda_i(x+\eta ))\Bigr ) \sum\limits_{j=1}^N
c_j(x)\Phi (\lambda -\lambda_j(x))
$$
$$-c(x,t)\sum\limits_{i=1}^N c_i(x)\Phi (\lambda -\lambda_i(x))=0.
$$
The cancellation of poles yields the following system of equations:
\beq\label{el4}
c_i(x+\eta )=\dot \lambda_i(x+\eta )\sum\limits_{j=1}^N c_j(x)\Phi (\lambda_i(x+\eta)-\lambda_j(x)),
\eeq
\beq\label{el5}
\begin{array}{c}
\displaystyle{ \dot c_i(x)=\dot \lambda_i(x)\sum\limits_{j:j\neq i}^N c_j(x)
\Phi (\lambda_i(x)-\lambda_j(x)) +c_i(x)\sum\limits_{j:j\neq i}^N\dot \lambda_j(x)
\zeta (\lambda_i(x)-\lambda_j(x))}
\\ \\
\displaystyle{
-c_i(x) \sum\limits_{j=1}^N \dot \lambda_j(x+\eta )\zeta (\lambda_i(x)-\lambda_j (x+\eta))+
c_i(x)c (x,t)}.
\end{array}
\eeq
Introducing the matrices $L$, $M$ as
\beq\label{L}
L_{ij}(x, z)=\dot \lambda_i(x+\eta )\Phi (\lambda_i (x+\eta )-\lambda_j(x), z),
\eeq
\beq\label{M}
\begin{array}{c}
M_{ij}(x, z)=(1-\delta_{ij})
\dot \lambda_i(x)\Phi (\lambda_i (x)-\lambda_j(x), z)
\\ \\
\displaystyle{
+\, \delta_{ij}\Bigl (\sum\limits_{k:k\neq i}^N\dot \lambda_k (x)\zeta (\lambda_i(x)-\lambda_k(x))-
\sum\limits_{k=1}^N\dot \lambda_k (x+\eta )
\zeta (\lambda_i(x)-\lambda_k(x+\eta ))+c(x,t)\Bigr )},
\end{array}
\eeq
one can write the system (\ref{el4}), (\ref{el5}) in the form
\beq\label{el45}
c_i(x+\eta )=\sum\limits_{j=1}^N L_{ij}(x)c_j(x), \quad
\dot c_i(x)=\sum\limits_{j=1}^N M_{ij}(x)c_j(x).
\eeq
Let us also introduce the matrices
\beq\label{el6}
\begin{array}{l}
A_{ij}^+(x)=\Phi (\lambda_i(x+\eta)-\lambda_j(x)),
\\ \\
A_{ij}^0(x)=(1-\delta_{ij})\Phi (\lambda_i(x)-\lambda_j(x))
\end{array}
\eeq
and diagonal matrices
\beq\label{el7}
\begin{array}{l}
\Lambda_{ij}(x)=\delta_{ij}\lambda_i(x),
\\ \\
\displaystyle{D_{ij}^0(x)=\delta_{ij}\sum\limits_{k:k\neq i}^N\dot \lambda_k(x)
\zeta (\lambda_i(x)-\lambda_k(x))},
\\ \\
\displaystyle{D_{ij}^\pm (x)=\delta_{ij}\sum\limits_{k=1}^N\dot \lambda_k(x\pm \eta )
\zeta (\lambda_i(x)-\lambda_k(x\pm \eta ))}.
\end{array}
\eeq
In terms of these matrices, the matrices $L$ and $M$ read:
\beq\label{el8}
L(x)=\dot \Lambda (x+\eta)A^+(x), \quad M(x)=\dot \Lambda (x)A^0(x)+D^0(x)-
D^+(x)+c(x, t)I,
\eeq
where $I$ is the unity matrix. The compatibility condition of the overdetermined system
(\ref{el4}), (\ref{el5}) is the semi-discrete zero curvature (Zakharov-Shabat) equation
\beq\label{el9}
R(x):=\dot L(x)+L(x)M(x)-M(x+\eta )L(x)=0.
\eeq
The matrices $L$, $M$ here depend on the spectral parameter $z$.
We have:
$$
R(x)=\ddot \Lambda (x+\eta )A^+(x)+\dot \Lambda (x+\eta )\Bigl (
S(x)+A^+(x)(D^0(x)-D^+(x))
$$
$$-(D^0(x+\eta )-D^+(x+\eta ))A^+(x)
+(c(x,t)-c(x+\eta , t))A^+(x)\Bigr ),
$$
where
$$
S(x)=\dot A^+(x)+A^+(x)\dot \Lambda (x)A^0(x)-A^0(x+\eta )\dot \Lambda (x+\eta )A^+(x).
$$
Using (\ref{id1}), (\ref{id2}), we calculate:
$$
\dot A^+_{ij}(x)=(\dot \lambda_i(x+\eta )-\dot \lambda_j(x))\Phi (\lambda_i(x+\eta )-\lambda_j(x))
$$
$$
\times \Bigl (
\zeta (\lambda_i(x+\eta )-\lambda_j(x)+{z} )-\zeta (\lambda_i(x+\eta )-\lambda_j(x))-
\zeta ({z} )\Bigr ) ,
$$
and
$$
\Bigl (A^+(x)\dot \Lambda (x)A^0(x)-A^0(x+\eta )\dot \Lambda (x+\eta )A^+(x)\Bigr )_{ij}
$$
$$
=\sum\limits_{k:k\neq j}^N\Phi (\lambda_i(x+\eta )-\lambda_k(x))\Phi (\lambda_k(x)-\lambda_j(x))
\dot \lambda_k(x)
$$
$$-
\sum\limits_{k:k\neq i}^N\Phi (\lambda_i(x+\eta )-\lambda_k(x+\eta))\Phi (\lambda_k(x+\eta)-\lambda_j(x))
\dot \lambda_k(x+\eta)
$$
$$
=\Phi (\lambda_i(x+\eta)-\lambda_j(x))\left (
\sum\limits_{k:k\neq j}^N\dot \lambda_k(x)\zeta (\lambda_i(x+\eta)-\lambda_k(x))-
\sum\limits_{k:k\neq j}^N\dot \lambda_k(x)\zeta (\lambda_j(x)-\lambda_k(x))\right )
$$
$$
-\Phi (\lambda_i(x+\eta)-\lambda_j(x))\left (
\sum\limits_{k:k\neq i}^N\dot \lambda_k(x+\eta)\zeta (\lambda_i(x+\eta)-\lambda_k(x+\eta))\right.
$$
$$-\left.
\sum\limits_{k:k\neq i}^N\dot \lambda_k(x+\eta)\zeta (\lambda_j(x)-\lambda_k(x+\eta))\right )
$$
$$
+(\dot \lambda_i(x+\eta )-\dot \lambda_j(x))\Phi (\lambda_i(x+\eta )-\lambda_j(x))
\Bigl (\zeta ({z} )-
\zeta (\lambda_i(x+\eta )-\lambda_j(x)+{z} )\Bigr ).
$$
In the calculation, the condition (\ref{el2}) was taken into account.
Therefore, we have:
\beq\label{el10}
S_{ij}(x)=\Phi (\lambda_i(x+\eta )-\lambda_j(x))\Bigl (
D_{ii}^-(x+\eta )+D_{jj}^+(x)-D_{jj}^0(x)-D_{ii}^0(x+\eta )\Bigr )
\eeq
and, combining everything together, we obtain the matrix identity
\beq\label{el11}
R(x)=\Bigl (\ddot \Lambda (x+\eta )\dot \Lambda^{-1}(x+\eta)+D^-(x+\eta)+
D^+(x+\eta )-2D^0(x+\eta )
$$
$$+(c(x,t)-c(x+\eta , t))I\Bigr ) L(x),
\eeq
from which we see that the compatibility condition is equivalent to vanishing of the
diagonal matrix in front of $L(x)$:
\beq\label{el12}
\ddot \Lambda (x+\eta )\dot \Lambda^{-1}(x+\eta)+D^-(x+\eta)+
D^+(x+\eta )-2D^0(x+\eta )+(c(x,t)-c(x+\eta , t))I=0.
\eeq
This results in the equations of motion
\beq\label{el13}
\begin{array}{c}
\displaystyle{
\ddot \lambda_i(x)+\sum\limits_{k=1}^N \Bigl (
\dot \lambda_i(x)\dot \lambda_k(x-\eta )\zeta (\lambda_i(x)-\lambda_k(x-\eta ))+
\dot \lambda_i(x)\dot \lambda_k(x+\eta )\zeta (\lambda_i(x)-\lambda_k(x+\eta ))\Bigr )}
\\ \\
\displaystyle{
-\, 2 \sum\limits_{k:k\neq i}^N\dot \lambda_i(x)\dot \lambda_k(x )
\zeta (\lambda_i(x)-\lambda_k(x))+(c(x-\eta , t)-c(x, t))\dot \lambda_i(x)=0.}
\end{array}
\eeq
Equations (\ref{el13})
resemble the Ruijsenaars-Schneider equations of motion (\ref{a16}) and provide
their field generalization. If $\lambda_i(x)=x_i(t)+x$, then
equations (\ref{el13})
become the equations of motion for the elliptic Ruijsenaars-Schneider system
with coordinates of particles $x_i$ (with
$c(x,t)=c(x+\eta , t)$).

The condition (\ref{el2b}) allows us to find the explicit form of the function
$c(x,t)$. Summing equations (\ref{el13}) over $i=1, \ldots , N$
and using (\ref{el2b}), we get
\beq\label{el14a}
c(x, t)=\Bigl (\sum\limits_{i=1}^N \dot \lambda_i(x)\Bigr )^{-1}
\sum\limits_{i,k=1}^N\dot \lambda_i(x)\dot \lambda_k(x+\eta )
\zeta (\lambda_i(x)-\lambda_k(x+\eta ))
\eeq
(up to an arbitrary function of $t$ and
an $\eta$-periodic function of $x$ which do not affect the equations of motion).

Let us also present the lattice version of equations (\ref{el13}) which are obtained from
them after the substitution $\lambda_i^k=\lambda_i (k\eta +x_0)$:
\beq\label{el13a}
\begin{array}{c}
\displaystyle{
\ddot \lambda_i^k+\sum\limits_{j=1}^N \Bigl (
\dot \lambda_i^k\dot \lambda_j^{k-1}\zeta (\lambda_i^k-\lambda_j^{k-1})+
\dot \lambda_i^k\dot \lambda_j^{k+1}\zeta (\lambda_i^k-\lambda_j^{k+1})\Bigr )}
\\ \\
\displaystyle{
-\, 2 \sum\limits_{j:j\neq i}^N\dot \lambda_i^k\dot \lambda_j^k
\zeta (\lambda_i^k-\lambda_j^k)+(c^{k-1}(t)-c^k(t))\dot \lambda_i^k=0}
\end{array}
\eeq
with
\beq\label{el14b}
c^k(t)=\frac{1}{\beta}\sum\limits_{i,j=1}^N\dot \lambda_i^k\dot \lambda_j^{k+1}
\zeta (\lambda_i^k-\lambda_j^{k+1}).
\eeq
Here $\displaystyle{\beta =\sum\limits_{i=1}^N \dot \lambda_i^k}$ may
be regarded as a constant since
$\displaystyle{\beta =\sum\limits_{i=1}^N \dot \lambda_i(x)}$
is an $\eta$-periodic
function of $x$ (see (\ref{el2b})).

\subsection{Equivalence to the equations of sections \ref{section:EM} and \ref{section:53}}

It is not difficult to see that with the
conditions (\ref{el2a})--(\ref{el2c}) equations (\ref{el13a}), (\ref{el14b})
with $\beta =1$
are equivalent to (\ref{b7091}).
Indeed, identifying $\lambda_i^k=q_i^k$ and passing to the
center of masses frame, we see that
$$
\begin{array}{l}
\lambda_i^k-\lambda_j^{k-1}=\bar q_i^k -\bar q_j^{k-1}+\alpha \eta /N,
\\ \\
\lambda_i^k-\lambda_j^{k+1}=\bar q_i^k -\bar q_j^{k+1}-\alpha \eta /N,
\end{array}
$$
where $\alpha$ is defined in (\ref{el2b}),
so that the arguments of the $\zeta$- and $E_1$-functions in (\ref{b7091}) and
(\ref{el13a}) coincide if $\alpha =-N$. Next, if one chooses the
periods to be $2\omega_1=1$, $2\omega_2=\tau$, the $\zeta (z)$- and $E_1(z)$-functions differ
by a term linear in $z$ (see (\ref{w1})). The corresponding contribution to $c^k(t)$ (\ref{el14b}) is
\beq\label{el16a}
\frac{1}{\beta}\sum\limits_{i=1}^N \dot \lambda_i^k \lambda_i^k \sum\limits_{j=1}^N \dot \lambda_j^{k+1}-
\frac{1}{\beta}\sum\limits_{i=1}^N \dot \lambda_i^{k+1}\lambda_i^{k+1}\sum\limits_{j=1}^N \dot \lambda_j^{k}=
\sum\limits_{i=1}^N \dot \lambda_i^k \lambda_i^k -
\sum\limits_{i=1}^N \dot \lambda_i^{k+1}\lambda_i^{k+1}
\eeq
since $\displaystyle{\sum\limits_{j=1}^N \dot \lambda_j^k =\beta}$. Writing equations (\ref{el13a}) as
\beq\label{el13b}
\begin{array}{c}
\displaystyle{
\frac{\ddot \lambda_i^k}{\dot \lambda_i^k}+\sum\limits_{j=1}^N \Bigl (
\dot \lambda_j^{k-1}\zeta (\lambda_i^k-\lambda_j^{k-1})+
\dot \lambda_j^{k+1}\zeta (\lambda_i^k-\lambda_j^{k+1})\Bigr )-
}
\\ \ \\
\displaystyle{
- 2 \sum\limits_{j:j\neq i}^N\dot \lambda_j^k
\zeta (\lambda_i^k-\lambda_j^k)+c^{k-1}(t)-c^k(t)=0},
\end{array}
\eeq
we find the corresponding contribution from the sums over $j$ to be
$$
\sum\limits_{j=1}^N \dot \lambda_j^{k-1}(\lambda_i^k -\lambda_j^{k-1})+
\sum\limits_{j=1}^N \dot \lambda_j^{k+1}(\lambda_i^k -\lambda_j^{k+1})-2
\sum\limits_{j=1}^N \dot \lambda_j^{k}(\lambda_i^k -\lambda_j^{k})=
$$
\beq\label{el15}
=\lambda_i^k
\underbrace{\sum\limits_{j=1}^N (\dot \lambda_j^{k-1}+
\dot \lambda_j^{k+1}-2\dot \lambda_j^k)}_{\mbox{$=0$ due to (\ref{el2a})}}-
\sum\limits_{j=1}^N (\dot \lambda_j^{k-1}\lambda_j^{k-1}+\dot \lambda_j^{k+1}\lambda_j^{k+1}-2
\dot \lambda_j^{k}\lambda_j^{k}),
\eeq
so this contribution cancels with the one coming from $c^k(t)$ (\ref{el16a}).

The $L-M$ pair (\ref{b463}), (\ref{b67}) discussed in sections
\ref{section:EM}, \ref{section:ZS} is also equivalent to the $L-M$ pair
(\ref{L}), (\ref{M}). Indeed, one can straightforwardly check that with the
identification of the $L$- and $M$-matrices
\beq\label{el16}
\tilde L_{ij}^k(z)=-h_{k-1, k}e^{E_1(z)(\lambda_i^{k-1}-\lambda_j^k)}
L_{ji}^{k-1}(-z),
\eeq
\beq\label{el17}
\tilde M_{ij}^k(z)=e^{E_1(z)(\lambda_i^{k}-\lambda_j^k)}M_{ji}^{k}(-z)-
\delta_{ij}(E_1(z)\dot \lambda_i^k +\tilde c^k)
\eeq
the semi-discrete Zakharov-Shabat equations (\ref{b47}) and (\ref{el9}) become
equivalent. In the right hand sides of (\ref{el16}), (\ref{el17}) the matrices
$L^k$, $M^k$ are given by (\ref{L}), (\ref{M}) under the identification
$x=k\eta +x_0$: $L^k =L(k\eta +x_0)$, $M^k =M(k\eta +x_0)$. For the modified
Lax pair (\ref{b46333}), (\ref{b4736}) the relations (\ref{el16}), (\ref{el17})
slightly simplify:
\beq\label{el18}
{L'}_{ij}^k(z)=-e^{E_1(z)(\lambda_i^{k-1}-\lambda_j^k)}
L_{ji}^{k-1}(-z),
\eeq
\beq\label{el19}
{M'}_{ij}^k(z)=e^{E_1(z)(\lambda_i^{k}-\lambda_j^k)}M_{ji}^{k}(-z)-
\delta_{ij}E_1(z)\dot \lambda_i^k\,.
\eeq

\subsection{Hamiltonian structure}

Let us show that the equations (\ref{el13a}) with $c^k(t)$ given by
(\ref{el14b}) are Hamiltonian.
We fix the canonical Poisson brackets
\beq\label{h1}
\{p_i^k, p_j^l\}=\{\lambda_i^k, \lambda_j^l\}=0, \quad
\{\lambda_i^k, p_j^l\}=\delta_{ij}\delta_{kl}.
\eeq
The Hamiltonian is
\beq\label{h2}
{\cal H}=\frac{\beta}{\eta}\sum\limits_{k=1}^n \log H_k
\eeq
with
\beq\label{h3}
H_k=\sum\limits_{i=1}^N e^{\eta p_i^k}\,\frac{\prod\limits_{j=1}^N
\sigma (\lambda_i^k-\lambda_j^{k-1})}{\prod\limits_{j:j\neq i}^N
\sigma (\lambda_i^k-\lambda_j^k)}\,.
\eeq
The first set of Hamiltonian equations is
\beq\label{h4}
\dot \lambda_i^k=\frac{\p {\cal H}}{\p p_i^k}=\frac{\beta}{H_k}\,
 e^{\eta p_i^k}\frac{\prod\limits_{j=1}^N
\sigma (\lambda_i^k-\lambda_j^{k-1})}{\prod\limits_{j:j\neq i}^N
\sigma (\lambda_i^k-\lambda_j^k)}\,.
\eeq
Taking the time derivative of (logarithm of) this equation, we get
\beq\label{h5}
\eta \dot p_i^k=\frac{\ddot \lambda_i^k}{\dot \lambda_i^k}-\sum\limits_{j=1}^N
(\dot \lambda_i^k-\dot \lambda_j^{k-1})\zeta (\lambda_i^k-\lambda_j^{k-1})
+\sum\limits_{j:j\neq i}^N(\dot \lambda_i^k-\dot \lambda_j^k)\zeta (\lambda_i^k-\lambda_j^k)
+\p_t \log H_k\,.
\eeq
The second set of Hamiltonian equations is
\beq\label{h6}
\dot p_i^k=-\frac{\p {\cal H}}{\p \lambda_i^k}.
\eeq
The variation of the Hamiltonian is
$$
\eta \delta {\cal H}=\beta \sum\limits_{k=1}^n \frac{\delta H_k}{H_k}=
$$
$$
=\sum\limits_{k=1}^n \sum\limits_{i,l=1}^N \dot \lambda_i^k
\zeta (\lambda_i^k-\lambda_l^{k-1})
(\delta \lambda_i^k-\delta \lambda_l^{k-1})
-\sum\limits_{k=1}^n \sum\limits_{i,l:i\neq l}^N \dot \lambda_i^k
\zeta (\lambda_i^k-\lambda_l^{k})
(\delta \lambda_i^k-\delta \lambda_l^{k}).
$$
Changing the summation indices $k\to k+1$ and $i\leftrightarrow l$
when necessary, we have:
$$
\eta \delta {\cal H}=
\sum\limits_{k=1}^n \sum\limits_{i,l=1}^N \dot \lambda_i^k
\zeta (\lambda_i^k-\lambda_l^{k-1})\delta \lambda_i^k +
\sum\limits_{k=1}^n \sum\limits_{i,l=1}^N \dot \lambda_l^{k+1}
\zeta (\lambda_i^k-\lambda_l^{k+1})\delta \lambda_i^k-
$$
$$
-\sum\limits_{k=1}^n \sum\limits_{i,l:i\neq l}^N (\dot \lambda_i^k+\dot \lambda_l^k)
\zeta (\lambda_i^k-\lambda_l^{k})
\delta \lambda_i^k.
$$
From here we see that
\beq\label{h7}
\begin{array}{c}
\displaystyle{\eta
\dot p_i^k=-\sum\limits_{l=1}^N \dot \lambda_i^k
\zeta (\lambda_i^k-\lambda_l^{k-1}) -
\sum\limits_{l=1}^N \dot \lambda_l^{k+1}
\zeta (\lambda_i^k-\lambda_l^{k+1})
+\sum\limits_{l:l\neq i}^N (\dot \lambda_i^k+\dot \lambda_l^k)
\zeta (\lambda_i^k-\lambda_l^{k}).}
\end{array}
\eeq
Comparing with (\ref{h5}), we obtain:
$$
\frac{\ddot \lambda_i^{k}}{\dot \lambda_i^k}+\sum\limits_{l=1}^N \dot \lambda_l^{k+1}
\zeta (\lambda_i^k-\lambda_l^{k+1})+
\sum\limits_{l=1}^N \dot \lambda_l^{k-1}
\zeta (\lambda_i^k-\lambda_l^{k-1})-2\sum\limits_{l:l\neq i}^N \dot \lambda_l^k
\zeta (\lambda_i^k-\lambda_l^{k})+\p_t \log H_k=0.
$$
The calculation of $\p_t \log H_k =\dot H_k/H_k$ is straightforward using
(\ref{h3}) and (\ref{h7}).
The result is
\beq\label{h8}
\p_t \log H_k = c^{k-1}(t)-c^k(t),
\eeq
where $c_k(t)$ is given by (\ref{el14b}). Therefore, the equations of motion
(\ref{el13a}) are reproduced.

%Note that the equations of motion are symmetric
%under the change $\eta \to -\eta$.

\subsection{The limit to 1+1 Calogero-Moser field theory}

In this section we show that the $\eta \to 0$ limit of the field
Ruijsenaars-Schneider model yields the field Calogero-Moser model as it appears
in \cite{AKV02}.

Instead of the lattice version (\ref{h2}) it is convenient to work with the
equivalent $x$-dependent Hamiltonian density
\beq\label{c1}
{\cal H}(x)=\frac{\beta}{\eta}\, \log \left (
\sum\limits_{i=1}^N e^{\eta p_i(x)}\sigma (\lambda_i(x)-\lambda_i(x-\eta ))\prod\limits_{l:l\neq i}^N
\frac{\sigma (\lambda_i(x)-\lambda_l(x-\eta ))}{\sigma (\lambda_i(x)-\lambda_l(x))}
\right )
\eeq
and the canonical Poisson brackets
\beq\label{c2}
\{p_i(x), p_j(y)\}=\{\lambda_i(x), \lambda_j(y)\}=0, \quad
\{\lambda_i(x), p_j(y)\}=\delta_{ij}\delta (x-y).
\eeq
The Hamiltonian is $\displaystyle{\int {\cal H}(x)dx}$.
The Hamiltonian ((\ref{h2}) is a straightforward lattice version of it.
The calculation which is completely parallel to the one performed in
section 5.5 shows that this Hamiltonian generates the equations of motion
(\ref{el13}), (\ref{el14a}) (with the only change that the derivatives
with respect to the canonical variables $\lambda_i^k$,
$p_i^k$ on the lattice become variational derivatives
with respect to the canonical variables $\lambda_i(x)$, $p_i(x)$).

We are interested in the $\eta$-expansion of (\ref{c1}) as $\eta \to 0$.
We have:
$$
{\cal H}(x)=\frac{\beta}{\eta}\, \log \left [\eta \sum\limits_{i=1}^N \Bigl (1+\eta p_i+
\frac{1}{2}\, \eta^2 p_i^2+O(\eta^3)\Bigr )(\lambda_i'-
\frac{1}{2}\, \eta \lambda_i'' +
\frac{1}{6}\, \eta^2 \lambda_i'''+O(\eta^3)\Bigr )\right.
$$
$$
\left.
\times \exp \Bigl (\sum\limits_{j:j\neq i}^N\Bigl (
\eta \lambda_j' \zeta (\lambda_i-\lambda_j)-\frac{1}{2}\, \eta^2 \lambda_j''
\zeta (\lambda_i-\lambda_j)-\frac{1}{2}\, \eta^2 \lambda_j'{}^{2}
\wp (\lambda_i-\lambda_j)+O(\eta^3)\Bigr )\Bigr )\right ],
$$
where prime denotes the $x$-derivative. Equation (\ref{el2b}) implies that
\beq\label{c3}
\sum\limits_{i=1}^N \lambda_i'(x)=\alpha .
\eeq
Since $\alpha$ is in general an $\eta$-periodic function of $x$,
in the limit $\eta \to 0$ it is natural to assume
that $\alpha$
is a constant. In the limit $\eta \to 0$ the vector from the marked point
$P_0$ to the marked point $P_{\infty}$ on the Riemann surface $\Gamma$
becomes the tangent vector at $P_{\infty}$. This means that
the $x$-flow tends to the $t_1$-flow, and so
the limit of $\alpha$ as $\eta \to 0$ is equal to $\beta$.
Therefore, the first few terms of the $\eta$-expansion of
${\cal H}(x)$ are
\beq\label{c4}
{\cal H}(x)=\mbox{const} \, +(1+O(\eta ))H_1^{\rm CM}(x)-\frac{\eta}{2}\,
H_2^{\rm CM}(x) +O(\eta^2),
\eeq
where
\beq\label{c5}
H_1^{\rm CM}(x)=\sum\limits_{i=1}^N \tilde p_i\lambda_i'
\eeq
is the first Hamiltonian density of the field Calogero-Moser model
(a field analogue of
the total momentum) and
\beq\label{c6}
\begin{array}{c}
\displaystyle{
H_2^{\rm CM}(x)=-\sum\limits_{i=1}^N \tilde p_i^2 \lambda_i'+
\frac{1}{4}\sum\limits_{i=1}^N \frac{\lambda_i''{}^2}{\lambda_i'}-
\frac{1}{3}\sum\limits_{i=1}^N \lambda_i'''
+\frac{1}{\beta}\Bigl (\sum\limits_i \tilde p_i \lambda_i'\Bigr )^2}
\\ \\
\displaystyle{-\frac{1}{2}\sum\limits_{i,j:i\neq j}^N(\lambda_i''\lambda_j'-
\lambda_j''\lambda_i')
\zeta (\lambda_i-\lambda_j)+\frac{1}{2}\sum\limits_{i,j:i\neq j}^N
(\lambda_i'\lambda_j'{}^2 +\lambda_j'\lambda_i'{}^2)\wp (\lambda_i-\lambda_j)}
\end{array}
\eeq
is the second (standard) Hamiltonian density. The Hamiltonian of the model is
\beq\label{c7}
{\cal H}_2^{\rm CM}=\int H_2^{\rm CM}(x)dx.
\eeq
Up to a total derivative and the canonical transformation
\beq\label{c8}
p_i\longrightarrow \tilde p_i=p_i-\frac{\lambda_i''}{2\lambda_i'}+\sum\limits_{j:j\neq i}^N
\lambda_j'\zeta (\lambda_i-\lambda_j)
\eeq
the Hamiltonian density (\ref{c6}) coincides with the Hamiltonian density
for the field Calo\-ge\-ro-Mo\-ser model presented in \cite{AKV02}.

The fact that the transformation (\ref{c8}) is canonical can be verified
straightforwardly. The only nontrivial calculation that
is required is to show that
$\{\tilde p_i(x), \tilde p_j(y)\}=0$. This can be done using the identities
\beq\label{c9}
f(x)\delta ''(x-y)-f(y)\delta ''(y-x)=-\Bigl (f'(x)\delta '(x-y)-
f'(y)\delta ' (y-x)\Bigr ),
\eeq
\beq\label{c10}
f(x)\delta '(x-y)+f(y)\delta '(y-x)=-f'(x)\delta (x-y)
\eeq
for the delta-function and its derivatives.

\section{Fully discrete version}

The fully discrete (or difference) version of the above construction can be obtained by
considering elliptic families of solutions to the
Hirota bilinear difference equation \cite{Hirota}
for the tau-function $\tau^{l,m}(x)$, where $l,m$ are discrete times:
\beq\label{disc1}
\tau^{l,m}(x+\eta )\tau^{l+1,m+1}(x)-\kappa \tau^{l,m+1}(x+\eta )\tau^{l+1,m}(x)+
(\kappa -1)\tau^{l+1,m}(x+\eta )\tau^{l,m+1}(x)=0.
\eeq
Here $\kappa$ is a parameter.
This equation is known to provide an integrable time discretization of the 2D Toda equation.
One of the auxiliary linear problems for the equation (\ref{disc1}) is \cite{KWZ98}
\beq\label{disc3}
\psi^{m+1}(x)=\psi^m(x+\eta)-\kappa
\frac{\tau^m(x)\tau^{m+1}(x+\eta )}{\tau^{m+1}(x)\tau^{m}(x+\eta )}\, \psi^m(x),
\eeq
where the index $l$ is supposed to be fixed.

The elliptic families of solutions with elliptic parameter $\lambda$
are given by
\beq\label{disc2}
\tau^{l,m}(x)=\rho^{l,m}(x)e^{c_1\lambda +c_2\lambda^2}\prod\limits_{j=1}^N \sigma (\lambda -
\lambda_j^{l,m}(x)),
\eeq
where $\rho^{l,m}(x)$ is some function which does not depend on $\lambda$
and $c_1, c_2$ are constants.
If the constraint
\beq\label{disc4}
\sum\limits_{j=1}^N  \Bigl (\lambda^{m+1}_j(x+\eta )-\lambda^{m+1}_j(x)\Bigr )=
\sum\limits_{j=1}^N  \Bigl (\lambda^{m}_j(x+\eta )-\lambda^{m}_j(x)\Bigr )
\eeq
is satisfied,
then the coefficient in front of the second term in the right hand side of
(\ref{disc3}) is an elliptic function of $\lambda$ and we can find double-Bloch
solutions of the form
\beq\label{el3a}
\psi^m (x)=\sum\limits_{i=1}^N  c_i^m(x)\Phi (\lambda -\lambda_i^m(x), z ).
\eeq

The substitution of (\ref{el3a}) into (\ref{disc3}) yields:
$$
\sum\limits_{i=1}^N c_i^{m+1}(x)\Phi (\lambda -\lambda_i^{m+1}(x))-
\sum\limits_{i=1}^N c_i^{m}(x+\eta )\Phi (\lambda -\lambda_i^{m}(x+\eta ))+
$$
$$
+\, \kappa_m(x)\frac{\prod\limits_{j=1}^N\sigma (\lambda -\lambda_j^{m}(x))
\sigma (\lambda -\lambda_j^{m+1}(x+\eta ))}{\prod\limits_{j=1}^N\sigma (\lambda -\lambda_j^{m+1}(x))
\sigma (\lambda -\lambda_j^{m}(x+\eta ))}
\sum\limits_{i=1}^N c_i^{m}(x)\Phi (\lambda -\lambda_i^{m}(x))=0,
$$
where
$$
\kappa_m(x)=\kappa
\frac{\rho^m(x)\rho^{m+1}(x+\eta )}{\rho^{m+1}(x)\rho^{m}(x+\eta )}.
$$
It is enough to cancel poles in the left hand side at $\lambda = \lambda_i^{m+1}(x)$ and
$\lambda = \lambda_i^{m}(x+\eta )$.
A direct calculation shows that the conditions of cancellation of the poles read
\beq\label{disc5}
c_i^{m}(x+\eta )=f_i^m(x)\sum\limits_{j=1}^N c_j^m(x)
\Phi (\lambda_i^{m}(x+\eta )-\lambda_j^m(x)),
\eeq
\beq\label{disc6}
c_i^{m+1}(x)=g_i^m(x)\sum\limits_{j=1}^N c_j^m(x)\Phi (\lambda_i^{m+1}(x)-\lambda_j^m(x)),
\eeq
where
\beq\label{disc7}
f_i^m(x)=\kappa_m(x)\frac{\prod\limits_{j=1}^N\sigma (\lambda_i^{m}(x+\eta )-\lambda_j^m(x))
\sigma (\lambda_i^{m}(x+\eta )-\lambda_j^{m+1}(x+\eta ))}{\prod\limits_{j=1}^N
\sigma (\lambda_i^{m}(x+\eta )-\lambda_j^{m+1}(x))
\prod\limits_{j:j\neq i}^N\sigma (\lambda_i^{m}(x+\eta )-\lambda_j^{m}(x+\eta ))},
\eeq
\beq\label{disc8}
g_i^m(x)=-\kappa_m(x)\frac{\prod\limits_{j=1}^N\sigma (\lambda_i^{m+1}(x)-\lambda_j^m(x))
\sigma (\lambda_i^{m+1}(x)-\lambda_j^{m+1}(x+\eta ))}{\prod\limits_{j=1}^N
\sigma (\lambda_i^{m+1}(x)-\lambda_j^{m}(x+\eta ))
\prod\limits_{j:j\neq i}^N\sigma (\lambda_i^{m+1}(x)-\lambda_j^{m+1}(x ))}.
\eeq
Note that
\beq\label{disc9}
\sum\limits_{i=1}^N (f_i^m(x)-g_i^m(x))=0
\eeq
as the sum of residues of the elliptic function
$$
\varphi (\lambda )=\frac{\prod\limits_{j=1}^N\sigma (\lambda -\lambda_j^{m}(x))
\sigma (\lambda -\lambda_j^{m+1}(x+\eta ))}{\prod\limits_{j=1}^N\sigma (\lambda -\lambda_j^{m+1}(x))
\sigma (\lambda -\lambda_j^{m}(x+\eta ))}.
$$
Introduce the matrices $L^m(x)$, $M^m(x)$:
\beq\label{disc10}
L^m_{ij}(x, z)=f_i^m(x)\Phi \Bigl (\lambda_i^{m}(x+\eta )-\lambda_j^m(x), z\Bigr ),
\eeq
\beq\label{disc11}
M^m_{ij}(x, z)=g_i^m(x)\Phi \Bigl (\lambda_i^{m+1}(x)-\lambda_j^m(x), z\Bigr ).
\eeq
They depend on the spectral parameter $z$.
In terms of these matrices, the equations (\ref{disc5}), (\ref{disc6}) read
\beq\label{disc56}
c_i^{m}(x+\eta )=
\sum\limits_{j=1}^N L_{ij}^m(x)c_j^m(x), \quad
c_i^{m+1}(x)=
\sum\limits_{j=1}^N M_{ij}^m(x)c_j^m(x).
\eeq

The compatibility condition of the linear problems (\ref{disc5}), (\ref{disc6})
has the form of the fully discrete zero curvature equation
\beq\label{disc12}
R^m(x):=L^{m+1}(x)M^m(x)-M^m (x+\eta )L^m(x)=0.
\eeq
We have:
\beq\label{disc13}
\begin{array}{c}
\displaystyle{R^m_{ij}(x)=f_i^{m+1}(x)\sum\limits_{k=1}^N g_k^m(x)\Phi (\lambda_i^{m+1}(x+\eta)-
\lambda_k^{m+1}(x))\Phi (\lambda_k^{m+1}(x)-
\lambda_j^{m}(x))}
\\ \\
\displaystyle{-\, g_i^{m}(x+\eta)\sum\limits_{k=1}^N f_k^m(x) \Phi (\lambda_i^{m+1}(x+\eta)-
\lambda_k^{m}(x+\eta ))\Phi (\lambda_k^{m}(x+\eta )-
\lambda_j^{m}(x))}.
\end{array}
\eeq
Cancellation of the leading singularity at $z =0$ leads to the condition
$$
f_i^{m+1}(x)\sum\limits_{k=1}^N g_k^m(x)-g_i^{m}(x+\eta)\sum\limits_{k=1}^N f_k^m(x)=0
$$
Taking into account (\ref{disc9}) we see that it is equivalent to
\beq\label{disc14}
f_i^{m+1}(x)=g_i^{m}(x+\eta).
\eeq

Now we are going to prove that if (\ref{disc14}) is satisfied, then
$R_{ij}^m(x)=0$, so the zero curvature equation is fulfilled.
The proof is along the lines of ref. \cite{KWZ98}.
Using the identity (\ref{id2}), we
rewrite (\ref{disc13}) as
$$
\begin{array}{c}
\displaystyle{R_{ij}^m(x)=\Phi (\lambda_i^{m+1}(x+\eta)-\lambda_j^m(x))
f_i^{m+1}(x)\sum\limits_{k=1}^N g_k^m(x)}
\\ \\
\displaystyle{\times \Bigl (\zeta (\lambda_i^{m+1}(x+\eta)-\lambda_k^{m+1}(x))
+\zeta (\lambda_k^{m+1}(x)-\lambda_j^m(x))
+ \zeta (\mu)-
\zeta (\lambda_i^{m+1}(x+\eta)-\lambda_j^m(x)\Bigr )}
\\ \\
\displaystyle{-\, \Phi (\lambda_i^{m+1}(x+\eta)-
\lambda_j^m(x))g_i^{m}(x+\eta)\sum\limits_{k=1}^N f_k^m(x)}
\\ \\
\displaystyle{\times \Bigl (\zeta (\lambda_i^{m+1}(x+\eta)\! -\! \lambda_k^{m}(x+\eta))
\! +\! \zeta (\lambda_k^{ m}(x+\eta)\! -\! \lambda_j^m(x))
\! + \! \zeta (\mu)\!-\!
\zeta (\lambda_i^{m+1}(x+\eta)\! -\! \lambda_j^m(x)\Bigr )}
\end{array}
$$
Using (\ref{disc9}) and (\ref{disc14}), we can represent it in the form
\beq\label{disc15}
R_{ij}^m(x)=g_i^m(x+\eta)\Phi (\lambda_i^{m+1}(x+\eta)-\lambda_j^m(x))G_{ij}^m(x),
\eeq
where
$$
G_{ij}^m(x)=\sum\limits_{k=1}^N \Bigl (g_k^m(x)\zeta (\lambda_i^{m+1}(x+\eta )-\lambda_k^{m+1}(x))+
g_k^m(x)\zeta (\lambda_k^{m+1}(x)-\lambda_j^{m}(x))
$$
$$\phantom{aaaaaaaaa}
-f_k^m(x)\zeta (\lambda_i^{m+1}(x+\eta )-\lambda_k^{m}(x+\eta))-
f_k^m(x)\zeta (\lambda_k^{m}(x+\eta)-\lambda_j^{m}(x))\Bigr ).
$$
But $G_{ij}^m(x)$ is the sum of residues of the elliptic function
$$
F(\lambda )=\Bigl (\zeta (\lambda_i^{m+1}(x+\eta)-\lambda )+
\zeta (\lambda -\lambda_j^m(x))\Bigr )\prod\limits_{j=1}^N
\frac{\sigma (\lambda -\lambda_j^{m}(x))
\sigma (\lambda -\lambda_j^{m+1}(x+\eta ))}{\sigma (\lambda -\lambda_j^{m+1}(x))
\sigma (\lambda -\lambda_j^{m}(x+\eta ))}
$$
and, therefore, $G_{ij}^m(x)=0$.

Finally, let us write down equations (\ref{disc14}) explicitly:
\beq\label{disc16}
\prod\limits_{j=1}^N \frac{\sigma (\lambda_i^m(x)-\lambda_j^{m}(x-\eta))
\sigma (\lambda_i^m(x)-\lambda_j^{m+1}(x))
\sigma (\lambda_i^m(x)-\lambda_j^{m-1}(x+\eta))}{\sigma
(\lambda_i^m(x)-\lambda_j^{m}(x+\eta))
\sigma (\lambda_i^m(x)-\lambda_j^{m-1}(x))
\sigma (\lambda_i^m(x)-\lambda_j^{m+1}(x-\eta))}=-
\frac{\kappa_{m-1}(x)}{\kappa_{m}(x\! -\! \eta)}.
\eeq
Note that
\beq\label{disc17}
\frac{\kappa_{m}(x\! -\! \eta)}{\kappa_{m-1}(x)}=
\frac{\rho^{m-1}(x+\eta)
\rho^m(x-\eta)\rho^{m+1}(x)}{\rho^{m-1}(x)\rho^m(x+\eta)\rho^{m+1}(x-\eta)}.
\eeq
This is the field analog of the doubly discrete Ruijsenaars-Schneider
system \cite{NRK}.
Similar equations were obtained in \cite{DNY15}.

\section{Conclusion}

In this paper we have introduced two integrable models one of which is a natural lattice
version of the other.
The first one is a finite-dimensional system which we call
the Ruijsenaars-Schneider chain. We show that it is gauge equivalent to a special case of the
homogenous classical elliptic (XYZ) spin chain when residues of all Lax matrices
in the chain are of rank one. The second one is the field analogue of the
Ruijsenaars-Schneider model with continuous space and time variables.
The definition of this model is the main result of the paper. This is a (1+1)-dimensional
model which admits a semi-discrete zero curvature (Zakharov-Shabat)
representation for elliptic Lax pair with spectral parameter.
This model is obtained through a multi-pole ansatz for general
elliptic solutions (elliptic families) of the 2D Toda lattice hierarchy. Then we show
that a natural space discretization
of this model coincides with the Ruijsenaars-Schneider chain.

The fully discrete version of the model (i.e. discrete in both space and time) is also
introduced. It is based on studying elliptic families of solutions to the Hirota
bilinear difference equation \cite{Hirota} which is known to provide the integrable
discretization of the 2D Toda equation. The corresponding
equations of motion for poles of the elliptic solutions
are very similar to those obtained in \cite{DNY15}
from a general ansatz for elliptic
$L$-$M$ pair.

We also discuss a limit of the model which coincides with
the field analogue of the Calogero-Moser system introduced in
\cite{Krichever02} and reproduced in \cite{AKV02} as a dynamical system for poles
of elliptic families of solutions to the Kadomtsev-Petviashvili equation.
Note that by construction the Ruijsenaars-Schneider chain is gauge equivalent to
the elliptic spin chain. A similar gauge equivalence exists between
the (1+1)-dimensional Calogero-Moser field theory and the continuous
Landau-Lifshitz equation \cite{LOZ,AtZ}.
The exact relation between the obtained (1+1)-dimensional field analogue
of the Ruijsenaars-Schneider model and the
(semi-discrete) equations of the Landau-Lifshitz type will be discussed elsewhere.

\section{Appendix A: Elliptic functions}
\def\theequation{A.\arabic{equation}}
\setcounter{equation}{0}

\paragraph{Weierstrass $\sigma$-, $\zeta$- and $\wp$-functions.}
The $\sigma$-function
with quasi-periods $2\omega_1$, $2\omega_2$ such that
${\rm Im} (\omega_2/ \omega_1 )>0$
is defined by the infinite product
\beq\label{w01}
\sigma (x)=\sigma (x |\, \omega_1 , \omega_2)=
x\prod_{s\neq 0}\Bigl (1-\frac{x}{s}\Bigr )\, e^{\frac{x}{s}+\frac{x^2}{2s^2}},
\quad s=2\omega_1 m_1+2\omega_2 m_2 \quad \mbox{with integer $m_1, m_2$}.
\eeq
It is connected with the Weierstrass
$\zeta$- and $\wp$-functions by the formulas $\zeta (x)=\sigma '(x)/\sigma (x)$,
$\wp (x)=-\zeta '(x)=-\p_x^2\log \sigma (x)$.
We also need the function $\Phi =\Phi (x, z )$ defined as
\beq\label{w02}
\Phi (x, z )=\frac{\sigma (x+z )}{\sigma (z )\sigma (x)}\,
e^{-\zeta (z )x}.
\eeq
It has a simple pole
at $x=0$ with residue $1$ and the expansion
\beq\label{w03}
\Phi (x, z )=\frac{1}{x}-\frac{1}{2}\, \wp (z ) x +\ldots , \qquad
x\to 0.
\eeq
The quasiperiodicity properties of the function $\Phi$ are
\beq\label{bloch}
\Phi (x+2\omega_{\alpha} , z )=e^{2(\zeta (\omega_{\alpha} )z -
\zeta (z )\omega_{\alpha} )}
\Phi (x, z ).
\eeq
In the main text we often suppress the second argument of $\Phi$ writing simply
$\Phi (x, z )=\Phi (x)$.
We will also need the $x$-derivative
$\Phi '(x, z )=\p_x \Phi (x, z )$.
The function $\Phi$ satisfies the following identities:
\beq\label{id1}
\Phi '(x)=\Phi (x)\Bigl (\zeta (x+z )-\zeta (x)-\zeta (z )\Bigr ),
\eeq
\beq\label{id2}
\Phi (x) \Phi (y)=\Phi (x+y)\Bigl (\zeta (x)+\zeta (y)+\zeta (z )-\zeta (x+y+z )\Bigr )
\eeq
which are used in the main text.

\paragraph{Theta-functions.}
The theta-functions with characteristics $a,b$ are defined as follows:
\beq\label{a24}
 \begin{array}{c}
  \displaystyle{
\theta{\left[\begin{array}{c}
a\\
b
\end{array}
\right]}(z|\, \tau ) =\sum_{j\in \z}
\exp\left(2\pi\imath(j+a)^2\frac\tau2+2\pi\imath(j+a)(z+b)\right)\,.
}
 \end{array}
 \eeq
In our paper, we consider the case of rational characteristics $a\,,b\in\frac{1}{N}\,\ZZ$.
 In particular, the odd theta function used in the paper ($\theta_1(z)$ in the Jacobi notation) is
 \beq\label{a25}
 \begin{array}{c}
  \displaystyle{
\vth(z)=\vth(z,\tau)\equiv-\theta{\left[\begin{array}{c}
1/2\\
1/2
\end{array}
\right]}(z|\, \tau )\,.
 }
 \end{array}
 \eeq
The following quasi-periodicity properties hold. For $a,b,a'\in(1/N)\ZZ$
\beq\label{c09}
 \begin{array}{c}
  \displaystyle{
\theta{\left[\begin{array}{c}
a\\
b
\end{array}
\right]}(z+1|\, \tau )
=\bfe(a)\,
\theta{\left[\begin{array}{c}
a\\
b
\end{array}
\right]}(z|\, \tau )\,,
 }
 \end{array}
 \eeq
\beq\label{c091}
 \begin{array}{c}
  \displaystyle{
\theta{\left[\begin{array}{c}
a+1\\
b
\end{array}
\right]}(z|\, \tau )
=
\theta{\left[\begin{array}{c}
a\\
b
\end{array}
\right]}(z|\, \tau )\,,
 }
 \end{array}
 \eeq
\beq\label{c101}
 \begin{array}{c}
  \displaystyle{
\theta{\left[\begin{array}{c}
a\\
b
\end{array}
\right]}(z+a'\tau|\, \tau )
=\bfe\Big( -{a'}^2\frac{\tau}{2}-a'(z+b) \Big)\,
\theta{\left[\begin{array}{c}
a+a'\\
b
\end{array}
\right]}(z|\, \tau )\,,
 }
 \end{array}
 \eeq
where
we denote
  \beq\label{c01}
  \begin{array}{c}
  \displaystyle{
\bfe(x):=\exp(2\pi\imath x)
 }
 \end{array}
 \eeq
 for brevity.

 %%%%%%%%%%%%%%%%%%%%%%%%%%%%%%%%%%%%%%%%%%
\paragraph{The Kronecker function and the function $E_1$.}
We also use the following set of $N^2$ functions:
 \beq\label{a08}
 \begin{array}{c}
  \displaystyle{
 \vf_a(z,\om_a+\eta)=\bfe (a_2z/N)\,\phi(z,\om_a+\eta)\,,\quad
 \om_a=\frac{a_1+a_2\tau}{N}\,,
 }
 \end{array}
 \eeq
where $a=(a_1, a_2)\in\ZZ_N\times\ZZ_N$ and
 \beq\label{a09}
  \begin{array}{l}
  \displaystyle{
 \phi(z,u)=\frac{\vth'(0)\vth(z+u)}{\vth(z)\vth(u)}
 }
 \end{array}
 \eeq
 is the Kronecker function. It has a simple pole at $z=0$ with residue 1:
 \beq\label{a095}
  \begin{array}{l}
  \displaystyle{
\res\limits_{z=0}\phi(z,u)=1\,.
 }
 \end{array}
 \eeq
 The quasi-periodicity properties are as follows:
 \beq\label{a096}
  \begin{array}{l}
  \displaystyle{
 \phi(z+1,u)= \phi(z,u)\,,\qquad  \phi(z+\tau,u)= \bfe (-u)\phi(z,u)\,.
 }
 \end{array}
 \eeq
 The expansion of $\phi(z,u)$ near $z=0$ has the form
 \beq\label{a091}
  \begin{array}{l}
  \displaystyle{
 \phi(z,u)=\frac{1}{z}+E_1(u)+z\,\frac{E_1^2(u)-\wp(u)}{2}+O(z^2),
 }
 \end{array}
 \eeq
 where
 \beq\label{a093}
  \begin{array}{l}
  \displaystyle{
 E_1(u)=\frac{\vth'(u)}{\vth(u)}.
 }
 \end{array}
 \eeq

 It follows from the definition (\ref{a09}) that
 \beq\label{a092}
  \begin{array}{l}
  \displaystyle{
 \p_z\phi(z,u)=(E_1(z+u)-E_1(z))\phi(z,u)\,,
 }
 \\ \ \\
   \displaystyle{
  \p_u\phi(z,u)=(E_1(z+u)-E_1(u))\phi(z,u)\,.
 }
 \end{array}
 \eeq
We also use a set of widely known addition formulae:
\beq\label{b151}
  \begin{array}{c}
  \displaystyle{
 \phi(z,u_1)\phi(z,u_2)=\phi(z,u_1+u_2)\Big(E_1(z)+E_1(u_1)+E_1(u_2)-E_1(z+u_1+u_2)\Big)\,,
 }
 \end{array}
 \eeq
\beq\label{b152}
  \begin{array}{c}
  \displaystyle{
  \phi(z_1, u_1) \phi(z_2, u_2) = \phi(z_1, u_1 + u_2) \phi(z_2 - z_1, u_2) + \phi(z_2, u_1 + u_2) \phi(z_1 - z_2, u_1)
 }
 \end{array}
 \eeq
and
\beq\label{b153}
  \begin{array}{c}
  \displaystyle{
  \phi(z,u_1-v)\phi(w,u_2+v)\phi(z-w,v)-\phi(z,u_2+v)\phi(w,u_1-v)\phi(z-w,u_1-u_2-v)=
 }
 \\ \ \\
   \displaystyle{
  =\phi(z,u_1)\phi(w,u_2)\Big( E_1(v)-E_1(u_1-u_2-v)+E_1(u_1-v)-E_1(u_2+v) \Big)\,.
 }
 \end{array}
 \eeq

 \paragraph{Relation to the Weierstrass functions.}
 The above definitions of the Weierstrass functions (\ref{w01})--(\ref{w03})
 are easily related to those given in terms of theta-functions (\ref{a09})--(\ref{a091})
if we choose the periods to be $2\omega_1=1$, $2\omega_2=\tau$:
\beq\label{w1}
\displaystyle{
\zeta(z)=E_1(z)+2\eta_0 z\,,\quad \eta_0=-\frac{1}{6}\frac{\vth'''(0)}{\vth'(0)}\,,
}
\eeq
\beq\label{w2}
\displaystyle{
\sigma(z)=\frac{\vth(z)}{\vth'(0)}\,e^{\eta_0 z^2}\,,
}
\eeq
\beq\label{w3}
\displaystyle{
\Phi(z,u)=\phi(z,u)\,e^{-z E_1(u)}\,,
}
\eeq
Under the substitution (\ref{w3}) the identities (\ref{id1})--(\ref{id2}) are transformed into
(\ref{a092}) and (\ref{b151}) respectively.
 The Weierstrass $\wp$-function appearing in (\ref{a091}) is
 \beq\label{a094}
  \begin{array}{l}
  \displaystyle{
 \wp(u)=-\p^2_u\log\vth(u)+\frac{1}{3}\frac{\vth'''(0)}{\vth'(0)}\,.
 }
 \end{array}
 \eeq

\paragraph{Some relations for
theta-functions with rational characteristics.}
Using definition (\ref{a24}), one cane rewrite
the set of functions (\ref{a08}) in a slightly different form.
Set
  \beq\label{d01}
  \begin{array}{c}
  \displaystyle{
 \theta_{\al}(z,\tau)=\theta{\left[\begin{array}{c}
 \frac{\al_2}{N}+\frac{1}{2}
  \\
  \frac{\al_1}{N}+\frac{1}{2}
\end{array}
\right]}(z,\tau)\,,\qquad \al\in\ZZ_N\times\ZZ_N
 }
 \end{array}
 \eeq
Then for any $\al$ we have
  \beq\label{d02}
  \begin{array}{c}
  \displaystyle{
 \frac{\theta_{\al}(z+\eta,\tau)}{\theta_{\al}(\eta,\tau)}=\bfe (\al_2z/N)
 \frac{\vth(z+\eta+\om_\al)}{\vth(\eta+\om_\al)}
 }
 \end{array}
 \eeq
and, therefore,
  \beq\label{d03}
  \begin{array}{c}
  \displaystyle{
 \vf_\al(z,\eta+\om_\al)=\frac{\vth'(0)\theta_{\al}(z+\eta,\tau)}{\vth(z)\theta_{\al}(\eta,\tau)}\,.
 }
 \end{array}
 \eeq
 Introduce also
  \beq\label{d04}
  \begin{array}{c}
  \displaystyle{
 \theta^{(j)}(z)=\theta{\left[\begin{array}{c}
\frac{1}{2}-\frac{j}{N}
  \\
  \frac{1}{2}
\end{array}
\right]}(z,N\tau)\,,\qquad j\in\ZZ_N\,.
 }
 \end{array}
 \eeq
 Then for $-\vth(z)$ we have
  \beq\label{d05}
  \begin{array}{c}
  \displaystyle{
 \theta{\left[\begin{array}{c}
\frac{1}{2}
  \\
  \frac{1}{2}
\end{array}
\right]}(z,\tau)=C(\tau)\prod\limits_{j=0}^{N-1}\theta^{(j)}(z)\,,
\qquad
  \displaystyle{
C(\tau)=\frac{\vth'(0,\tau)}{\vth'(0,N\tau)}\,\frac{1}{\prod\limits_{j=1}^{N-1}\theta^{(j)}(0)}\,,
  }
 }
 \end{array}
 \eeq
 so that the following relation holds:
  \beq\label{d06}
  \begin{array}{c}
  \displaystyle{
 \frac{\vth'(0,\tau)}{\vth(z,\tau)}\,\frac{\prod\limits_{j=0}^{N-1}\theta^{(j)}(z)}
 {\prod\limits_{j=1}^{N-1}\theta^{(j)}(0)}=-\vth'(0,N\tau)\,.
  }
 \end{array}
 \eeq

Consider the matrix
 \beq\label{a47}
 \begin{array}{c}
  \displaystyle{
X_{ij}(x_j)=
 \vth\left[  \begin{array}{c}
 \frac12-\frac{i}{N} \\ \frac N2
 \end{array} \right] \left(Nx_j\left.\right|N\tau\right)\,.
 }
 \end{array}
 \eeq
 Then the following determinant of the Vandermonde type formula holds \cite{Hasegawa}:
 \beq\label{a48}
 \begin{array}{c}
  \displaystyle{
 \det X=C_N(\tau)\,\vth(\sum\limits_{k=1}^N
 x_k)\prod\limits_{i<j}\vth(x_j-x_i)\,,\quad\quad
 C_N(\tau)=\frac{(-1)^{N}}{(\imath\eta(\tau))^{\frac{(N-1)(N-2)}{2}}}\,,
 }
 \end{array}
 \eeq
where $\eta(\tau)$ is the Dedekind eta-function:
 \beq\label{a49}
 \begin{array}{c}
  \displaystyle{
 \eta(\tau)=e^{\frac{\pi\imath\tau}{12}}\prod\limits_{k=1}^\infty (1-e^{2\pi\imath\tau
 k})=\Big(\frac{\vth'(0)}{2\pi}\Big)^{1/3}\,.
 }
 \end{array}
 \eeq

 \paragraph{Finite Fourier transformation on $\ZZ_N$.} For any $m\in\ZZ$ and $N\in\ZZ_+$
 \beq\label{a771}
 \begin{array}{c}
  \displaystyle{
 e^{2\pi\imath m\eta}\phi(N\eta,z+m\tau|N\tau)=
 \frac{1}{N}\sum\limits_{k=0}^{N-1}e^{-2\pi\imath m\frac{k}{N}}\phi(z,\eta+\frac{k}{N}|\tau)\,,
 }
 \end{array}
 \eeq
 \beq\label{a772}
 \begin{array}{c}
  \displaystyle{
\phi(z,\eta|\tau)=
\sum\limits_{k=0}^{N-1}e^{2\pi\imath zk}\phi(Nz,\eta+k\tau|N\tau)\,.
 }
 \end{array}
 \eeq

\section{Appendix B: Elliptic $R$-matrix and its properties}
\def\theequation{B.\arabic{equation}}
\setcounter{equation}{0}

\paragraph{Matrix basis.}
Consider the pair of $N\times N$ matrices
 \beq\label{a04}
 \begin{array}{c}
  \displaystyle{
 (Q_1)_{kl}=\delta_{kl}\exp(\frac{2\pi
 \imath}{{ N}}k)\,,\ \ \
 (Q_2)_{kl}=\delta_{k-l+1=0\,{\hbox{\tiny{mod}}}\,
 { N}}\,.
 }
 \end{array}
 \eeq
 They satisfy the properties
 \beq\label{a05}
 \begin{array}{c}
  \displaystyle{
 Q_2^{a_2} Q_1^{a_1}=\exp\left(\frac{2\pi\imath}{{ N}}\,a_1
 a_2\right)Q_1^{a_1} Q_2^{a_2}\,,\ a_{1,2}\in\ZZ;\qquad
 Q_1^{ N}=Q_2^{ N}=1_{{ N}\times { N}}\,,
 }
 \end{array}
 \eeq
 so that these matrices represent the generators of the Heisenberg group.
 Let us construct a special basis in ${\rm Mat}(N,\CC)$ in terms of
  (\ref{a04}) in the following way:
 \beq\label{a07}
 \begin{array}{c}
  \displaystyle{
 T_a=T_{a_1 a_2}=\exp\left(\frac{\pi\imath}{{ N}}\,a_1
 a_2\right)Q_1^{a_1}Q_2^{a_2}\,,\quad
 a=(a_1,a_2)\in\ZZ_{ N}\times\ZZ_{ N}\,.
 }
 \end{array}
 \eeq
 In particular, $T_0=T_{(0,0)}=1_N$. For the product we have
  \beq\label{a071}
 \begin{array}{c}
  \displaystyle{
T_\al T_\be=\kappa_{\al,\be} T_{\al+\be}\,,\quad
\kappa_{\al,\be}=\exp\left(\frac{\pi \imath}{N}(\be_1
\al_2-\be_2\al_1)\right)\,,\quad \al+\be=(\al_1+\be_1,\al_2+\be_2)
 }
 \end{array}
 \eeq
 Also
  \beq\label{a072}
 \begin{array}{c}
  \displaystyle{
\tr(T_\al T_\be)=N\delta_{\al+\be,(0,0)}\,.
 }
 \end{array}
 \eeq
%where $\al+\be=(\al_1+\be_1,\al_2+\be_2)$.
%

Let us perform the transformation relating the standard matrix
basis $E_{ij}$ in $\Mat$, given by $(E_{ij})_{kl}=\delta_{ik}\delta_{jl}$,
with (\ref{a07}).
  For the pair of matrices $Q_{1,2}$ (\ref{a04}) and integer numbers $a_1,a_2$ we have
  \beq\label{c02}
  \begin{array}{c}
  \displaystyle{
 Q_1^{a_1}=\sum\limits_{k=1}^N E_{kk}\,\bfe(\frac{k a_1}{N})\,,\qquad
 Q_2^{a_2}=\sum\limits_{k=1}^N E_{k-a_2,k}\,,

 }
 \end{array}
 \eeq
where in the last sum we assume the value of index $k-a_2$ modulo $N$. Then for the basis matrix $T_a$
(\ref{a07}) one gets
  \beq\label{c03}
  \begin{array}{c}
  \displaystyle{
T_a=\bfe(-\frac{a_1 a_2}{2N})
\sum\limits_{k=1}^N E_{k-a_2,k}\,\bfe(\frac{k a_1}{N})\,.
 }
 \end{array}
 \eeq
For an arbitrary matrix $B=\sum_{i,j=1}^N E_{ij}B_{ij}\in\Mat$ its
components $B_a=B_{(a_1,a_2)}$ in the basis $T_a$ can be found using
(\ref{a072}) and (\ref{c03}):
  \beq\label{c04}
  \begin{array}{c}
  \displaystyle{
B_a=\frac{1}{N}\,\tr(B T_{-a})=\frac{1}{N}\,\bfe(-\frac{a_1a_2}{2N})\sum\limits_{k=1}^N B_{k,k+a_2}\,
\bfe(-\frac{a_1 k}{N})\,.
 }
 \end{array}
 \eeq
Similarly, given a set of components $B_{(\al_1,\al_2)}$, $\al_{1,2}\in\ZZ_N$
for a matrix $B\in\Mat$ in the basis $\{T_\al\}$
we have the following expression for its components $B_{ij}$ in the standard basis:
 \beq\label{a491}
 \begin{array}{c}
 B_{ij}=\left\{
 \begin{array}{l}
   \displaystyle{
   \sum\limits_{\al_1=0}^{N-1}B_{(\al_1,j-i)}\bfe\Big(\frac{\al_1(i+j)}{2N}\Big)\,,\quad j\geq i\,,
   }
 \\
   \displaystyle{
    \sum\limits_{\al_1=0}^{N-1}B_{(\al_1,j-i+N)}\bfe\Big(\frac{\al_1(i+j-N)}{2N}\Big)\,,\quad j<i\,.
   }
 \end{array}
 \right.
 \end{array}
 \eeq

\paragraph{The Baxter-Belavin $R$-matrix.}
We use the Baxter-Belavin elliptic quantum $R$-matrix in the following form:
\beq\label{d07}
 \begin{array}{c}
  \displaystyle{
 R_{12}^\hbar(z)=\frac{1}{N}\sum\limits_{a\in\,\ZZ_{ N}\times\ZZ_{ N}} T_a\otimes T_{-a} \vf_a(z,\om_a+\hbar)
 \in{\rm Mat}(N,\CC)^{\otimes 2}\,.
 }
 \end{array}
 \eeq
 It
is equivalently written in the standard basis as follows \cite{RT}:
\beq\label{d08}
 \begin{array}{c}
  \displaystyle{
 R_{12}^\hbar(z)=\sum\limits_{i,j,k,l=1}^N R_{ij,kl}\, E_{ij}\otimes E_{kl}\,,
 }
 \end{array}
 \eeq
\beq\label{d09}
 \begin{array}{c}
  \displaystyle{
 R_{ij,kl}=-\vth'(0,N\tau)
 \frac{ \theta^{(i-k)}(z+N\hbar) }{ \theta^{(j-k)}(z)\theta^{(i-j)}(N\hbar) }\,
 \delta_{i+k=j+l\ {\rm mod}\ N}\,.
 }
 \end{array}
 \eeq
It is also convenient to write it in terms of the Kronecker function (\ref{a09}).
For this purpose we need the identity
\beq\label{d10}
 \begin{array}{c}
  \displaystyle{
 -\vth'(0,N\tau)
 \frac{ \theta^{(a+b)}(z+u) }{ \theta^{(a)}(z)\theta^{(b)}(u) }
 =\bfe\Big( \frac{ab\tau-au-bz}{N} \Big)\phi(z-a\tau,u-b\tau|N\tau)
 }
 \\ \ \\
   \displaystyle{
 =
 \bfe(-u\frac{a}{N})\,\vf_{(0,-b)}(z-a\tau,u-b\tau|N\tau)\,.
 }
 \end{array}
 \eeq
Then
\beq\label{d11}
 \begin{array}{c}
  \displaystyle{
 R_{ij,kl}=\delta_{i+k=j+l\ {\rm mod}\ N}\bfe\Big( \frac{(k-j)
 (j-i)\tau+(k-j)N\hbar+(j-i)z}{N} \Big)
  }
 \\ \ \\
   \displaystyle{
 \times\phi(z+(k-j)\tau,N\hbar+(j-i)\tau|N\tau)\,.
 }
 \end{array}
 \eeq
Equivalence between different representations can be shown by using the relation
between the bases $E_{ij}$, $T_a$ and the Fourier
formulae (\ref{a771})--(\ref{a772}).

%%%%%%%%%%%%%%%%%%%%%%%%%%%%%%%%%%%%%%
 %

\paragraph{IRF-Vertex correspondence.} The matrix (\ref{a21}) participates in the
IRF-Vertex relation
 \beq\label{a26}
 \begin{array}{c}
  \displaystyle{
g_2(z_2,q)\,
g_1(z_1,q+N\hbar^{(2)})\,R^{\hbox{\tiny{F}}}_{12}(\hbar,z_1-z_2|\,q)=R^\hbar_{12}(z_1-z_2)\,
g_1(z_1,q)\, g_2(z_2,q+N\hbar^{(1)})
 }
 \end{array}
 \eeq
 between the (vertex type) Baxter-Belavin $R$-matrix (\ref{a10}) and the (IRF-type)
 Felder's dynamical $R$-matrix \cite{Felder2}:
 \beq\label{a27}
 \begin{array}{c}
  \displaystyle{
 R^{\hbox{\tiny{F}}}_{12}(\hbar,z_1-z_2|\,q)= \sum\limits_{i,j:\,i\neq j}^N
 E_{ii}\otimes E_{jj}\, \phi(N\hbar,-q_{ij})+
 }
\\ \ \\
  \displaystyle{
+\sum\limits_{i,j:\,i\neq j}^N
 E_{ij}\otimes E_{ji}\, \phi(z_1-z_2,q_{ij})+\phi(N\hbar,z_1-z_2)\sum\limits_{i=1}^N
 E_{ii}\otimes E_{ii}\,.
 }
 \end{array}
 \eeq
 The shift of argument $g_1(z_1,q+N\hbar^{(2)})$ in (\ref{a26}) is understood as
  \beq\label{a28}
  \begin{array}{c}
  \displaystyle{
g_1(z_1,q+N\hbar^{(2)})=P_2^{N\hbar}\,
g_1(z_1,q) P_2^{-N\hbar} \,,\quad
P_2^\hbar=\sum\limits_{k=1}^N 1_{N\times N}\otimes E_{kk}
\exp(\hbar\frac{\p}{\p q_k})\,.
 }
 \end{array}
 \eeq

\paragraph{Properties of the intertwining matrix.}
The matrix $g(z,q)$ is degenerated at $z=0$ due to (\ref{a48}):
%The simple pole at $z=0$ in (\ref{a41}) comes from $g^{-1}(z,q^k)$ since
% the determinant is degenerated at $z=0$. Otherwise, we would have
%a pole at the center of mass coordinate.
 %
%That is, for the matrix (\ref{a22}) we have
 %
 \beq\label{a50}
 \begin{array}{c}
  \displaystyle{
 \det\Xi(z,q)=C_N(\tau)\,\vth(z)\prod\limits_{i<j}^N\vth(q_i-q_j)\,,
 }
 \end{array}
 \eeq
and the factor $\vth(z)$ comes from the fact that the sum of
coordinates (in the center of masses frame) equals zero.

%To summarize,
The matrix $g(z)$ (\ref{a21}) satisfies the following
properties (see \cite{VZ} for a review):

1. The matrix $g(z,q)$ is degenerated at $z=0$ (\ref{a50}).

2. The matrix $g(0,q)$ has one-dimensional kernel generated by
the vector-column $\rho$:
 \beq\label{b30}
 \begin{array}{c}
  \displaystyle{
 g(0,q)\rho=0\,,\quad \rho=(1,1,...,1)^T\in\CC^N\,.
 }
 \end{array}
 \eeq
Properties of this type were described in \cite{RT}. Their proof can be also found in \cite{LOZ}.

Let us consider $g^{-1}(z,q)$ near $z=0$:
 \beq\label{b31}
 \begin{array}{c}
  \displaystyle{
 g^{-1}(z,q)=\frac1z\,{\breve g}(0,q)+A(q)+O(z)\,,\quad {\breve
 g}(0,q)=\res\limits_{z=0}\,g^{-1}(z,q)\,.
 }
 \end{array}
 \eeq
Then the matrix ${\breve g}(0)$ is of rank one\footnote{Locally, in some basis $g(z,q)$ is represented in the form
${\rm diag}(z,1,...,1)$. Therefore, ${\breve g}(0,q)$ has $N-1$ zero eigenvalues.}
 \beq\label{b32}
 \begin{array}{c}
  \displaystyle{
 {\breve g}(0)=\rho\otimes\upsilon\,,\quad \upsilon=\frac1N\,\rho^T{\breve g}(0,q)\in\CC^N\,.
 }
 \end{array}
 \eeq
Below we derive an explicit expression for the inverse of the matrix $g(z,q)$.

\paragraph{IRF-Vertex correspondence for semidynamical $R$-matrix.}
In \cite{ACF} the following (semidynamical) $R$-matrix was used for quantization
of the Ruijsenaars-Schneider model:
 \beq\label{b27}
 \begin{array}{c}
  \displaystyle{
 R^{\hbox{\tiny{ACF}}}_{12}(\hbar,z_1,z_2|\,q)=\sum\limits_{i,j:\,i\neq j}^N
 E_{ii}\otimes E_{jj}\, \phi(N\hbar,q_{ij})+\sum\limits_{i\neq j}
 E_{ij}\otimes E_{ji}\, \phi(z_1-z_2,q_{ij})-
 }
\\ \ \\
  \displaystyle{
-\sum\limits_{i,j:\,i\neq j}^N
 E_{ij}\otimes E_{jj}\, \phi(z_1+N\hbar,q_{ij})+\sum\limits_{i,j:\,i\neq j}^N
 E_{jj}\otimes E_{ij}\, \phi(z_2,q_{ij})+
 }
\\ \ \\
  \displaystyle{
+\Big(E_1(N\hbar)+E_1(z_1-z_2)+E_1(z_2)-E_1(z_1+N\hbar)\Big)\sum\limits_{i=1}^N
 E_{ii}\otimes E_{ii}\,,
 }
 \end{array}
 \eeq
 where $E_1$ is defined in (\ref{a091}).
This $R$-matrix satisfies the quantum Yang-Baxter equation with shifted spectral parameters. Following
\cite{SeZ} let us write down its relation to the Baxter-Belavin $R$-matrix (\ref{a10})
in the form of type (\ref{a26}):
 \beq\label{b28}
  \begin{array}{c}
  \displaystyle{
 R^\hbar_{12}(z_1-z_2)=g_1(z_1+N\hbar,q)\,
 g_2(z_2,q)\,
 R^{\hbox{\tiny{ACF}}}_{12}(\hbar,z_1,z_2|\,q)\, g_2^{-1}(z_2+N\hbar,q)\,
 g_1^{-1}(z_1,q)\,.
 }
 \end{array}
 \eeq
Multiplying both sides by $g_2^{-1}(z_2,q)$ and evaluating residue
at $z_2=0$, we get the following useful formula
\footnote{Note that in the $N=1$ case (\ref{b01}) boils down to the definition (\ref{a09})
of the Kronecker function. A similarity of the quantum $R$-matrix (\ref{a10})
with the Kronecker function
underlies the so-called associative Yang-Baxter equation. See \cite{SeZ} and references therein.}:
%
%In what follows we also use the formula obtained in \cite{SeZ} from the IRF-Vertex type
%relations
%
  \beq\label{b01}
  \begin{array}{c}
  \displaystyle{
 %\frac1N\,
 {\breve g}_2(0,q)\,R^\hbar_{12}(z)=g_1(z+N\hbar,q)\,
 \mathcal O_{12}\, g_2^{-1}(N\hbar,q)\,
 g_1^{-1}(z,q)\,,
 }
 \end{array}
 \eeq
 where  ${\breve g}(0)$ is given by (\ref{b31})
%   \beq\label{b02}
%  \begin{array}{c}
%  \displaystyle{
%{\breve g}(0,q)=\res\limits_{z=0}\,g^{-1}(z)\in\Mat
% }
% \end{array}
% \eeq
and
   \beq\label{b03}
  \begin{array}{c}
  \displaystyle{
 \mathcal O_{12}=\sum\limits_{i,j=1}^N E_{ii}\otimes
E_{ji}\,.
 }
 \end{array}
 \eeq
For an arbitrary matrix
$T=\sum_{i,j}E_{ij}T_{ij}\in\Mat$ we have
  \beq\label{b07}
  \begin{array}{c}
  \displaystyle{
\tr_2 \left(\mathcal O_{12}T_2\right)=\sum\limits_{i=1}^N
E_{ii}\sum\limits_{j=1}^N T_{ij}\,.
 }
 \end{array}
 \eeq
Besides (\ref{b01}), we use its degeneration (see the classical limit (\ref{b05}) below)
$\hbar\rightarrow 0$\footnote{ Relation (\ref{b011}) appears in the $\hbar^0$
order, while in $\hbar^{-1}$ order one has
${\breve g}_2(0)=g_1(z){\mathcal O}_{12}\,g_1^{-1}(z)\,{\breve
 g}_2(0)$, which is true due to the property (\ref{b32}).}:
  \beq\label{b011}
  \begin{array}{c}
  \displaystyle{
 {\breve g}_2(0,q)\,r_{12}(z)=g_1'(z)\, \mathcal O_{12}\, {\breve g}_2(0)\, g_1^{-1}(z)+
 g_1(z)\, \mathcal O_{12}\,A_2\, g_1^{-1}(z)\,,
 }
 \end{array}
 \eeq
where $A$ comes from the expansion (\ref{b31}) and $g'(z)$ is the
derivative of $g(z)$ with respect to $z$.

\section{Appendix C: Explicit change of variables}
\def\theequation{C.\arabic{equation}}
\setcounter{equation}{0}

\paragraph{Change of variables.}
Here we show how to obtain (\ref{b40}) using the
factorization formula (\ref{a31}) for the Ruijsenaars-Schneider Lax matrix (\ref{a12}).
Let us compute the $a=(a_1, a_2)$-component of the Lax matrix (\ref{a33})
  \beq\label{c05}
  \begin{array}{l}
  \displaystyle{
 \mL^\eta_{ij}(z)=\frac{\vth'(0)}{\vth(\eta)}\sum\limits_{m=1}^N
 \Xi_{im}(z+N\eta,q)e^{p_m/c}\,\Xi_{mj}^{-1}(z,q)\,.
 }
 \end{array}
 \eeq
Plugging it into (\ref{c04}), we get
  \beq\label{c06}
  \begin{array}{l}
  \displaystyle{
 \mL^\eta_{a}(z)=\frac{1}{N}\,\bfe(-\frac{a_1a_2}{2N})\frac{\vth'(0)}{\vth(\eta)}\sum\limits_{k,m=1}^N
 \Xi_{km}(z+N\eta,q)e^{p_m/c}\,\Xi_{m,k+a_2}^{-1}(z,q)\,\bfe(-\frac{a_1 k}{N})\,.
 }
 \end{array}
 \eeq
From (\ref{b040}) we know that $\mL^\eta_{a}(z)=S_a\vf_a(z,\om_a+\eta)$.
Therefore, we could find $S_a$
from $\mL^\eta_{a}(z)$, which we are going to compute. Let us represent (\ref{c06}) in the form
  \beq\label{c07}
  \begin{array}{l}
  \displaystyle{
 \mL^\eta_{a}(z)=\sum\limits_{m=1}^N e^{p_m/c}\,\mL^\eta_{a;m}(z)\,,
 }
 \end{array}
 \eeq
  \beq\label{c08}
  \begin{array}{l}
  \displaystyle{
 \mL^\eta_{a;m}(z)=\frac{1}{N}\,\bfe(-\frac{a_1a_2}{2N})\frac{\vth'(0)}{\vth(\eta)}\sum\limits_{k=1}^N
 \Xi_{km}(z+N\eta,q)\Xi_{m,k+a_2}^{-1}(z,q)\,\bfe(-\frac{a_1 k}{N})\,.
 }
 \end{array}
 \eeq
 Our aim now is to evaluate the latter expression. For this purpose we
 need the properties (\ref{c09})--(\ref{c101}).
  Using explicit form of the matrix $\Xi$ (\ref{a22}), it easy to see from (\ref{c09}) that
  \beq\label{c11}
  \begin{array}{l}
  \displaystyle{
 \Xi_{km}(z+N\eta+a_1,q)=(-1)^{a_1}\bfe(-\frac{a_1 k}{N})\,\Xi_{km}(z+N\eta,q)\,.
 }
 \end{array}
 \eeq
  Therefore,
  \beq\label{c12}
  \begin{array}{l}
  \displaystyle{
 \mL^\eta_{a;m}(z)=\frac{1}{N}\,\bfe(-\frac{a_1a_2}{2N})\,(-1)^{a_1}\frac{\vth'(0)}{\vth(\eta)}\sum\limits_{k=1}^N
 \Xi_{km}(z+N\eta+a_1,q)\Xi_{m,k+a_2}^{-1}(z,q)\,.
 }
 \end{array}
 \eeq
  Next, add and subtract $a_2\tau$ to the argument of $\Xi_{km}(z+N\eta+a_1,q)$. Then using (\ref{c101})
  with $a'=-a_2/N$ we obtain
 \beq\label{c13}
 \begin{array}{c}
  \displaystyle{
\Xi_{km}(z+N\eta+a_1,q)=
 \vth\left[  \begin{array}{c}
 \frac12-\frac{k}{N} \\ \frac N2
 \end{array} \right] \left(z-N{\bar q}_m+N\eta+a_1+a_2\tau -\frac{a_2}{N}\,N\tau
 \left.\right|N\tau\right)
 }
 \\ \ \\
   \displaystyle{
   =\bfe\Big( -\frac{a_2^2}{2N}\tau+\frac{a_2}{N}(z-N{\bar q}_m+N\eta+a_1+a_2\tau+\frac{N}{2}) \Big)\,
   \Xi_{k+a_2,m}(z+N(\eta+\om_a),q)\,,
   }
 \end{array}
 \eeq
 where the notation $\om_a$ (\ref{a08}) is used.
  Plugging it into (\ref{c12}), we arrive at
  \beq\label{c14}
  \begin{array}{c}
  \displaystyle{
 \mL^\eta_{a;m}(z)=\frac{1}{N}\,\bfe(\frac{a_1a_2}{2N}+\frac{a_2^2}{2N}\tau)\,(-1)^{a_1+a_2}
 \bfe(a_2(\eta-{\bar q}_m))\bfe(z\frac{a_2}{N})
 }
 \\ \ \\
   \displaystyle{
 \times\frac{\vth'(0)}{\vth(\eta)}\sum\limits_{k=1}^N
 \Xi_{m,k+a_2}^{-1}(z,q)\Xi_{k+a_2,m}(z+N(\eta+\om_a),q)\,.
 }
 \end{array}
 \eeq
  Finally, we use  (\ref{a31}) for the Ruijsenaars-Schneider Lax matrix (\ref{a12})--(\ref{a13}).
   Namely, we need the $i=j=m$ diagonal element of (\ref{a12}) with $\eta$ being
   replaced by $\eta+\om_a$ (except for the common factor $\vth'(0)/\vth(\eta)$). This yields
  \beq\label{c15}
  \begin{array}{c}
  \displaystyle{
 \mL^\eta_{a;m}(z)=\frac{1}{N}\,\bfe(\frac{a_2}{2}\,\om_a)\,(-1)^{a_1+a_2}
 }
 \\ \ \\
   \displaystyle{
 \times\bfe(a_2(\eta-{\bar q}_m))\vf_a(z,\om_a+\eta)\frac{\vth(\eta+\om_{a})}{\vth(\eta)}
 \prod\limits_{l:\,l\neq m}^N\frac{\vth(q_m-q_l-\eta-\om_a)}{\vth(q_m-q_l)}\,.
 }
 \end{array}
 \eeq
 % \sum\limits_{m=1}^N e^{p_m/c}
Returning back to (\ref{c07}) and canceling $\vf_a(z,\om_a+\eta)$, we find the final answer
  \beq\label{c16}
  \begin{array}{c}
  \displaystyle{
 S_a=\frac{(-1)^{a_1+a_2}}{N}\,\bfe(\frac{a_2}{2}\,\om_a)\,
 \sum\limits_{m=1}^N e^{p_m/c}\bfe(a_2(\eta-{\bar q}_m))
 \frac{\vth(\eta+\om_{a})}{\vth(\eta)}\prod\limits_{l:\,l\neq m}^N\frac{\vth(q_m-q_l-\eta-\om_a)}{\vth(q_m-q_l)}\,.
 }
 \end{array}
 \eeq

\paragraph{Inverse of the
matrix $\Xi(z,q)$.} Consider the set of matrices with components (\ref{c15}) in the basis $T_a$:
  \beq\label{c17}
  \begin{array}{c}
  \displaystyle{
 \mL_{;m}^\eta(z)=\sum\limits_{a}\mL_{a;m}^\eta(z)T_a\in\Mat\,,\quad m=1,...,N\,.
 }
 \end{array}
 \eeq
It follows from its initial definition (\ref{c05}), (\ref{c07}) that the matrix elements in the standard basis
are of the form:
  \beq\label{c18}
  \begin{array}{l}
  \displaystyle{
 \mL^\eta_{ij;m}(z)=\frac{\vth'(0)}{\vth(\eta)}
 \Xi_{im}(z+N\eta,q)\Xi_{mj}^{-1}(z,q)\,.
 }
 \end{array}
 \eeq
Therefore,
  \beq\label{c19}
  \begin{array}{c}
  \displaystyle{
 \Xi_{mj}^{-1}(z,q)=\frac{\vth(\eta)}{\vth'(0)}\frac{\mL^\eta_{ij;m}(z)}{\Xi_{im}(z+N\eta,q)}\,.
 }
 \end{array}
 \eeq
 To get an explicit expression, we need to compute the matrices
 $\mL_{;m}^\eta(z)$, $m=1,...,N$ in the standard basis. For this purpose substitute (\ref{c15}) into (\ref{a491}) with $B_a=\mL^\eta_{a;m}(z)$. Both cases in the r.h.s. of (\ref{a491})
 provide the same answer (the latter is verified directly using the transformation
 properties (\ref{c09})--(\ref{c101}) for the theta-function (\ref{a25})):
  \beq\label{c20}
  \begin{array}{c}
  \displaystyle{
 \frac{\vth(\eta)}{\vth'(0)}\mL^\eta_{ij;m}(z)=\frac{1}{N}\sum\limits_{a_1=0}^{N-1}
 \bfe\Big(\frac{a_1}{2N}(i+j)\Big)\bfe\Big(\frac{j-i}{2}\,\om_{(a_1,j-i)}\Big)\,(-1)^{a_1+j-i}
 }
 \\ \ \\
   \displaystyle{
 \times\bfe\Big((j-i)(\eta-{\bar q}_m)\Big)\bfe\Big(z\frac{j-i}{N}
 \Big)\frac{\vth(z+\eta+\om_{(a_1,j-i)})}{\vth(z)}
 \prod\limits_{l:\,l\neq m}^N\frac{\vth(q_m-q_l-\eta-\om_{(a_1,j-i)})}{\vth(q_m-q_l)}\,,
 }
 \end{array}
 \eeq
 where
$
\om_{(a_1,j-i)}=\frac{a_1+(j-i)\tau}{N}
$. Dividing this expression by $\Xi_{im}(z+N\eta,q)$ we obtain $\Xi_{mj}^{-1}(z,q)$ (\ref{c19}).
Notice that
by construction the r.h.s. of (\ref{c19}) is independent of $\eta$, so we put $\eta=0$ in the final answer since
all entering functions are regular in $\eta$. Also, the r.h.s. of (\ref{c19}) is independent of index $i$. We fix it as $i=N$. Finally, using $\Xi_{Nm}(z+N\eta,q)=-\vth(z+\frac{N-1}{2}+N\eta-N{\bar q}_m|N\tau)$ we obtain
  \beq\label{c21}
  \begin{array}{c}
  \displaystyle{
 \Xi_{ij}^{-1}(z,q)=\frac{(-1)^{j+1}}{N\vth(z+\frac{N-1}{2}-N{\bar q}_i|N\tau)}
 }
 \\ \ \\
  \displaystyle{\times
  \sum\limits_{a_1=0}^{N-1}
 (-1)^{a_1}\bfe\Big(\frac{a_1j}{2N}+j\,\frac{a_1+j\tau}{2N}\Big)
 \bfe(-j{\bar q}_i)\bfe\Big(z\frac{j}{N}\Big)\frac{\vth(z+\frac{a_1+j\tau}{N})}{\vth(z)}
 \prod\limits_{l:\,l\neq i}^N\frac{\vth(q_i-q_l-\frac{a_1+j\tau}{N})}{\vth(q_i-q_l)}\,.
 }
 \end{array}
 \eeq
Equivalently, for the matrix $g(z,q)$ (\ref{a21}) we have
  \beq\label{c22}
  \begin{array}{c}
  \displaystyle{
 g_{ij}^{-1}(z,q)=\frac{(-1)^{j+1}}{N\vth(z+\frac{N-1}{2}-N{\bar q}_i|N\tau)}
 }
 \\ \ \\
  \displaystyle{\times
  \sum\limits_{a_1=0}^{N-1}(-1)^{a_1}
 \bfe\Big(\frac{a_1j}{2N}+j\,\frac{a_1+j\tau}{2N}\Big)
 \bfe(-j{\bar q}_i)\bfe\Big(z\frac{j}{N}\Big)\frac{\vth(z+\frac{a_1+j\tau}{N})}{\vth(z)}
 \prod\limits_{l:\,l\neq i}^N\vth \Big (q_i-q_l-\frac{a_1+j\tau}{N}\Big )\,.
 }
 \end{array}
 \eeq

\paragraph{Derivation of $S=\xi\otimes \psi$.} The matrix ${\breve g}(0,q)$ (\ref{b31}) is easily calculated from
(\ref{c22}):
  \beq\label{c23}
  \begin{array}{c}
  \displaystyle{
 {\breve g}_{ij}(0,q)=\frac{(-1)^{j}}{N\vth(\frac{N-1}{2}-N{\bar q}_i|N\tau)}
 }
 \\ \ \\
  \displaystyle{\times
  \bfe\Big(\frac{j^2\tau}{2N}\Big)\bfe(-j{\bar q}_i)\frac{1}{\vth'(0)}\sum\limits_{a_1=0}^{N-1}(-1)^{a_1}
 \bfe\Big(\frac{a_1j}{N}\Big)
 \prod\limits_{l=1}^N\vth \Big (q_i-q_l-\frac{a_1+j\tau}{N}\Big )\,.
 }
 \end{array}
 \eeq
Note that due to (\ref{b32}) the r.h.s. of (\ref{c23}) is independent of the index $i$, so that
the functions
  \beq\label{c24}
  \begin{array}{c}
  \displaystyle{
 f_j(x,q)=\frac{\bfe(-jx)}{\vth(\frac{N-1}{2}-Nx|N\tau)}
  \sum\limits_{a_1=0}^{N-1}(-1)^{a_1}
 \bfe\Big(\frac{a_1j}{N}\Big)
 \prod\limits_{l=1}^N\vth \Big (x-{\bar q}_l-\frac{a_1+j\tau}{N}\Big )
 }
 \end{array}
 \eeq
obey the property
  \beq\label{c25}
  \begin{array}{c}
  \displaystyle{
 f_j(q)=f_j(q_i,q)=f_j(q_k,q)\quad \mbox{for all $i,j,k$}\,.
 }
 \end{array}
 \eeq
 In this notation
  \beq\label{c26}
  \begin{array}{c}
  \displaystyle{
 {\breve g}_{ij}(0,q)=\frac{(-1)^{j}}{N\vth'(0)}\,\bfe\Big(\frac{j^2\tau}{2N}\Big)f_j(q)\,.
 }
 \end{array}
 \eeq
 Finally, using (\ref{b37})-(\ref{b38}), we find the change of variables (\ref{c16}) in the form
 $S=\xi\otimes\psi$ with
  \beq\label{c27}
  \begin{array}{c}
  \displaystyle{
  \xi_i=\frac{\vth'(0)}{\vth(\eta)}\sum\limits_{k=1}^N
 g_{ik}(N\eta)\,e^{p_k/c}\,,\qquad
 \psi_j=\frac{(-1)^{j}}{N\vth'(0)}\,\bfe\Big(\frac{j^2\tau}{2N}\Big)f_j(q)\,.
 }
 \end{array}
 \eeq
The normalization can be chosen in a different way since (\ref{c27}) is defined up to
 $\xi_i\rightarrow \lambda\xi_i$, $\psi_j\rightarrow \psi_j/\lambda$. Let us also mention
 that in the rational case $\psi_j$ are elementary symmetric functions of coordinates (see \cite{LOZ8}), so that
 (\ref{c27}) provides its elliptic analogue.

%%%%%%%%%%%%%%%%%%%%%%%%%%%%%%%%%%%%%%%%%%%%%%%%%%%%%%%%%%%%%%%%%%%%%%%%%%%%%%%%%%%%%%%%%%%%%%%%%%%%%%
%%%%%%%%%%%%%%%%%%%%%%%%%%%%%%%%%%%%%%%%%%%%%%%%%%%%%%%%%%%%%%%%%%%%%%%%%%%%%%%%%%%%%%%%%%%%%%%%%%%%%%

\section*{Acknowledgments}

\addcontentsline{toc}{section}{\hspace{6mm}Acknowledgments}

The work of A. Zabrodin has been funded within the framework of the
HSE University Basic Research Program.
%The work of A. Zotov was performed at the Steklov International Mathematical Center and supported by the Ministry of %Science and Higher Education of the Russian Federation (agreement no. 075-15-2019-1614).

\end{document}